\documentclass[final]{siamltex}
\usepackage[utf8]{inputenc}
\usepackage[active]{srcltx}
\usepackage{amsfonts,amsmath,amssymb,graphicx}
\usepackage{multirow}
\usepackage{array}
\usepackage{enumitem}
\usepackage{type1cm}
\usepackage{bbm}
\usepackage{verbatim}
\newcommand{\uu}{\mathbf{u}}
\newcommand{\nn}{\mathbf{n}}

\newcommand{\nab}{\boldsymbol{\nabla}}

\newcommand{\RR}{\mathbb{R}}
\newcommand{\TT}{\mathbb{T}}
\newcommand{\HH}{\mathbf{H}}
\newcommand{\VV}{\mathbf{V}}

\newcommand{\kk}{\mathbf{k}}

\newcommand{\xx}{\mathbf{x}}
\newcommand{\yy}{\mathbf{y}}
\newcommand{\Rey}{\mathrm{Re}}

\newcommand{\ttau}{\boldsymbol{\tau}}
\newcommand{\pphi}{{\mathbf{u}}}

\title{
Approximation of the Laplace and Stokes operators with Dirichlet boundary conditions through volume penalization: a spectral viewpoint
\footnotemark[1]\ \footnotemark[2]
}

\author{R. Nguyen van yen \footnotemark[3]\ \footnotemark[4] \and D. Kolomenskiy \footnotemark[5]\ \footnotemark[6]  \and K. Schneider \footnotemark[5]\ \footnotemark[7]} 
\begin{document}
\maketitle

\footnotetext[1]{
The authors would like to thank Claude Bardos and Zhou-Ping Xin for fruitful discussions, 
and the French Federation for Fusion Studies as well as the PEPS program of CNRS-INSMI for financial support.
}
\footnotetext[2]{
This work, 
supported by the European Communities under the contract of Association between EURATOM, 
CEA and the French Research Federation for fusion studies, 
was carried out within the framework of the European Fusion Development Agreement. 
The views and opinions expressed herein do not necessarily reflect those of the European Commission.
}
\footnotetext[3]{Institut f\"ur Mathematik, Freie Universit\"at Berlin, Arnimallee 6, 14185 Berlin, Germany. 
RNVY thanks the Humboldt foundation for a post-doctoral fellowship.}
\footnotetext[4]{LMD--CNRS, \'Ecole Normale Sup\'erieure, 24 rue Lhomond, 75231 Paris cedex 05, France}
\footnotetext[5]{M2P2--CNRS, Universit\'e d'Aix-Marseille, 39, rue Frédéric Joliot-Curie, 13453 Marseille cedex 13, France}
\footnotetext[6]{CERFACS, 42 Avenue Gaspard Coriolis, 31057 Toulouse cedex 01, France}
\footnotetext[7]{CMI, Universit\'e d'Aix-Marseille, 39 rue Frédéric Joliot-Curie, 13453 Marseille cedex 13, France}


\begin{abstract}
We report the results of a detailed study of the spectral properties of Laplace and Stokes operators, 
modified with a volume penalization term designed to approximate Dirichlet conditions in the limit when a penalization parameter, $\eta$, tends to zero.
The eigenvalues and eigenfunctions are determined either analytically or numerically as functions of $\eta$, 
both in the continuous case and after applying Fourier or finite difference discretization schemes.
For fixed $\eta$, we find that only the part of the spectrum corresponding to eigenvalues $\lambda \lesssim \eta^{-1}$ approaches
Dirichlet boundary conditions, while the remainder of the spectrum is made of uncontrolled, spurious wall modes.
The penalization error for the controlled eigenfunctions is estimated as a function of $\eta$ and $\lambda$.
Surprisingly, in the Stokes case, we show that the eigenfunctions approximately satisfy, with a precision $O(\eta)$, Navier slip boundary conditions 
with slip length equal to $\sqrt{\eta}$.
Moreover, for a given discretization, we show that there exists a value of $\eta$, corresponding to a balance between penalization and discretization errors,
below which no further gain in precision is achieved.
These results shed light on the behavior of volume penalization schemes when solving the Navier-Stokes equations,
outline the limitations of the method, 
and give indications on how to choose the penalization parameter in practical cases.
\end{abstract}







\section{Introduction}

Penalization approaches, as reviewed for example in Ref.~\cite{Peskin2002,Mittal2005},
are nowadays commonly employed to solve boundary or initial-boundary value problems, as encountered for example in computational fluid dynamics.
They consist in embedding the original, possibly complex spatial domain inside a bigger domain having a simpler geometry, for example a torus,
while keeping the boundary conditions approximately enforced thanks to new terms that are added to the equations.
One particular example is the volume penalization method which, 
inspired by the physical intuition that a solid wall in contact with a fluid is similar to a vanishingly porous medium,
uses the Brinkman-Darcy drag force \cite{Brinkman1949,Arquis1984} as penalization term.
Due to this analogy, the method is sometimes called Brinkman penalization.
It has been mathematically justified in academic cases \cite{Angot1999,Carbou2003},
and its use has rapidly spread to various domains of scientific computing.

The main advantage of such penalized equations is that they can be discretized independently of the geometry of the original problem, 
as the latter has been encoded into the penalization terms.
Such a simplification permits a massive reduction in solver development time,
since it avoids the issues associated to the design and management of the grid,
allowing for example the use of simple spectral solvers in Cartesian geometries \cite{Forestier2000,Kevlahan2001,Schneider2005a}.
The gain becomes even more substantial when the geometry is time-dependent, as in the case of moving obstacles \cite{Kolomenskiy2009,Kolomenskiy2011}, 
or when fluid-structure interaction is taken into account.
However, depending on the desired accuracy, it may happen that the numerical resolution, and hence computational time,
required when using volume penalization grow disproportionately and even completely offset
the gains.
To afford a reasonable way of deciding whether one should implement a penalization method and, if so, how to tune its parameters, 
realistic order estimates are therefore of primary importance, taking into account not only the penalization error but also the discretization error.

Even in the most simple case, which is the penalized Poisson equation, such estimates are either still lacking or not widely understood.
Indeed, to see where the main issues come from, consider the following one-dimensional example:
\begin{equation}
  -w''(x) = m^2 \sin mx,
\label{eq:poisson_eq}
\end{equation}
where the unknown $w$ is a twice differentiable function on $\Omega = ]0,\pi[$ satisfying homogeneous Dirichlet boundary conditions,
$
  w(0) = w(\pi)= 0
\label{eq:poisson_dirichlet_boco}
$, 
and $m$ is an integer.
The exact solution is
\begin{equation}
  w(x) = \sin mx,
\label{eq:poisson_exact_solution}
\end{equation}
but forgetting it for a moment, let us approximate it using the volume penalization method.
To this end, we extend the original domain $\Omega$ by using its natural embedding into the one-dimensional torus $\mathbb{T} = \mathbb{R}/2\pi\mathbb{Z}$,
and then look for a function on $\mathbb{T}$ satisfying
\begin{equation}
  -v'' + \frac{1}{\eta} \chi v = m^2 \sin mx,
\label{eq:penalized_poisson_eq}
\end{equation}
where $\eta > 0$ is the penalization parameter, and the mask function $\chi$ is defined by
\begin{equation}
  \chi(x) = \left\{
  \begin{array}{ll}
  0  & \mathrm{\,if\,} x \in ]0,\pi[ \\
  1/2  & \mathrm{\,if\,} x=0 \mathrm{\,or\,} x=\pi \\
  1  & \mathrm{\,otherwise.}
  \end{array}
  \right.
\label{eq:mask}
\end{equation}
The discontinuity of $\chi$ is the essential mathematical feature of such penalized problems,
implying in particular that the unique weak $H^1$ solution to (\ref{eq:penalized_poisson_eq}) is not a classical solution,
which in turn makes the problem hard to discretize efficiently.
Note that, since $\chi = 0$ on $]0,\pi[$, the restriction to $]0,\pi[$ of any solution to (\ref{eq:penalized_poisson_eq}) solves the original Poisson equation (\ref{eq:poisson_eq}) exactly.
Its discrepancy with the exact solution of the full Dirichlet problem is therefore entirely due to
its non-zero boundary values.

Taking advantage of the fact that the solution can be written analytically in this simple case (see Appendix A),
one can show that the leading order $L^2$ error $\varepsilon_1$ with respect to the solution (\ref{eq:poisson_exact_solution}) of the Dirichlet problem is:
\begin{equation}
  \displaystyle \varepsilon_1 \sim \frac{m \sqrt{2-(-1)^m}}{\sqrt{6}} \sqrt{\eta} \quad \mathrm{as}
  \quad \eta \to 0
  \label{eq:penalization_error_norm_asy0}
\end{equation}
where the $\sqrt{\eta}$ behavior is consistent with previous studies \cite{Angot1999,Carbou2003}.

In order to discretize (\ref{eq:penalized_poisson_eq}), it is possible to use a Fourier pseudo-spectral method,
where collocation on a regular grid of $N$ points is used to evaluate the product $\chi v$.
The $L^2$ error between the discrete solution and the exact solution of the penalized problem can then be estimated using 
a mixture of analytical and numerical techniques (see Appendix A), and yields
\begin{equation}
  \varepsilon_2 \sim K \frac{m \pi^{3/2}}{3\sqrt{2}} \frac{1}{\sqrt{\eta} N^2} \quad \mathrm{as} \quad \eta \to 0,\quad \sqrt{\eta}N^2 \to \infty,
\label{eq:norm_error_asy0}
\end{equation}
where $K=2$ for $m$ even and $K \approx 3.8423$ for $m$ odd. 
The $N^{-2}$ behavior  is related to the regularity of the exact penalized solution, 
and was indeed observed for the same reason in a study of collocation schemes for elliptic equations with discontinuous coefficients \cite{Min2003}.

Combining the two estimates (\ref{eq:penalization_error_norm_asy0})  and (\ref{eq:norm_error_asy0}) thanks to the triangle inequality,
we obtain a bound for the total error $\varepsilon$ between the discrete-penalized solution and the exact solution of the original Dirichlet boundary-value problem:
\begin{equation}
  \varepsilon \leq \varepsilon_1 + \varepsilon_2 \sim \frac{m \sqrt{2-(-1)^m}}{\sqrt{6}} \sqrt{\eta} + K \frac{m \pi^{3/2}}{3\sqrt{2}} \frac{1}{\sqrt{\eta} N^2} \quad \mathrm{as} \quad \eta \to 0,\quad \sqrt{\eta}N^2 \to \infty.
\label{eq:total_norm_error_asy0}
\end{equation}
When $\eta$ is chosen with the right order of magnitude, i.e. $\eta \propto 1/N^2$, in order to optimize 
the preceding estimate, then the resulting error is
\begin{equation}
  \varepsilon \propto \frac{1}{N},
\label{eq:total_norm_error_final}
\end{equation}
which suggests that the penalization method with Fourier discretization is, for practical matters, a first order method,
a simple fact that has been mostly overlooked by the literature on the subject.

Motivated by the results obtained for the 1D Poisson problem, we have sought to extend them and understand
whether they should be seen as general properties of penalized equations or merely as a special case.
For example, one may ask what the influence of the terms on the left-hand side of the equation, namely the Laplace operator, is
on the accuracy of the penalization scheme.
Since one of the main domains of application of penalization methods is fluid dynamics, 
we have also oriented ourselves towards vector-valued penalized problems,
and more specifically the penalized Stokes equation,
which plays a role in the penalized incompressible Navier-Stokes equations. 

Another question concerns the dependency on the right hand side of the equation.
To obtain a general answer, which characterizes the discrepancy between the original and penalized operators over the full range of scales,
our approach in the following is to compare their eigenvalues and eigenfunctions.
Note that the eigenvalue problem in itself may also be of interest for some applications.
In fact, the spectral analysis of the penalized Laplace operator
is analog to the analysis of the semi-classical limit for the Schr\"odinger operator \cite{Berry1972,Cohen-Tannoudji1977}
in a finite potential well, familiar from quantum mechanics textbooks.
The penetration of the eigenfunctions within the penalized region corresponds to the quantum tunnel effect.
By computing the convergent modes and monitoring their convergence rates,
we obtain precise quantitative information on the effective accuracy of penalization methods,
as well as on the properties of the inevitable spurious modes fostered by the extension of the domain.
Then, by carrying out the same procedure after discretization, the properties of several schemes can be compared,
providing realistic error estimates that can be used to evaluate the benefits of a penalization scheme.

To achieve these objectives, the present work is organized as follows.
First, the eigenvalue problems themselves are formalized, and some important notations are introduced.
Then the eigenvalues and eigenfunctions for the penalized Laplace operator in a 1D interval are derived using a semi-analytical method.
A numerical study of their behavior as the penalization parameter tends to zero is then conducted.
After that, a similar analysis of the penalized Stokes operator, defined for simplicity on a straight 2D channel, is performed.
A relationship between the penalized eigenfunctions and Navier boundary conditions is outlined. 
Subsequently, several discretization schemes are compared by studying the eigenfunctions and eigenvalues of the discretized-penalized operators numerically.
Thanks to an analysis of the penalization and discretization errors,
efficient ways of choosing the parameters for realistic applications of the penalization method are proposed.
Finally, the practical relevance of the preceding results is further supported by analyzing the convergence 
of the penalization method for a 2D Navier-Stokes test problem.

\section{Mathematical setting}

For $\Omega$ a regular bounded open subset of $\RR^d$,
we shall denote respectively by $\Lambda^o$ and $A^o$
the Laplace and Stokes operators on $\Omega$, with their classical definitions:
\begin{equation}
\label{eq:laplace_operator}
\Lambda^o: \begin{cases} 
	D(\Lambda^o) := H^2(\Omega) \cap H^1_0(\Omega) \to L^2(\Omega)\\
	\psi \to -\Delta \psi
     \end{cases}
\end{equation}
and, with $\HH_0$ and $\VV_0$ being the closures in $L^2(\Omega)^d$ and $H^1(\Omega)^d$ (respectively) of 
$
\{ \uu \in \mathcal{C}_0^\infty(\Omega) \mid \nab \cdot \uu = 0 \}
$,
\begin{equation}
\label{eq:stokes_operator}
A^o: \begin{cases} 
	D(A^o) := (H^2(\Omega))^d \cap \VV_0 \to \HH_0 \\
	\uu \to -\mathbb{P}_0 \Delta \uu,
     \end{cases}
\end{equation}
where $\mathbb{P}_0$, called the Leray projector \cite{Foias2001}, is the
orthogonal projector on $\HH_0$ seen as a sub-Hilbert space of $L^2(\Omega)^d$.

$\Lambda^o$ and $A^o$ are positive, self-adjoint operators, with compact inverses,
and therefore one can construct Hilbert bases in $L^2(\Omega)$  with their eigenfunctions,
which we denote respectively by $(\psi_{i}^o)_{i\in \mathbb{N}^*}$ and $(\pphi_{i}^o)_{i\in \mathbb{N}^*}$.
The corresponding eigenvalues, sorted in increasing order, are denoted by $(\lambda_{i}^o)_{i\in \mathbb{N}^*}$ and $(\mu_{i}^o)_{i\in \mathbb{N}^*}$.

Now assume that, thanks to a suitable similarity transformation, 
$\Omega$ has been identified with a subset of the $d$-dimensional torus $\TT^d = \left(\RR/\mathbb{Z}\right)^d$.
The penalization method then consists in approximating
$\Lambda^o$ by the operator $\Lambda^*_\eta$ defined on $\TT^d$ as follows:
\begin{equation}
\label{eq:penalized_laplace_operator}
\Lambda^*_{\eta}: 
\begin{cases} 
D(\Lambda^*_{\eta}) :=  H^2(\TT^d)  \to L^2(\TT^d) \\
 \psi \to -\Delta \psi + \frac{1}{\eta}\chi \psi,
 \end{cases}
\end{equation}
where $\eta>0$ is the penalization parameter, and $\chi$ is the indicator function of $S = \TT^d \backslash \Omega$,
also called mask function.
The term $\frac{1}{\eta}\chi \psi$, with $\eta \ll 1$, is called penalization term.
Note that, here and in the following, the $*$ in the exponent serves to distinguish penalized operators and related quantities,
while the $o$ stands for Dirichlet boundary conditions.
Letting 
\begin{equation}
\HH  =  \{ \uu \in L^2(\TT^d)^d \mid \nab \cdot \uu = 0 \} ,\quad
\VV  =  \HH \cap H^1(\TT^d)^d,
\end{equation}
the penalized Stokes operator is defined in a similar way as follows:
\begin{equation}
\label{eq:penalized_stokes_operator}
A^*_{\eta}: 
\begin{cases} 
D(A^*_{\eta}) :=  (H^2(\TT^d))^d \cap \VV \to \HH \\
 \uu \to -\Delta \uu + \frac{1}{\eta} \mathbb{P}(\chi \uu),
 \end{cases}
\end{equation}
where $\mathbb{P}$ is the Leray projector from $L^2(\TT^d)^d$ onto $\HH$.

Like their exact counterparts $\Lambda^o$ and $A^o$, $\Lambda^*_\eta$ and $A^*_\eta$ are linear mappings with compact inverses and thus enjoy Hilbert bases of eigenfunctions, 
denoted respectively $(\psi^*_{\eta,i})_{i\in \mathbb{N}^*}$ and $(\pphi^*_{\eta,i})_{i\in \mathbb{N}^*}$, while $(\lambda^*_{\eta,i})_{i\in \mathbb{N}^*}$ and $(\mu^*_{\eta,i})_{i\in \mathbb{N}^*}$ are the corresponding eigenvalues sorted in increasing order.
Whereas, according to classical theorems, the eigenfunctions of the Laplace and Stokes operators are analytic in $\Omega$ if $\Omega$ is regular enough, 
this property is not enjoyed by the penalized eigenfunctions, due to the discontinuity of the mask function.
Indeed, although by construction the eigenfunctions are in $H^2(\TT^d)^d$, they are not in $C^2(\TT^d)^d$ as we shall see below.
In the upcoming sections, the relationship between the spectra of $\Lambda^o$ and $\Lambda^*_\eta$ on the one hand,
and of $A^o$ and $A^*_\eta$ on the other hand, are studied using analytical and numerical techniques.

\section{Spectral analysis of penalized operators}

	\subsection{Laplace operator}

In the case of the Laplace operator, a 1D analysis is sufficient to obtain a qualitative understanding of the behavior for sufficiently simple domains.
We therefore proceed to compute the eigenfunctions of $\Lambda^*_\eta$ for $d=1$ and $\Omega = ]0,\pi[$.
As a side remark, we may note the analogy with the classical eigenvalue problem for the Schr\"odinger equation (see \cite[p. 67]{Cohen-Tannoudji1977}):
\begin{equation}
 -\frac{\hbar^2}{2m} \Delta \psi + V \psi = E \psi,
\end{equation}
where $\frac{\hbar^2}{2m}$ corresponds to $\eta$, and the potential $V$ to the mask function $\chi$.
The limit $\eta \to 0$ that we study here is then analogous to the semi-classical limit $\hbar \to 0$.

\subsubsection{Semi-analytic computation of eigenvalues and eigenfunctions}

In the 1D case, the eigenvalue problem reduces to a reconnection problem for ordinary differential equations.
Indeed, the equations satisfied by $\psi^*_{\eta,i}$ and $\lambda^*_{\eta,i}$ can be rewritten as follows:
\begin{equation}
\label{eq:laplace_1d_eq}
\begin{cases}
-\psi'' = \lambda \psi \mathrm{\ on \ } \left]0,\pi\right[ \\
-\psi'' = \left(\lambda-\frac{1}{\eta}\right) \psi \mathrm{\ on \ } \left]\pi,2\pi\right[.
\end{cases}
\end{equation}

From the form of the second equation, it appears immediately that the 
behavior will be entirely different according to whether $\lambda < \eta^{-1}$ or $\lambda > \eta^{-1}$.
Let us first assume that $\lambda < \eta^{-1}$.
Applying the classical theory, we look for $C^1$ solutions under the form  
\begin{equation}
\label{eq:laplace_1d_eigenfunctions}
\psi(x) =
\begin{cases}
A_0 \sin\left(q_0\left(x-\frac{\pi}{2}\right)\right) + B_0 \cos\left(q_0\left(x-\frac{\pi}{2}\right)\right) \mathrm{\ for\ } x \in \left]0,\pi\right[ \\ 
A_1 \left(e^{q_1\left(x - 2\pi\right)} - e^{q_1\left(\pi-x\right)} \right) +  B_1 \left(e^{q_1\left(x - 2\pi\right)} + e^{q_1\left(\pi-x\right)}\right) \mathrm{\ for\ } x \in \left]\pi,2\pi\right[,
\end{cases}
\end{equation}
where $q_0 = \sqrt{\lambda}$, $q_1 = \sqrt{\eta^{-1}-\lambda}$, and the particular form of the terms has been tailored to avoid ill-conditionning of upcoming matrices.
Since the symmetry with respect to $x=\frac{\pi}{2}$ leaves the problem unchanged, 
it should either leave the eigenfunctions invariant or switch their sign,
which means that we can isolate two families of solutions, having either $B_0 = B_1 = 0$ or $A_0 = A_1 = 0$.
 
The $C^1$ reconnection conditions at $x=\pi$ yield the following systems of equations for 
the antisymmetric and symmetric cases, respectively:
\begin{subequations}
\label{eq:laplace_1d_system}
\begin{align}
(S_-) : {} &
\begin{cases}
A_0 \sin\left(\frac{\pi q_0}{2}\right) + A_1 \left(1-e^{-\pi q_1 }\right) = 0\\
A_0 q_0 \cos\left(\frac{\pi q_0}{2}\right) - A_1 q_1 \left(1 + e^{-\pi q_1} \right) = 0,
\end{cases} \\
(S_+) : {} &
\begin{cases}
B_0 \cos\left(\frac{\pi q_0}{2}\right) - B_1 \left( 1 + e^{-\pi q_1 }\right)  = 0\\
-B_0 q_0 \sin\left(\frac{\pi q_0}{2}\right) + B_1 q_1 \left(1-e^{-\pi q_1 } \right) = 0.
\end{cases}
\end{align}
\end{subequations}
Explicit computation of the determinants of these systems yield the following eigenvalue equations:
\begin{subequations}
\begin{align}
F^*_-(\lambda,\eta) ={} & \sin\left(\frac{\pi\sqrt{\lambda}}{2}\right) + \sqrt{\frac{\eta\lambda}{1-\eta\lambda}} \cos\left(\frac{\pi\sqrt{\lambda}}{2}\right) \mathrm{tanh}\left(\frac{\pi}{2}\sqrt{\frac{1}{\eta}-\lambda}\right) = 0 , \\
F^*_+(\lambda,\eta) ={} & \cos\left(\frac{\pi\sqrt{\lambda}}{2}\right) + \sqrt{\frac{\eta\lambda}{1-\eta\lambda}} \sin\left(\frac{\pi\sqrt{\lambda}}{2}\right) \mathrm{cotanh}\left(\frac{\pi}{2}\sqrt{\frac{1}{\eta}-\lambda}\right) = 0 ,
\end{align}
\end{subequations}
and the vanishing of $F^*_{\pm}(\lambda,\eta)$ is then equivalent to the existence of a non-zero $C^1$ solution to (\ref{eq:laplace_1d_eq}),
or in other words to $\lambda$ being an eigenvalue of $\Lambda^*_\eta$.
For fixed $\eta$, $F_{\pm}(\lambda,\eta)$ are oscillatory functions of $\lambda$ which both admit an infinite sequence of positive roots.
After elimination of the root $\lambda = 0$ of $F^*_-$ which leads to a vanishing $\psi$ and is therefore not an eigenvalue,
the remaining strictly positive roots constitute the eigenvalues $\lambda^*_{\eta,i}$ that we were looking for.
With $\lambda$ solution of $F^*_{\pm}(\lambda,\eta) = 0$, admissible values of $A_0,A_1$ or $B_0, B_1$ are then readily 
obtained by solving (\ref{eq:laplace_1d_system}).
In the present case, this can be done analytically very simply, but it is not useful to give the resulting expressions here,
since we are mostly interested in the behavior as $\eta \to 0$ which is studied further down.
The final values of the constants are obtained after renormalizing the functions in the $L^2$ norm,
finally yielding $\psi^*_{\eta,i}$.

In the case $\lambda \geq \eta^{-1}$, the Ansatz (\ref{eq:laplace_1d_eigenfunctions}) remains valid but with a complex value of $q_1$,
$q_1 = \iota \sqrt{\lambda-\eta^{-1}}$, where $\iota = \sqrt{-1}$.
Therefore, the corresponding eigenfunctions oscillate in $]\pi,2\pi[$ as well as in $]0,\pi[$, cannot approximate the Dirichlet eigenfunctions,
and should be considered as spurious modes generated by the penalization method.

\subsubsection{Behavior of eigenvalues and eigenfunctions as $\eta \to 0$}

As expected, when $\eta \to 0$ for a fixed $\lambda$, the eigenvalue problem for the Laplace operator in $\Omega = ]0,\pi[$ is recovered,
with the determinants
\begin{subequations}
\begin{align}
F^o_{-}(\lambda) = {} & \lim_{\eta \to 0} F^*_{-}(\lambda,\eta) = \mathrm{sin}\left(\frac{\pi \sqrt{\lambda}}{2}\right) , \\
F^o_{+}(\lambda) = {} & \lim_{\eta \to 0} F^*_{+}(\lambda,\eta) = \mathrm{cos}\left(\frac{\pi \sqrt{\lambda}}{2}\right) ,
\end{align}
\end{subequations}
which, excluding $\lambda = 0$ as above, admit the sequence of roots $\lambda^o_i = (i+1)^2$ $(i \in \mathbb{N})$.
Moreover, using symbolic computation, we have obtained that for all $\lambda > 0$:
\begin{equation}
\frac{\partial F_{\pm}}{\partial \sqrt{\eta}} \left(\frac{\partial F_{\pm}}{\partial {\lambda}}\right)^{-1} \xrightarrow[\eta\to 0]{} \frac{4}{\pi}\lambda,
\end{equation}
which, by the implicit function theorem, implies that
\begin{equation}
\label{eq:laplace_eigenvalue_convergence}
\lambda^*_{\eta,i} = \lambda^o_i \left( 1 - \frac{4}{\pi} \sqrt{\eta} + O(\eta) \right) \mathrm{\ when\ } \eta \to 0.
\end{equation}

This estimate can then be used to derive the error on the eigenfunctions.
For this, we first establish the following useful lemma:
\begin{lemma}
\label{lemma:expansion}
 Let $f : \RR \times \overline{\Omega} \to \RR^d$ be a continuous function which is $C^1$ in the first variable.
When $\xi \to \xi_0$, the following asymptotic expansion holds:
{\small
\begin{equation}
\left\Vert \frac{f(\xi,\cdot)}{\Vert f(\xi,\cdot) \Vert} - \frac{f(\xi_0,\cdot)}{\Vert f(\xi_0,\cdot) \Vert} \right\Vert = \vert \xi - \xi_0 \vert \left(\frac{ \left \Vert \partial_\xi f(\xi_0,\cdot)\right\Vert^2 }{\Vert f(\xi_0,\cdot) \Vert^2} - \frac{\left\langle f(\xi_0,\cdot) \mid \partial_\xi f(\xi_0,\cdot) \right\rangle^2 }{\Vert f(\xi_0,\cdot) \Vert^{{4}}}\right)^{\frac{1}{2}}+ o(\vert \xi - \xi_0 \vert)
\end{equation}
}
\end{lemma}
\textit{Proof.}
{ \small 
The relationship follows immediately from a direct computation:
letting $\varepsilon = \xi-\xi_0$, we have
\begin{subequations}
\begin{align}
\Vert f(\xi,\cdot) \Vert^2 -\Vert f(\xi_0,\cdot) \Vert^2 &=  \varepsilon \int \partial_\xi (f^2)(\xi_0,\cdot) + o(\varepsilon) \\
\Vert f(\xi,\cdot) \Vert^{-1} \Vert f(\xi_0,\cdot) \Vert^{-1}  &= - \frac{\varepsilon}{2}  \Vert f(\xi_0,\cdot) \Vert^{-3} \int \partial_\xi (f^2)(\xi_0,\cdot)+ o(\varepsilon)\\
\frac{f(\xi,x)}{\Vert f(\xi,\cdot) \Vert} - \frac{f(\xi_0,x)}{\Vert f(\xi_0,\cdot) \Vert} &=  \varepsilon \frac{ \partial_\xi f(\xi_0,\cdot) \Vert f(\xi_0,\cdot) \Vert^2 - \frac{1}{2} f(\xi_0,x) \int \partial_\xi (f^2)(\xi_0,\cdot) }{\Vert f(\xi_0,\cdot) \Vert^{{3}}}+ o(\varepsilon) \qquad \endproof
\end{align}
\end{subequations}
}

Since in the antisymmetric case we have $\partial_{q_0} \psi = A_0 \left(x-\frac{\pi}{2}\right) \cos \left(q_0 \left(x-\frac{\pi}{2}\right) \right)$, we obtain from 
Lemma~\ref{lemma:expansion} and (\ref{eq:laplace_eigenvalue_convergence}) that
\begin{align}
\label{eq:laplace_eigenfunctions_convergence}
\Vert \psi^*_{\eta,i} -  \psi^o_{i} \Vert &= \sqrt{\lambda^o_i \eta} \left( \frac{\pi}{6} + \frac{1}{\pi\lambda^o_i}\left(1 - \frac{2}{\pi^2}\right) \right)^{\frac{1}{2}}.
\end{align}

\begin{figure}
\includegraphics[width=0.32\columnwidth]{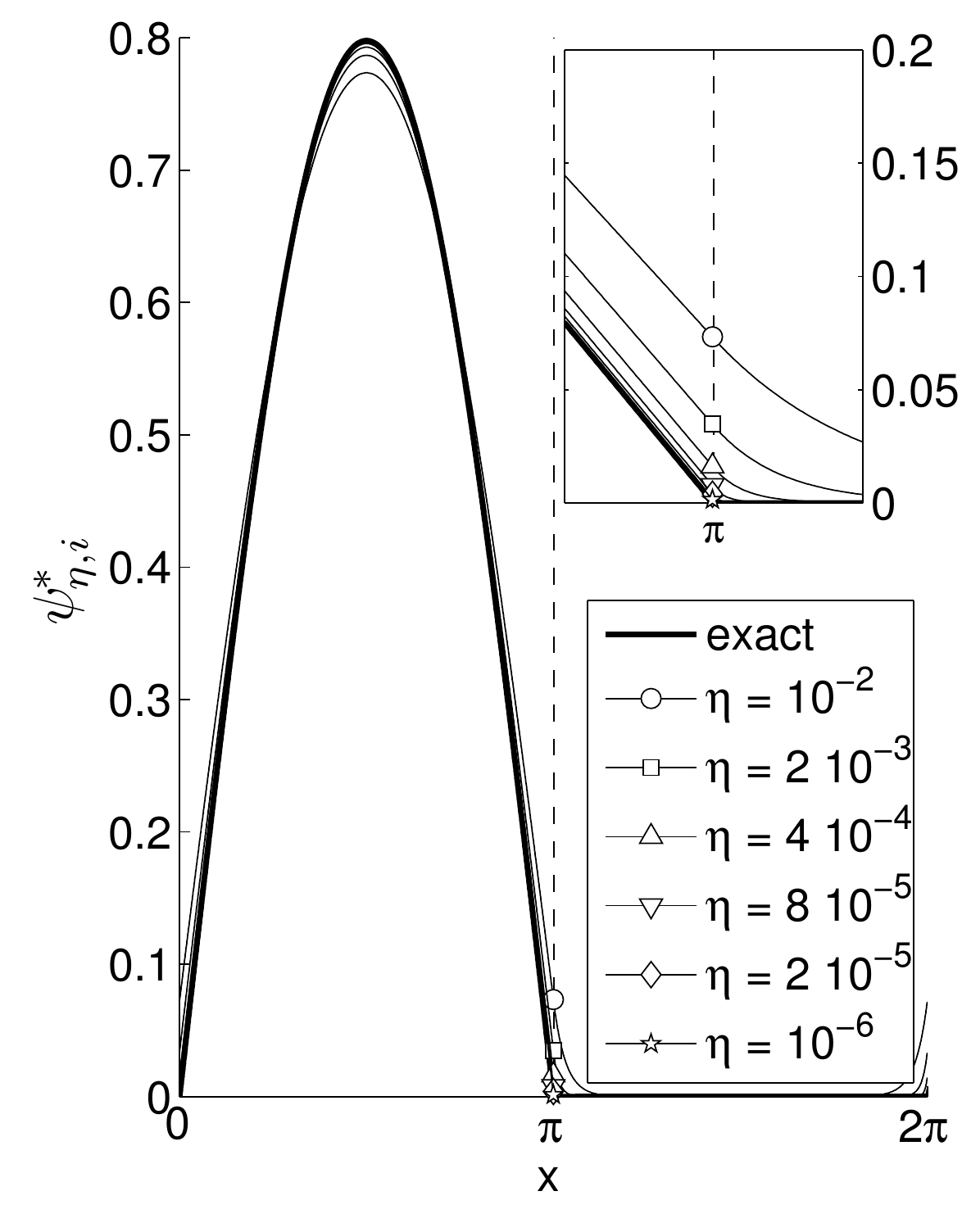} 
\includegraphics[width=0.32\columnwidth]{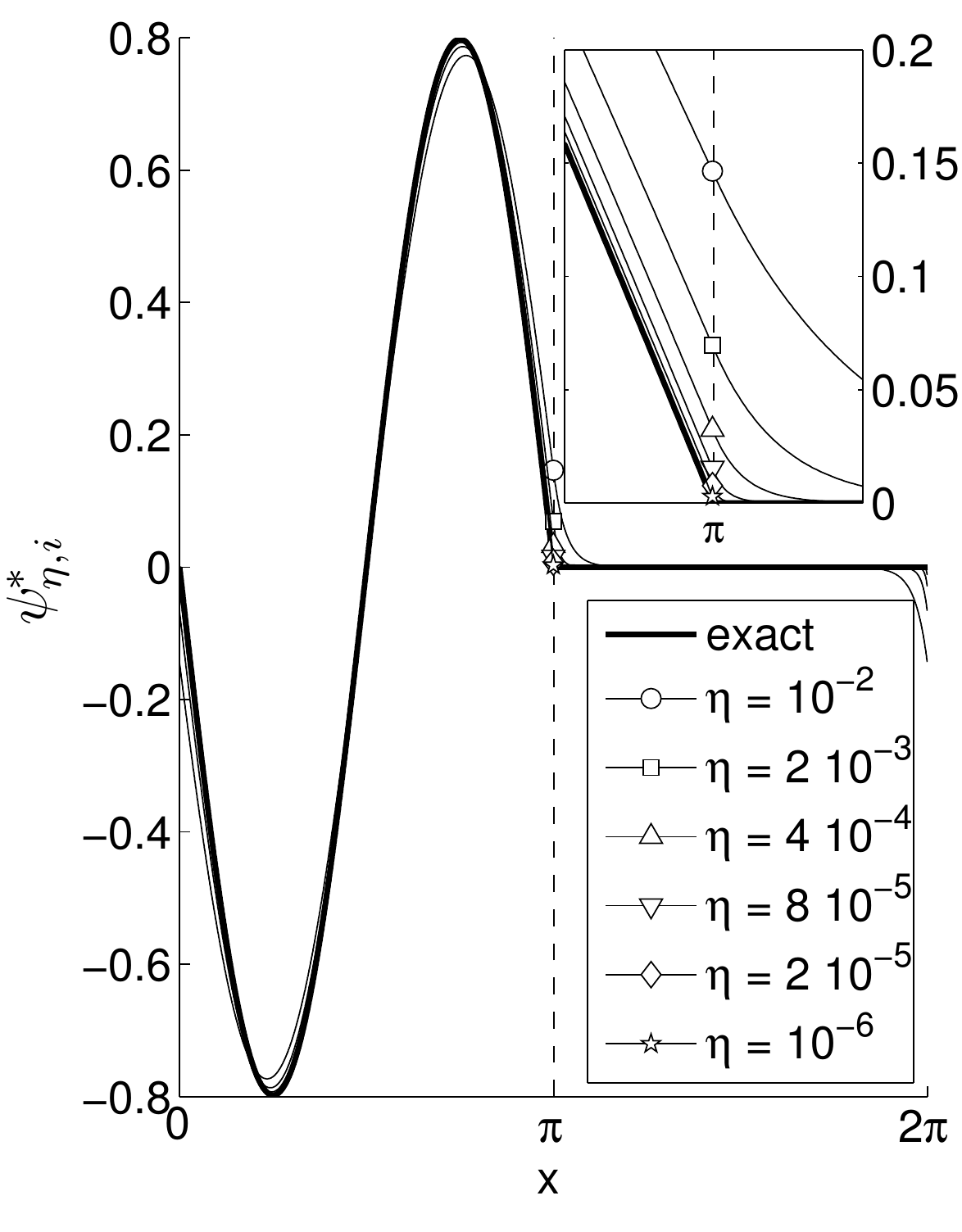} 
\includegraphics[width=0.32\columnwidth]{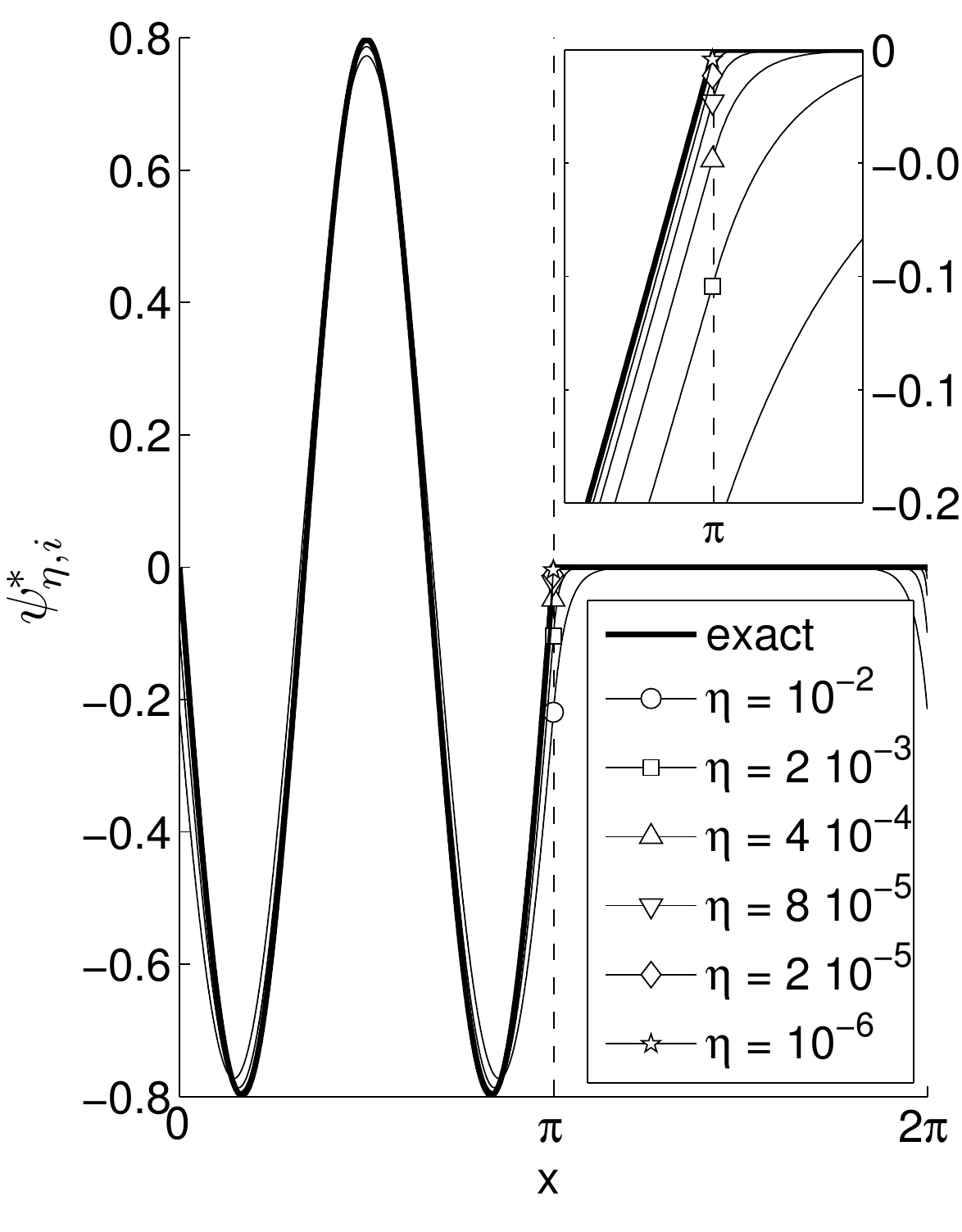} 
\caption{
\label{fig:convergence_laplace_ground_state_eta}
Eigenfunctions of the penalized Laplace operator with $\Omega = ]0,\pi[$, for $i=1,2,3$ (left to right) and varying $\eta$.
Zooms close to the location of the wall are shown in the insets.
}
\end{figure}

\begin{figure}
\begin{center}
\includegraphics[width=0.32\columnwidth]{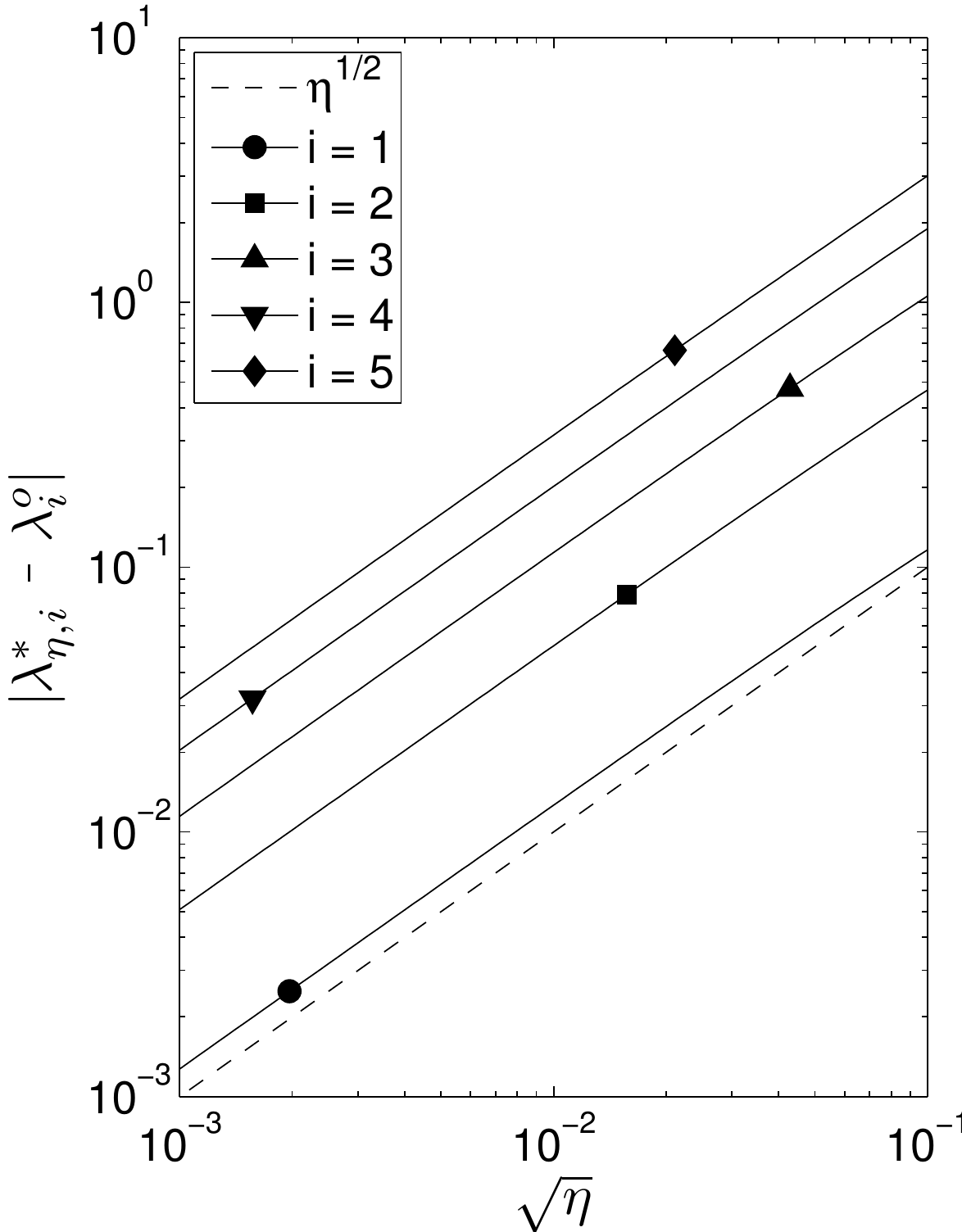}
\includegraphics[width=0.32\columnwidth]{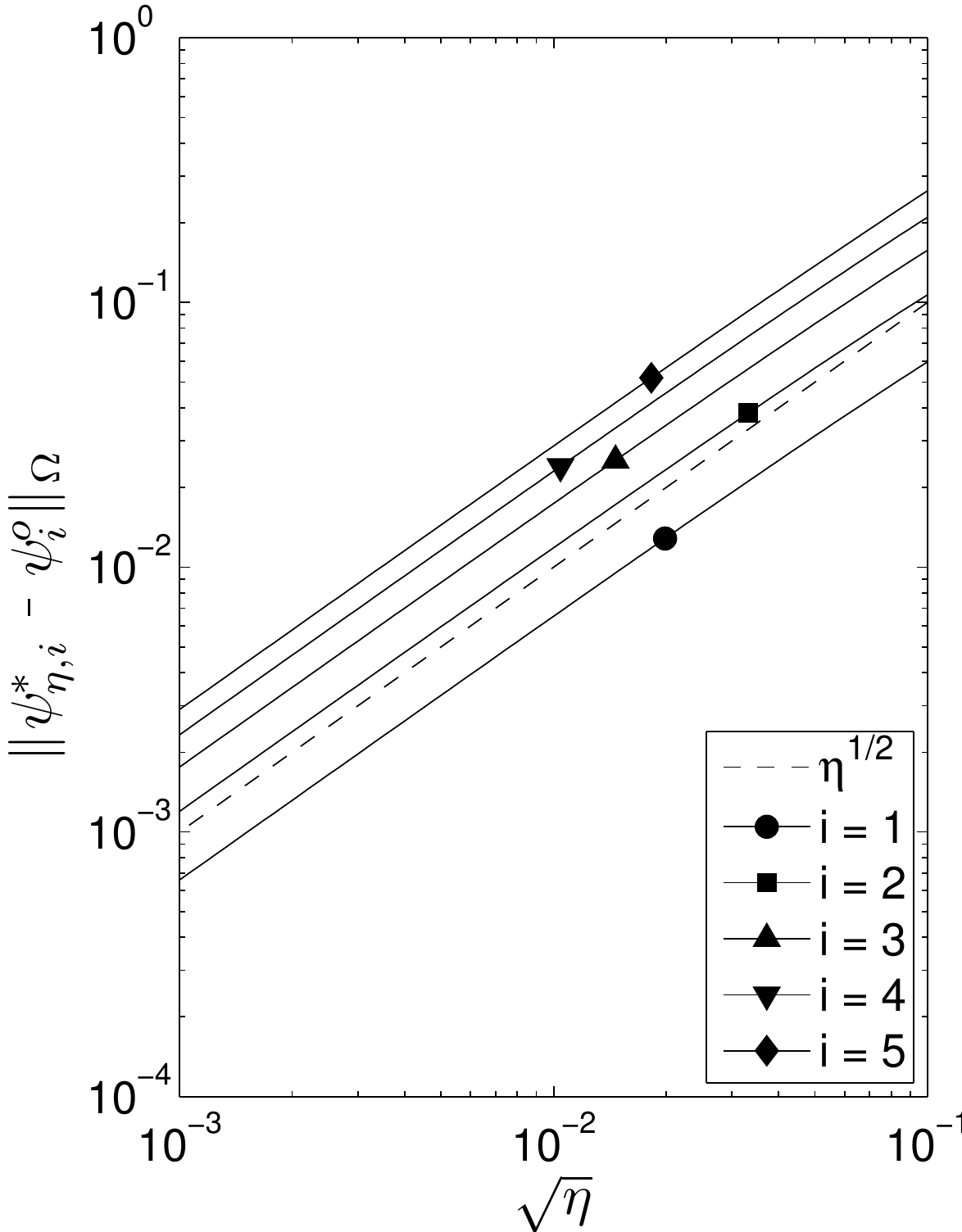}
\includegraphics[width=0.32\columnwidth]{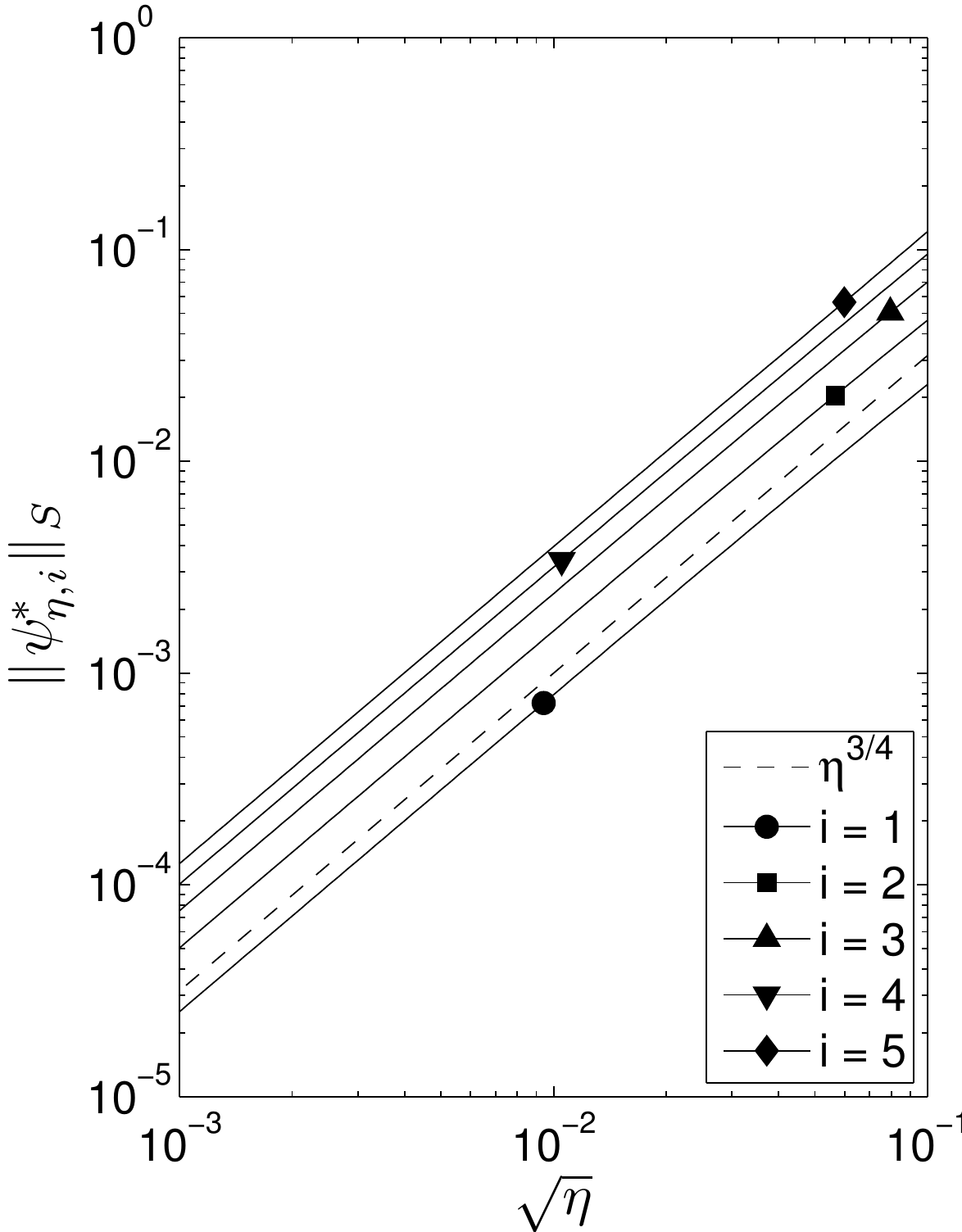}
\end{center}
\caption{
\label{fig:convergence_laplace_eigenvalues}
\label{fig:convergence_laplace_eigenfunctions}
Left: distance between eigenvalues of the penalized and exact Laplace operators with $\Omega=]0,\pi[$, as a function of $\sqrt{\eta}$.
Middle: $L^2$ distance between eigenfunctions of the penalized and exact Laplace operators with $\Omega=]0,\pi[$ as a function of $\sqrt{\eta}$,
measured in $\Omega = ]0,\pi[$ 
Right: same in $S = [\pi, 2\pi]$.
}
\end{figure}

By evaluating $\vert \lambda^*_{\eta,i} - \lambda^o_i \vert$ numerically using the values of $\lambda^*_{\eta,i}$
provided by Newton's algorithm, we recover, as expected, the scaling (\ref{eq:laplace_eigenvalue_convergence}),
see Fig.~\ref{fig:convergence_laplace_eigenvalues} (left).
We observe that, for low mode numbers and sufficiently small $\eta$, the eigenfunctions $\psi^*_{\eta,i}$ 
of the penalized Laplace operator closely approximate the corresponding eigenfunctions of the exact Laplace operator.
The cases $i = 1,2,3$ are presented in Fig.~\ref{fig:convergence_laplace_ground_state_eta}.
In fact, it follows from Ansatz (\ref{eq:laplace_1d_eigenfunctions}) that the eigenfunctions extend into the penalized region 
over a boundary layer of characteristic width $\sqrt{\eta}\left(1-\eta\lambda^*_{\eta,i}\right)^{-1/2}$.
Quantitative estimates of the difference can be measured in the $L^2$ norm by adaptive Gauss-Lobatto quadrature.
As predicted by (\ref{eq:laplace_eigenfunctions_convergence}), the error inside $\Omega$ behaves as $\eta^{1/2}$  (Fig.~\ref{fig:convergence_laplace_eigenfunctions}, middle),
and increases with the order of the eigenvalue.
Inside $S$, the $L^2$ error scales like $\eta^{3/4}$ (Fig.~\ref{fig:convergence_laplace_eigenfunctions}, right).
Both scalings are consistent with the predictions of a boundary layer analysis \cite{Boyer2006}.

The qualitative picture one can get from these results is that penalization
controls the eigenfunctions associated to sufficiently small eigenvalues,
namely those such that $\lambda < \eta^{-1}$, by making them converge to 
the corresponding eigenfunctions of the exact operator with Dirichlet boundary condition.
The convergence is however not uniform in $\lambda$, 
since it breaks down proportionally to $\lambda^o_i$, as shown by (\ref{eq:laplace_eigenvalue_convergence}).

Due to the second equation in (\ref{eq:laplace_1d_eq}), the remaining eigenfunctions of the penalized Laplace operator, namely those whose associated eigenvalues satisfy $\lambda > \eta^{-1}$,
oscillate both in $\Omega$ and in $S$.
They are not related to eigenfunctions of the exact Laplace operator, but are spurious modes introduced by the penalization method.
Note that $\left(\lambda^o_{i}\right)^{-1/2}$ has the dimension of a length and, for sufficiently simple domains, 
corresponds to a characteristic variation length of the associated eigenfunction $\psi^o_{i}$.
Therefore, penalization allows us to control what happens over length scales coarser than $\sqrt{\eta}$, and not finer.
We come back on the issue of the choice of the right cut-off to eliminate these spurious modes further down
in the section dedicated to discretization.

\subsection{Stokes operator}

The Stokes operator makes sense for dimensions greater or equal to $2$.
For simplicity, we consider a 2D channel embedded in $\TT^2$,
$\Omega = \left]0,\pi\right[ \times \TT $,
and proceed to compute the eigenfunctions of the Stokes operator in this geometry.
For convenience, the first and second coordinates in $\TT^2$ are denoted $x$ and $y$,
and the corresponding partial derivative operators by $\partial_x$ and $\partial_y$.

	\subsubsection{Semi-analytic computation of eigenvalues and eigenfunctions}

The method for the computation of the eigenfunctions is very similar to the one
which we applied in the 1D case above, with some additional complexities
which we briefly describe here.
The equation $A_{\eta}^*\uu = \mu \uu$ can be rewritten:
\begin{equation}
\begin{cases}
-\Delta\uu + \nab p = \left(\mu  - \frac{1}{\eta} \chi\right) \uu \\
\nab \cdot \uu  = 0,
\end{cases}
\end{equation}
where the effect of the Leray projector has been taken into account by introducing a pressure field $p$,
which is by definition in $H^1(\TT^2)$.

Since the problem is translation invariant in the $y$ direction, 
the first step is to expand $\uu$ in terms of Fourier modes with respect to $y$.
Denoting by $\widehat{\uu}$ and $\widehat{p}$ the partial Fourier tranforms of the dependent variables with respect to $y$, and by $k$ the corresponding wavenumber, we obtain
\begin{subequations}
\begin{align}
\label{eq:stokes_u_x}
(\partial_x^2-k^2)\widehat{u}_x - \partial_x \widehat{p} & = \left(-\mu+ \frac{1}{\eta} \chi\right) \widehat{u}_x \\
\label{eq:stokes_u_y}
(\partial_x^2-k^2)\widehat{u}_y - \iota k \widehat{p} & = \left(-\mu+ \frac{1}{\eta} \chi\right) \widehat{u}_y \\
\label{eq:stokes_divergence}
\partial_x \widehat{u}_x + \iota k \widehat{u}_y & = 0,
\end{align}
\end{subequations}
where we have used the fact that $\chi$ does not depend on $y$.
Note that the partial derivatives with respect to $x$ can now be understood as ordinary derivatives, since the dependency on $y$ has been eliminated.

In the special case $k=0$, (\ref{eq:stokes_u_y}) reduces to the eigenvalue problem for the penalized Laplace operator in $]0,\pi[$,
which we have already treated in the previous section, and (\ref{eq:stokes_divergence}) implies that $\widehat{u}_x$ vanishes identically.
Assuming from now on that $k > 0$, we use (\ref{eq:stokes_divergence}) to solve for $\widehat{u}_y$:
\begin{equation}
\label{eq:stokes_u_y_expression}
\widehat{u}_y = \frac{\iota}{k} \partial_x \widehat{u}_x,
\end{equation}
and then substitute in (\ref{eq:stokes_u_y}) to obtain the Fourier coefficients of the pressure $\widehat{p}$:
\begin{equation}
\label{eq:stokes_pressure_expression}
\widehat{p} = \left(\partial_x^2-k^2+\mu-\frac{1}{\eta} \chi\right) \frac{1}{k^2} \partial_x \widehat{u}_x.
\end{equation}
We now restrict our attention to either one of the two intervals $]0,\pi[$ and $]\pi,2\pi[$ over
which $\chi$ takes a constant value which we denote by $\kappa$, where $\kappa = 0$ or $1$ respectively.
We then insert the expression for the pressure into (\ref{eq:stokes_u_x}):
\begin{equation}
\left(\partial_x^2-k^2+\mu-\frac{1}{\eta} \kappa \right) \widehat{u}_x- \frac{1}{k^2}\left(\partial_x^2-k^2+\mu-\frac{1}{\eta} \kappa \right)  \partial_x^2 \widehat{u}_x = 0,
\end{equation}
so as to finally obtain a fourth order ordinary differential equation for $ \widehat{u}_x$:
\begin{equation}
\label{eq:stokes_reduced_eigenproblem}
\partial_x^4  \widehat{u}_x - \left(2k^2-\mu+\frac{1}{\eta} \kappa \right)\partial_x^2 \widehat{u}_x  + k^2\left(k^2-\mu+\frac{1}{\eta} \kappa \right) \widehat{u}_x = 0.
\end{equation}

By writing the solutions under the form $e^{\iota mx}$ we obtain:
\begin{equation}\label{eq:characteristic}
m^4 + \left(2k^2-\mu+\frac{1}{\eta} \kappa\right)m^2 + k^2\left(k^2-\mu+\frac{1}{\eta} \kappa \right) = 0.
\end{equation}
Both $k$ and $-k$ are roots of (\ref{eq:characteristic}) whatever the value of $\kappa$,
and for $\kappa=0$, there are two other conjugate purely imaginary other roots, which we denote by $\iota q_0$ and $-\iota q_0$.
Now assuming that $\frac{1}{\eta} > \mu - k^2$, for $\kappa=1$ the two opposite remaining roots are real,
and will be denoted by $q_1$ and $-q_1$.
The values of $q_0$ and $q_1$ are easily obtained by factorizing the polynomial equation (\ref{eq:characteristic}):
\begin{equation}
\label{eq:stokes_q_equations}
q_0 = \sqrt{\mu - k^2} \quad \mathrm{and} \quad q_1 = \sqrt{\eta^{-1} - \mu + k^2 }.
\end{equation}

\begin{figure}
\includegraphics[width=0.32\columnwidth]{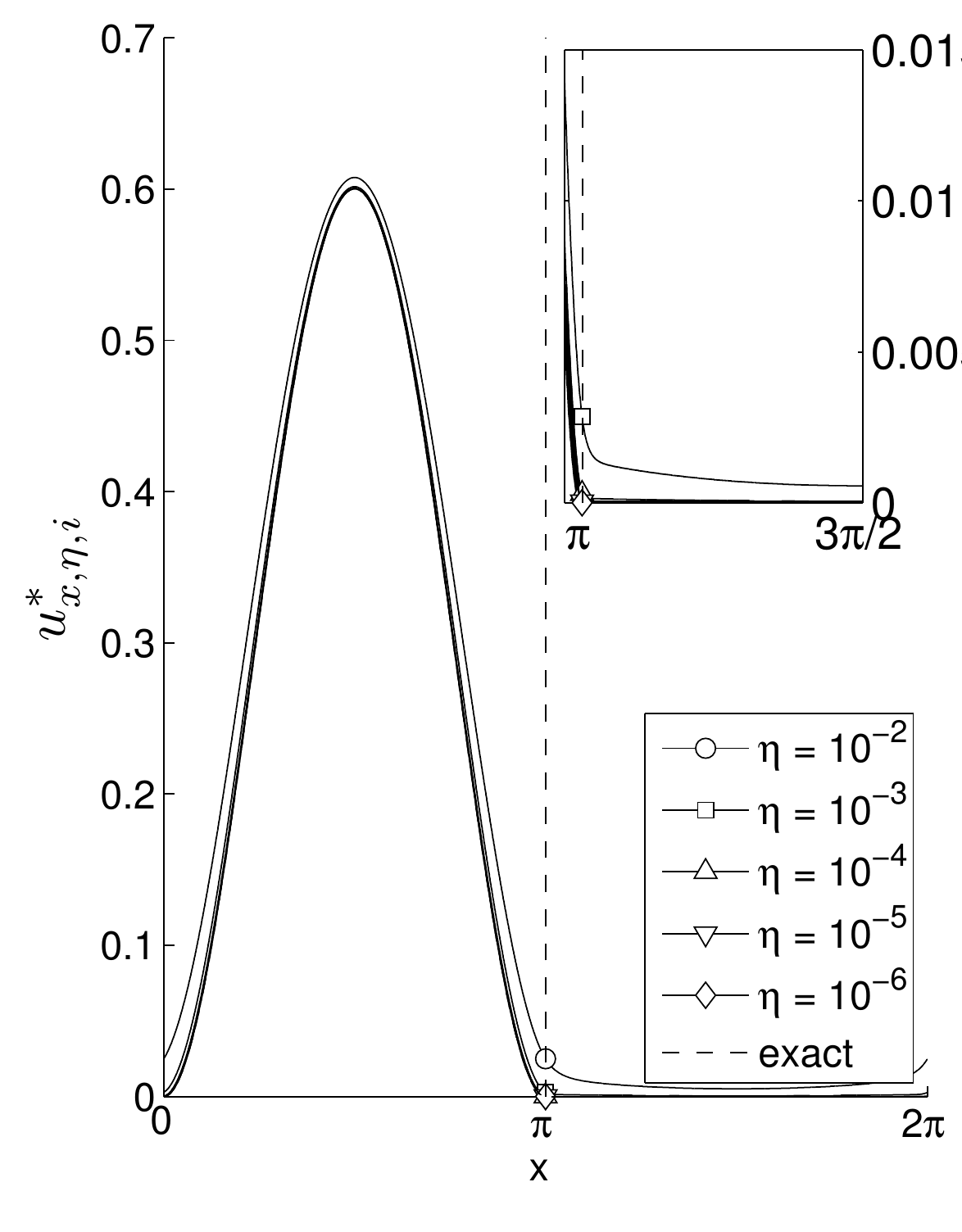} 
\includegraphics[width=0.32\columnwidth]{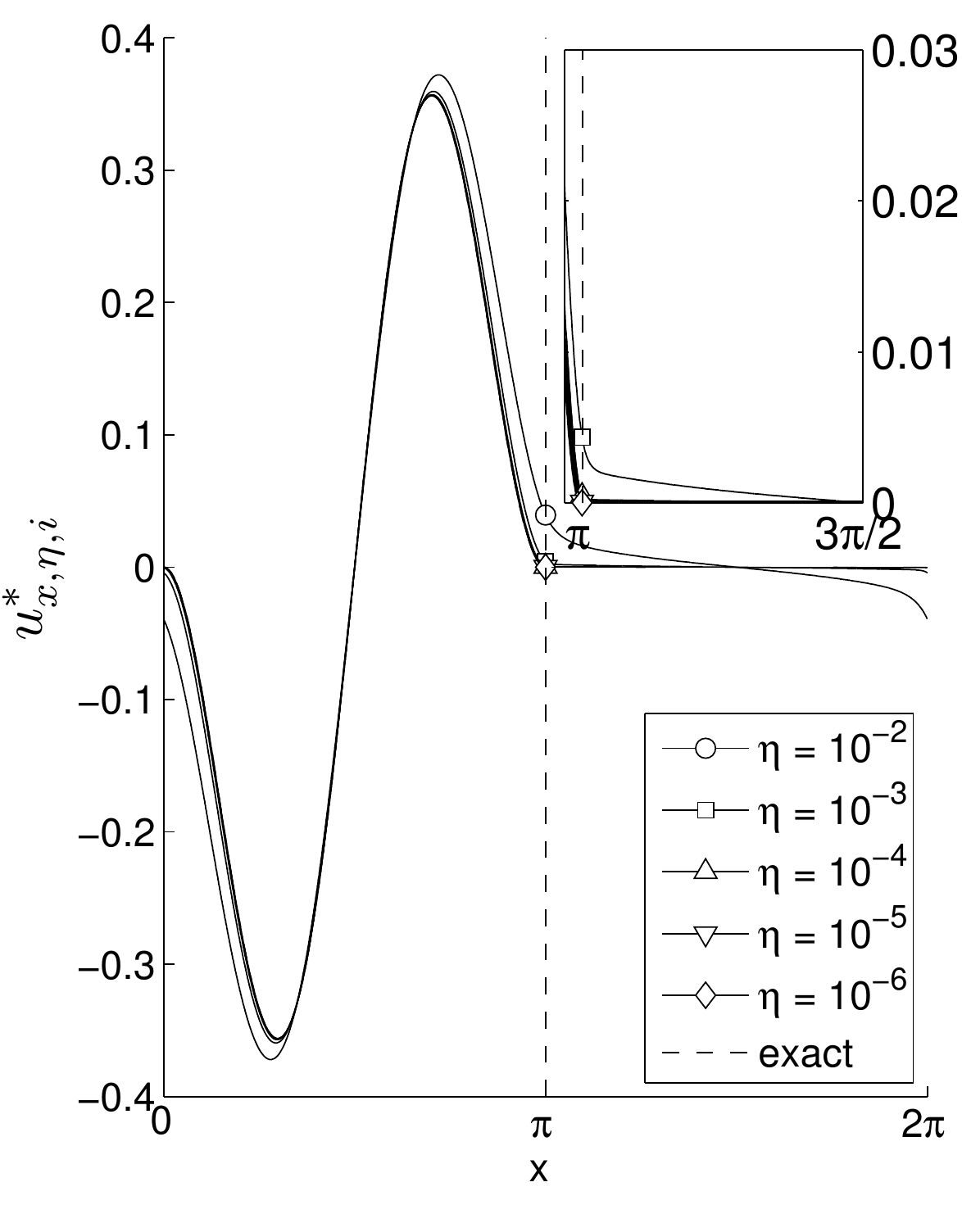}
\includegraphics[width=0.32\columnwidth]{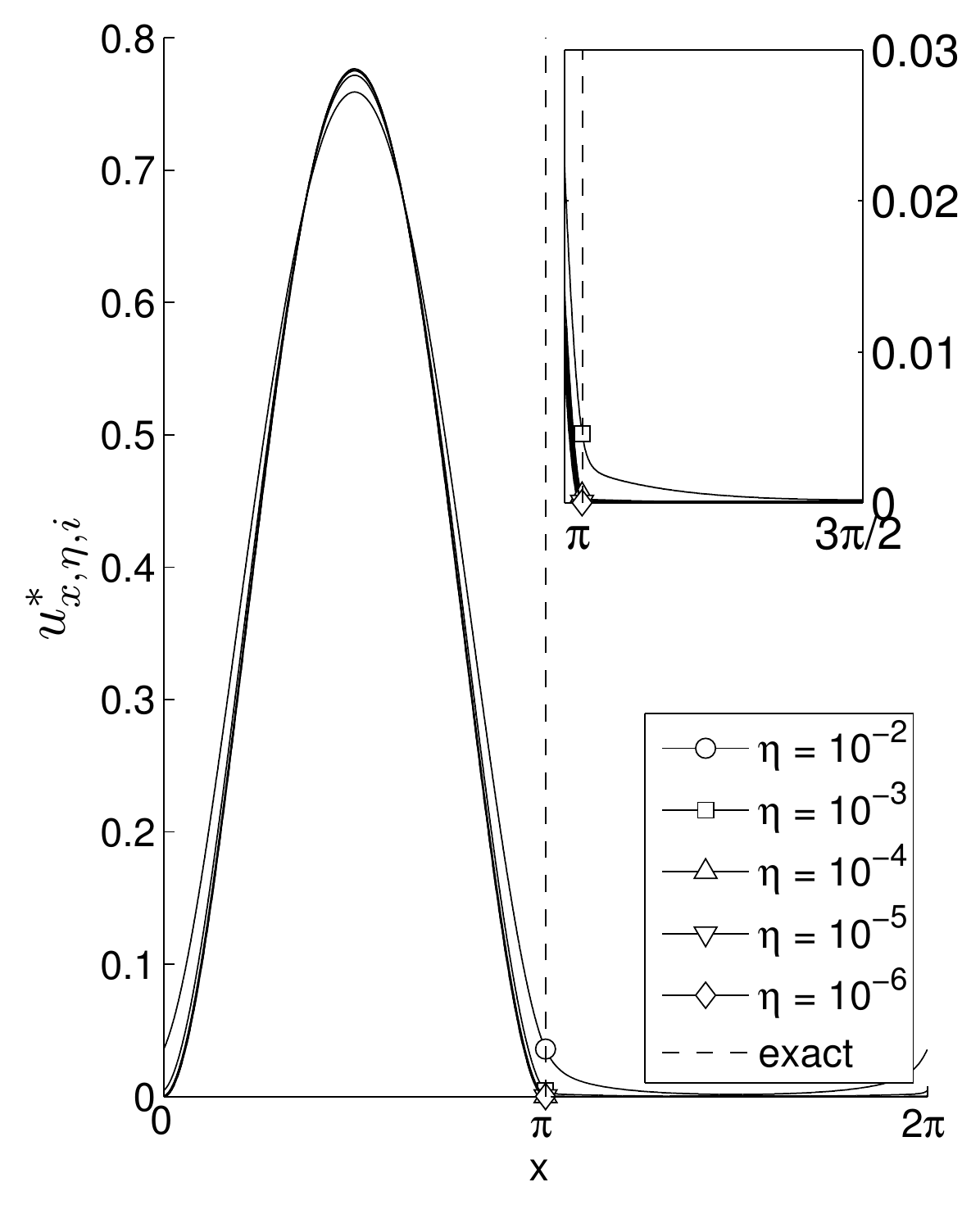}
\caption{
\label{fig:convergence_stokes_ground_state_eta}
Cross-sections through the first component $u_x$ of eigenfunctions of the penalized Stokes operator, for varying $\eta$.  
Left: $k=1$, $l=1$. Middle: $k=1$, $l=2$. Right: $k=2$, $l=1$. 
Zooms close to the location of the wall are shown in the insets.
}
\end{figure}

As in the Laplace case above, the eigenfunctions have to be either symmetric or antisymmetric under the transformation $x \to \frac{\pi}{2} - x$ in $\TT$,
since the latter leaves $\chi$ invariant.
Therefore we look for solutions under the form
\begin{subequations}
\label{eq:stokes_profile}
\begin{align}
 \widehat{u}_x^-(x) &=
 \begin{cases}
 A_0\sin\left(q_0 \left( x-\frac{\pi}{2}\right)\right) + C_0\sinh\left(k \left(x-\frac{\pi}{2}\right)\right) \mathrm{\ for\ } 0 < x < \pi \\
 A_1 \left(e^{q_1 \left(x-2\pi\right)} - e^{q_1 \left(\pi-x\right)} \right) + C_1\sinh\left(k \left(x-\frac{\pi}{2}\right)\right) \mathrm{\ for\ } \pi < x < 2\pi ,
 \end{cases} \\
 \widehat{u}_x^+(x) &=
 \begin{cases}
 B_0\cos\left(q_0\left( x-\frac{\pi}{2}\right)\right) + D_0\cosh\left(k \left(x-\frac{\pi}{2}\right)\right) \mathrm{\ for\ } 0 < x < \pi \\
 B_1 \left(e^{q_1 \left(x-2\pi\right)} + e^{q_1 \left(\pi-x\right)}\right) + D_1\cosh\left(k \left(x-\frac{\pi}{2}\right)\right) \mathrm{\ for\ } \pi < x < 2\pi .
 \end{cases}
\end{align}
\end{subequations}
The constants are determined by the $C^1$ reconnection conditions for $\uu$ at $x=\pi$, and by the continuity of $p$.
From the expressions of $u_y$ and $p$ as functions of $u_x$ (see Eqs.~\ref{eq:stokes_u_y_expression} and \ref{eq:stokes_pressure_expression} above),
these conditions may all be reformulated in terms of $u_x$ only:
\begin{subequations}
\label{stokes_linear_system}
\begin{align}
u_x(\pi^+) = {} & u_x(\pi^-) \\ 
\partial_x u_x(\pi^+) = {} & \partial_x u_x(\pi^-) \\ 
\partial_x^2 u_x(\pi^+) = {} & \partial_x^2 u_x(\pi^-) \\ 
\partial_x^3 u_x(\pi^+) ={}  & \partial_x^3 u_x(\pi^-) + \frac{1}{\eta}\partial_x u_x(\pi^-).
\end{align}
\end{subequations}
As before, the existence of a non-zero solution is equivalent to the vanishing of a determinant $G^*_{\pm}(k,\mu,\eta)$, whose expressions in the antisymmetric and symmetric cases (respectively) are
\begin{subequations}
\label{eq:stokes_penalized_determinant}
\begin{multline}
G^*_{-}(k,\mu,\eta) = k  \,\mathrm{cotanh}\!\left(\frac{\pi  k}{2}\right) - \left(1-2\eta\mu\right)  \left(1-\eta\mu\right) q_0\,\mathrm{cotan}\!\left(\frac{\pi  q_0}{2}\right) \\ + \mu \sqrt{\eta(1-\eta q_0^2)} \left(1 - 2  \eta\mu \right)\mathrm{cotanh}\left(-\frac{\pi q_1}{2}\right) = 0
\end{multline}
\begin{multline}
G^*_{+}(k,\mu,\eta) = k  \,\mathrm{tanh}\!\left(\frac{\pi  k}{2}\right) + \left(1-2\eta\mu\right)  \left(1-\eta\mu\right) q_0\,\mathrm{tan}\!\left(\frac{\pi  q_0}{2}\right) \\ + \mu \sqrt{\eta(1-\eta q_0^2)} \left(1 - 2  \eta\mu \right)\mathrm{tanh}\left(-\frac{\pi q_1}{2}\right) = 0.
\end{multline}
\end{subequations}
The roots of these determinants can be identified with an increasing subsequence of eigenvalues for $A^*_\eta$,
which we index from now on by a non-zero integer denoted $l$.
The complete sequence of eigenvalues of $A^*_\eta$ is thus indexed by the two integers $k$ and $l$,
or equivalently by $i$ if they are sorted in increasing order as a whole.
For each of these eigenvalues, the corresponding under-determined system (\ref{stokes_linear_system}) can then be solved to find the constants, as in the 1D case.
Each eigenfunction is thus computed with the desired accuracy.
The results are shown for some values of $k$ and $l$ in Fig.~\ref{fig:convergence_stokes_ground_state_eta}.

\subsubsection{Behavior of eigenvalues and eigenfunctions when $\eta \to 0$}

When $\eta \to 0$, the eigenvalue equations in the antisymmetric and symmetric cases respectively simplify to
\begin{subequations}
\label{eq:stokes_exact_determinant}
 \begin{align}
G^o_{-}(k,\mu) & {} = k \,\mathrm{cotanh}\!\left(\frac{\pi  k}{2}\right) - q_0 \,\mathrm{cotan}\!\left(\frac{\pi  q_0}{2}\right)  =   0, \\
G^o_{+}(k,\mu) & {} = k \,\mathrm{tanh}\!\left(\frac{\pi  k}{2}\right) + q_0 \tan\!\left(\frac{\pi  q_0}{2}\right)  =  0,
\end{align}
\end{subequations}
whose solutions, the eigenvalues of $A^o$, are determined numerically using Newton's algorithm.
The corresponding eigenfunctions are computed using Ansatz (\ref{eq:stokes_profile}) as above,
but enforcing $A_1 = C_1 = B_1 = D_1 = 0$, and using the Dirichlet boundary conditions at $x=\pi$ to find the values of $A_0, C_0$ or $B_0, D_0$.
The eigenvalues and eigenfunctions of $A^o$ can then be used as reference to study the convergence of $\mu^*_{\eta,i}$ and $\uu^*_{\eta,i}$, respectively.

\begin{figure}
\begin{center}
\includegraphics[width=0.32\columnwidth]{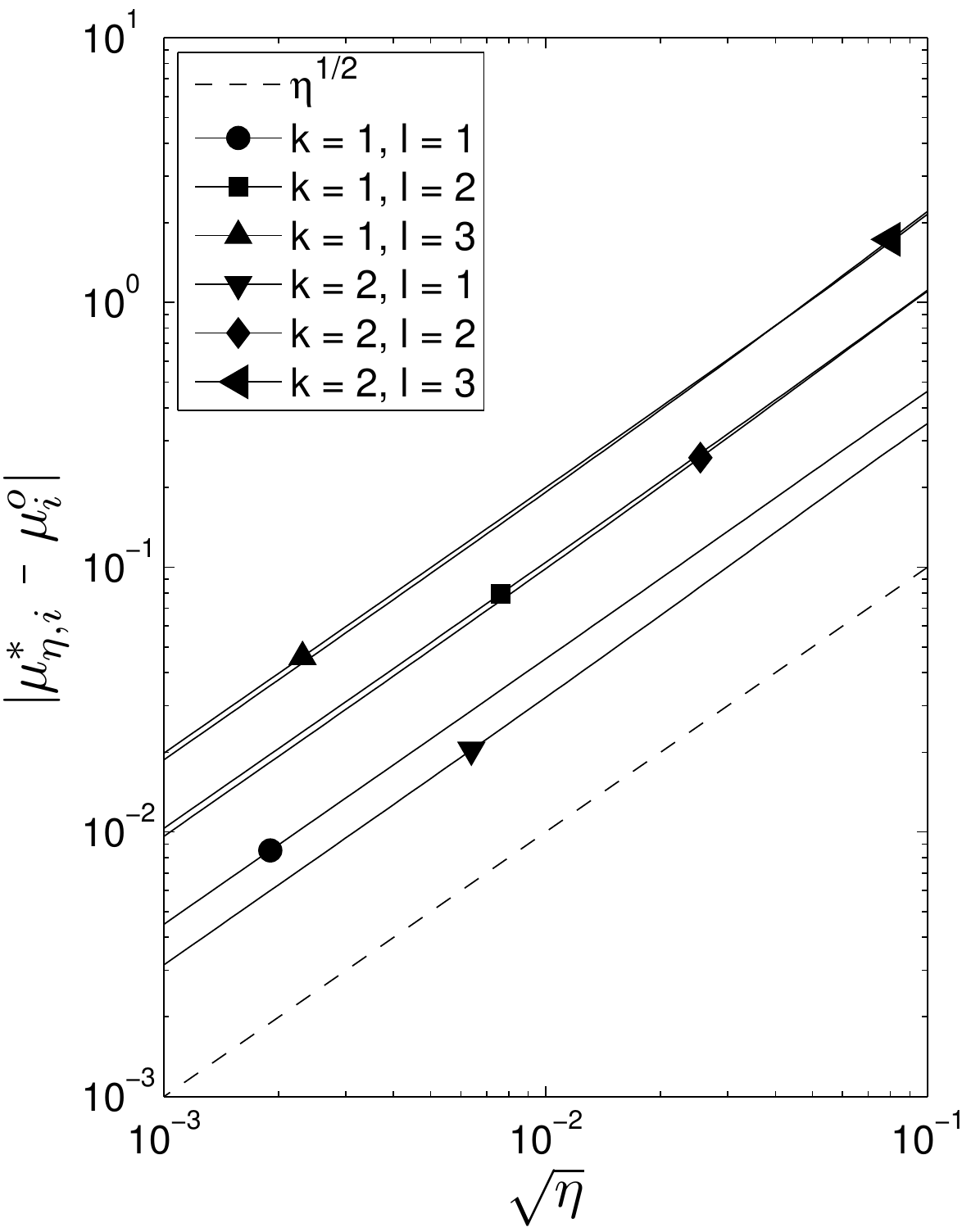}
\includegraphics[width=0.32\columnwidth]{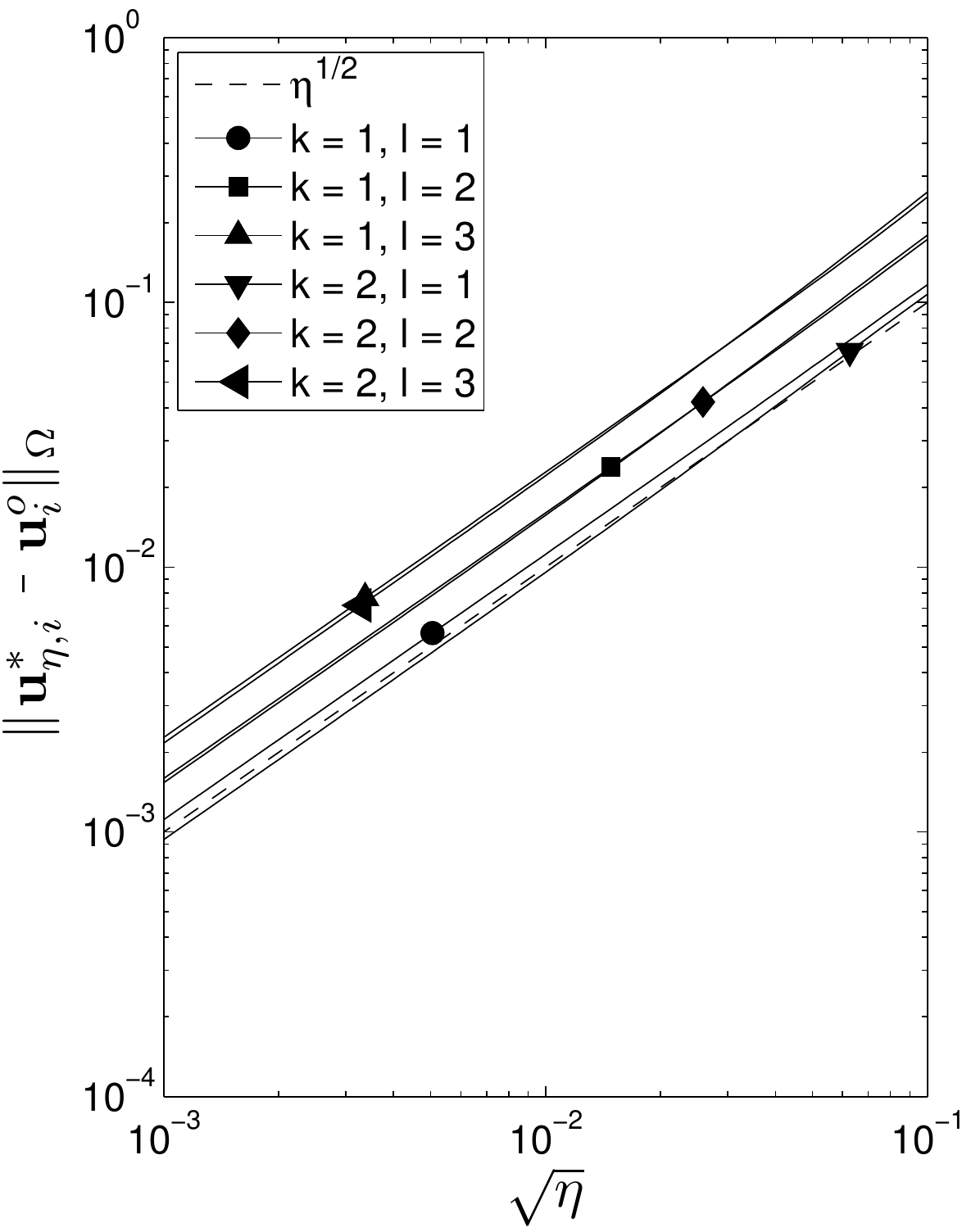}
\includegraphics[width=0.32\columnwidth]{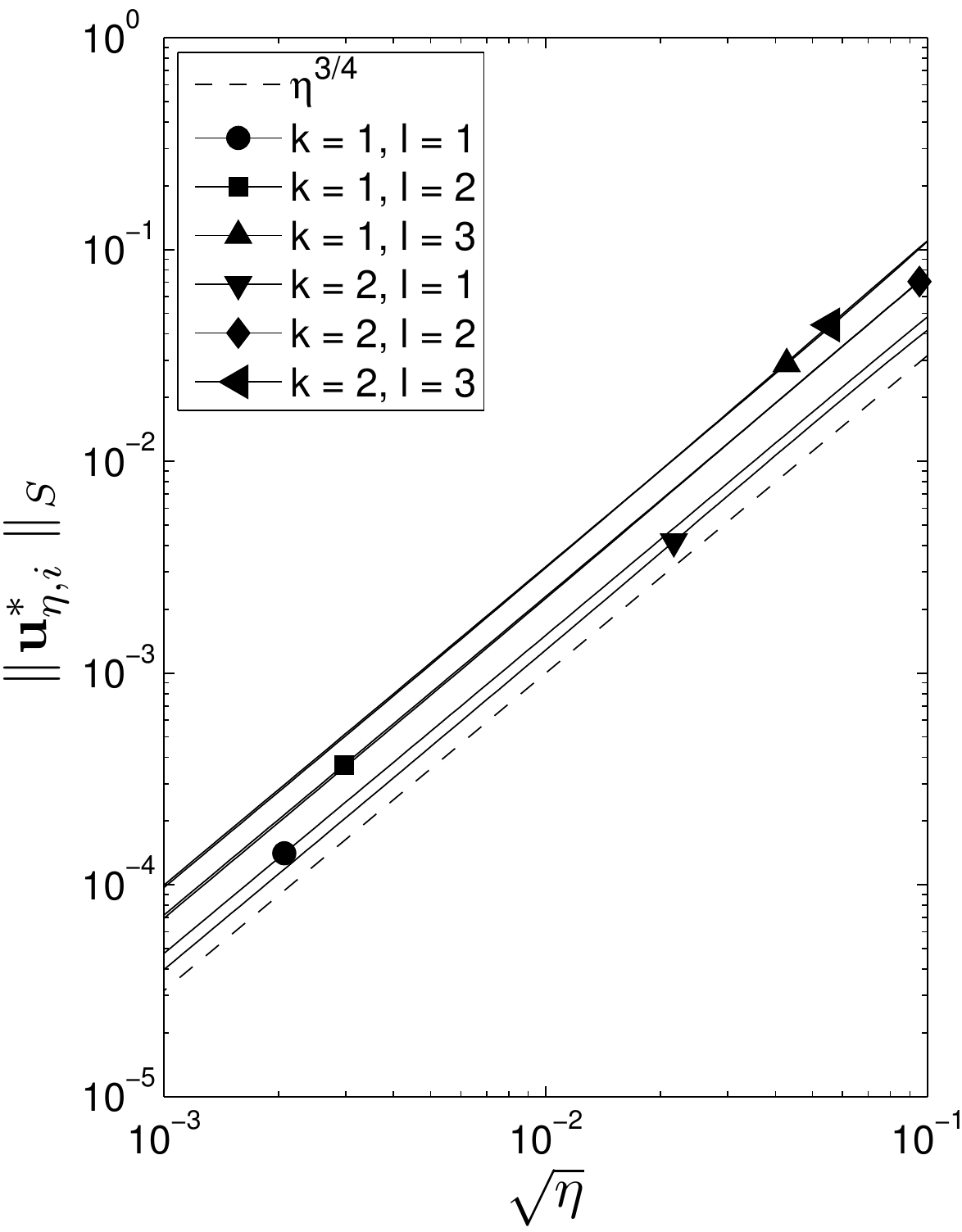}
\end{center}
\caption{
\label{fig:convergence_stokes}
Convergence of the penalized Stokes operator in a channel $A^*_\eta$, as function of $\sqrt{\eta}$, for $k=1,2$ and $l=1,2,3$,
to the exact Stokes operator with Dirichlet boundary conditions $A^o$.
Left: distance between eigenvalues. 
Middle: distance between eigenfunctions in $L^2(\Omega)$.
Right: distance between eigenfunctions in $L^2(S)$.
}
\end{figure}

As before we get by the implicit function theorem that, as $\eta \to 0$,
\begin{equation}
\label{eq:stokes_eigenvalue_convergence}
\mu^*_{\eta,k,l} = \mu^o_{k,l} - \frac{4}{\pi}(\mu^o_{k,l}-k^2)\beta_{k,l}\sqrt{\eta} + O({\eta}),
\end{equation}
where
\begin{equation}
\beta_{k,l} =
\begin{cases}
\left(1- \frac{2k}{\pi \mu^o_{k,l}} \mathrm{cotanh}\left(\frac{\pi k}{2}\right) + \frac{k^2}{\mu^o_{k,l}\mathrm{sinh}^2\left(\frac{\pi k}{2}\right)}\right)^{-1} \quad \mathrm{for\,}\, l \,\mathrm{\,even}\\
\left(1- \frac{2k}{\pi\mu^o_{k,l}} \mathrm{tanh}\left(\frac{\pi k}{2}\right) - \frac{k^2}{\mu^o_{k,l}\mathrm{cosh}^2\left(\frac{\pi k}{2}\right)}\right)^{-1} \quad \mathrm{for\,}\, l \,\mathrm{\,odd.}
\end{cases}
\end{equation}
The convergence of eigenvalues thus occurs at a rate $\sqrt{\eta}$,
as examplified in Fig.~\ref{fig:convergence_stokes} (left).
Similarly to the Laplace case (Eq.~\ref{eq:laplace_eigenvalue_convergence}), the prefactor degrades for increasing mode number.
To evaluate the convergence of the eigenfunctions themselves inside $\Omega$,
the $L^2$ error is computed by numerical quadrature from the Ansatz (\ref{eq:stokes_profile})
using the available numerical values of the parameters.
The observed scalings with $\eta$ are very similar to what we observed for the Laplace eigenfunctions,
namely $\eta^{1/2}$ in $\Omega$ (Fig.~\ref{fig:convergence_stokes}, middle),
and $\eta^{3/4}$ in $S$ (Fig.~\ref{fig:convergence_stokes}, right).

An interesting consequence of the terms $C_1$ and $D_1$ appearing in the Ansatz (\ref{eq:stokes_profile})
is that the characteristic penetration length of the penalized eigenfunctions inside the penalized region
does not, strictly speaking, go to zero with $\eta$, in contrast to what we have seen above in the case of the Laplace operator.
In the case of the Stokes operator, the penalization error in the boundary is not limited to a vanishingly small boundary layer. 
This is due to the non-local effect of the pressure field.
The effect is barely apparent when looking at the profiles of the ground state for varying $\eta$ (Fig.~\ref{fig:convergence_stokes_ground_state_eta}),
because the eigenfunctions actually behave like $O(\eta)$ deep inside the boundary.

\subsubsection{Interpretation of the leading order error as a residual slip}

\begin{figure}
\begin{center}
\includegraphics[width=0.32\columnwidth]{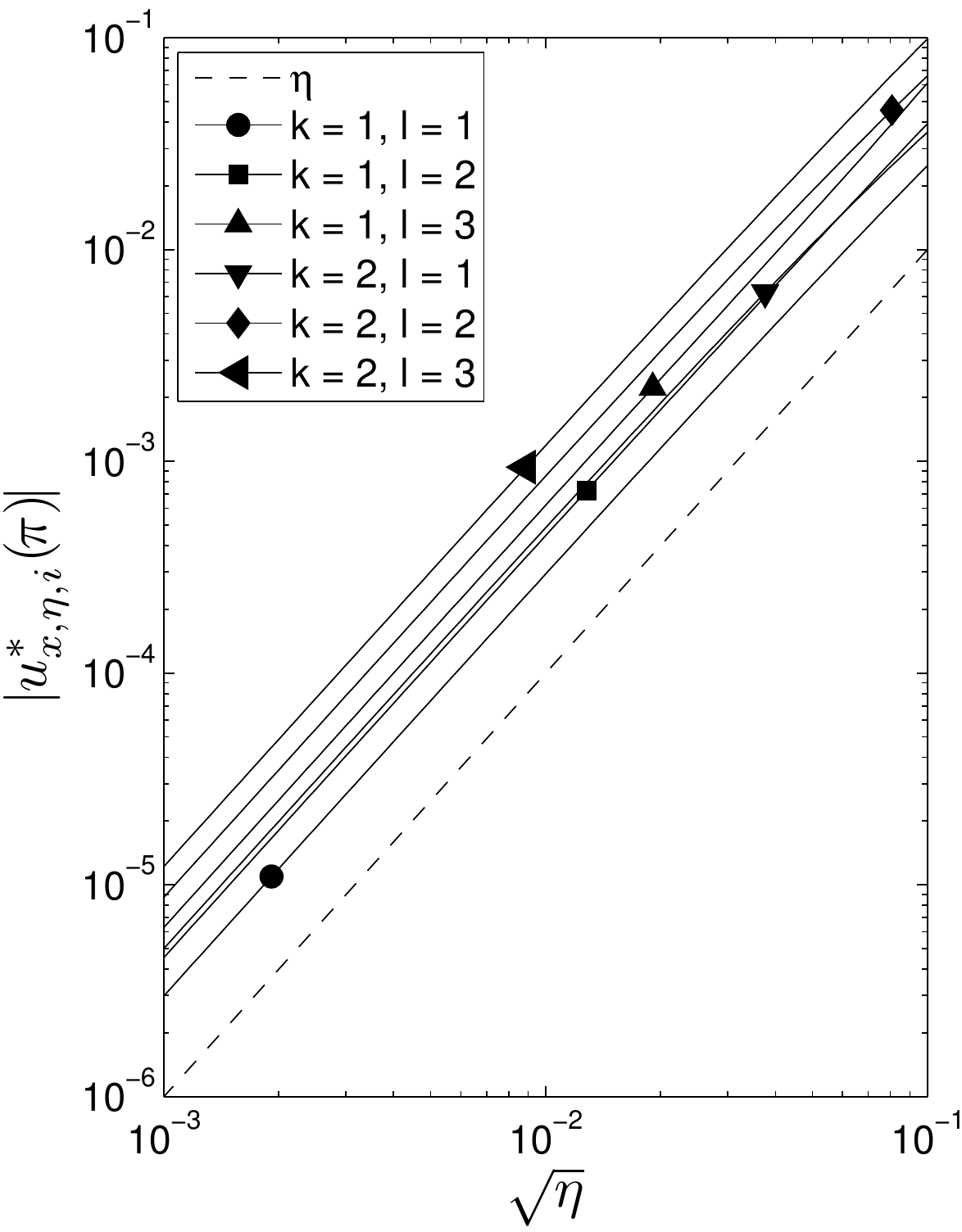}
\hspace{1cm}
\includegraphics[width=0.32\columnwidth]{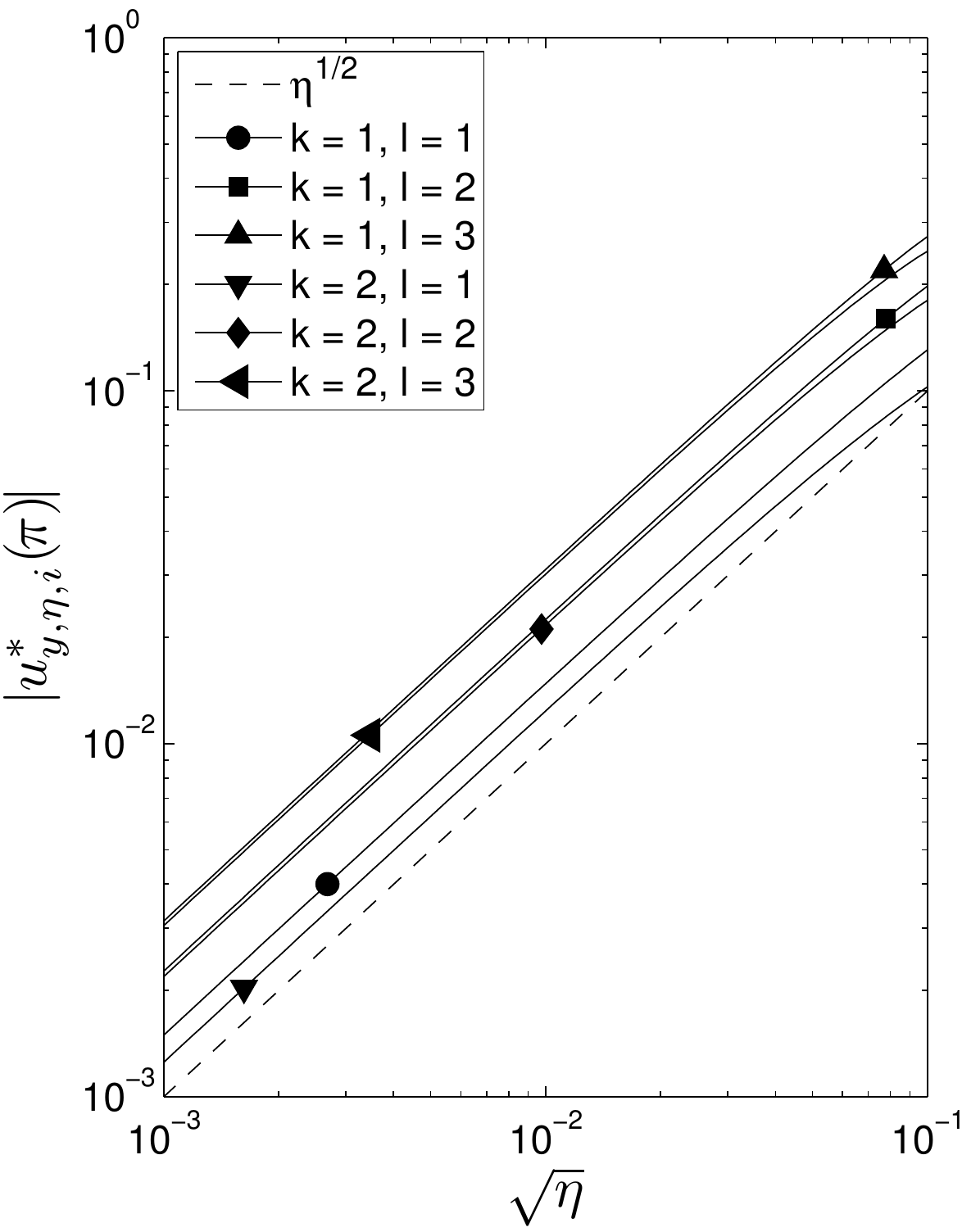}
\end{center}
\caption{
\label{fig:convergence_u_x_u_y_pi}
$\vert u^*_{x,\eta,i}(\pi) \vert $ (left) and $\vert u^*_{y,\eta,i}(\pi) \vert $ (right) as functions of $\sqrt{\eta}$, for $k=1,2$ and $l=1,2,3$.
}
\end{figure}

Since the eigenfunctions of the penalized Stokes problem also satisfy an eigenvalue equation for the Stokes operator inside $\Omega$,
their discrepancy with the exact eigenfunctions of the latter is due to their nonzero boundary values, which we now analyze.
In our case, the simple geometry allows us to eliminate the dependency on the wall parallel coordinate $y$ simply by square-averaging in this direction.
The resulting values of $\vert u^*_{x,\eta,i}(\pi) \vert $ and $\vert u^*_{y,\eta,i}(\pi) \vert $ are shown in Fig.~\ref{fig:convergence_u_x_u_y_pi}.
It appears that $u_x(\pi)$ scales like $\eta$ while $u_y(\pi)$ scales like $\sqrt{\eta}$.
In other words, the normal boundary condition of the limiting problem is approached 
much more rapidly in the penalized problem than the tangential boundary condition.
This observation, as well as some observations that we made when solving the Navier-Stokes equations \cite{Nguyenvanyen2011a}, 
has prompted us to introduce the Stokes eigenvalue problem with Navier boundary conditions, as we now present.

\begin{figure}
\begin{center}
\includegraphics[width=0.32\columnwidth]{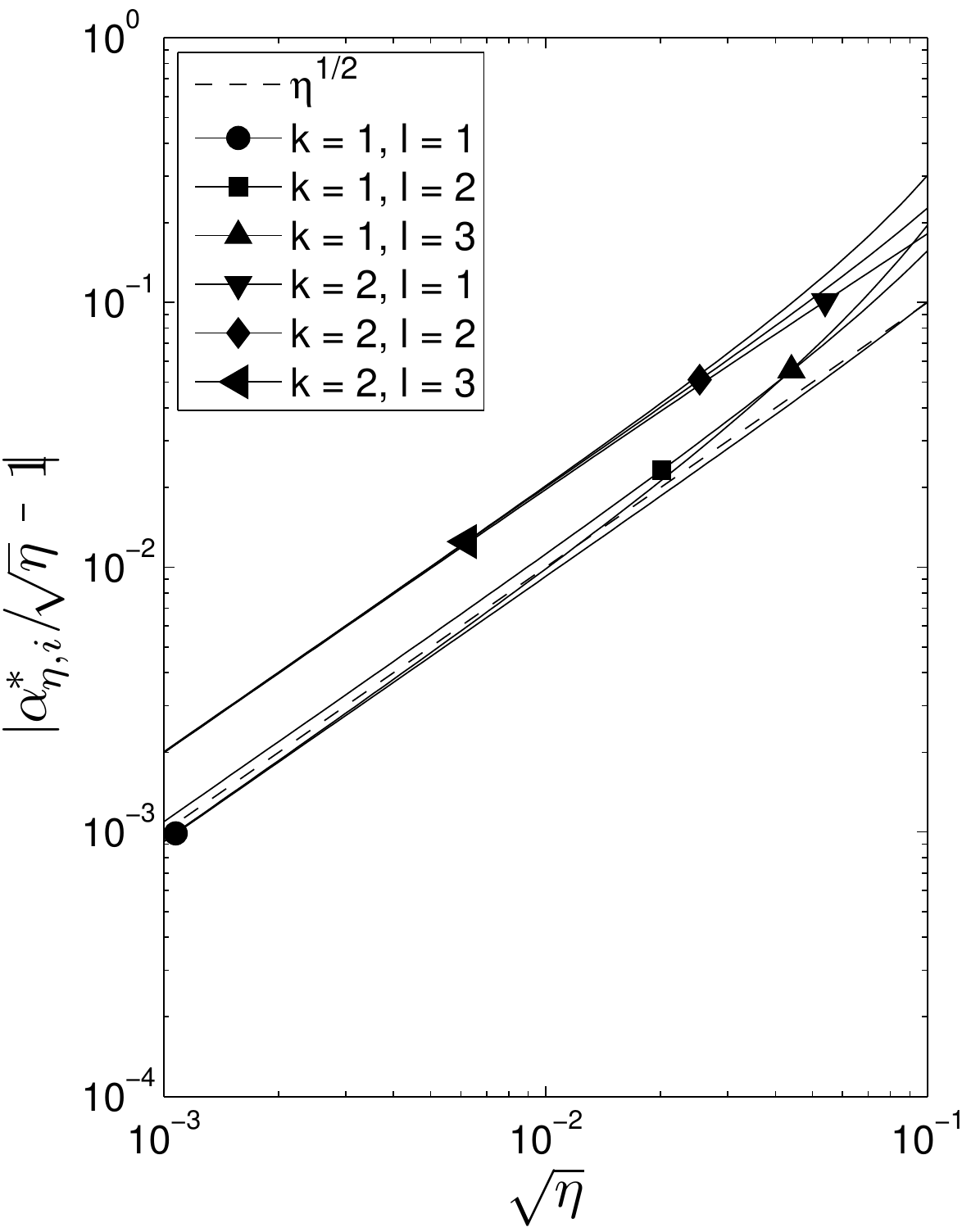}
\includegraphics[width=0.32\columnwidth]{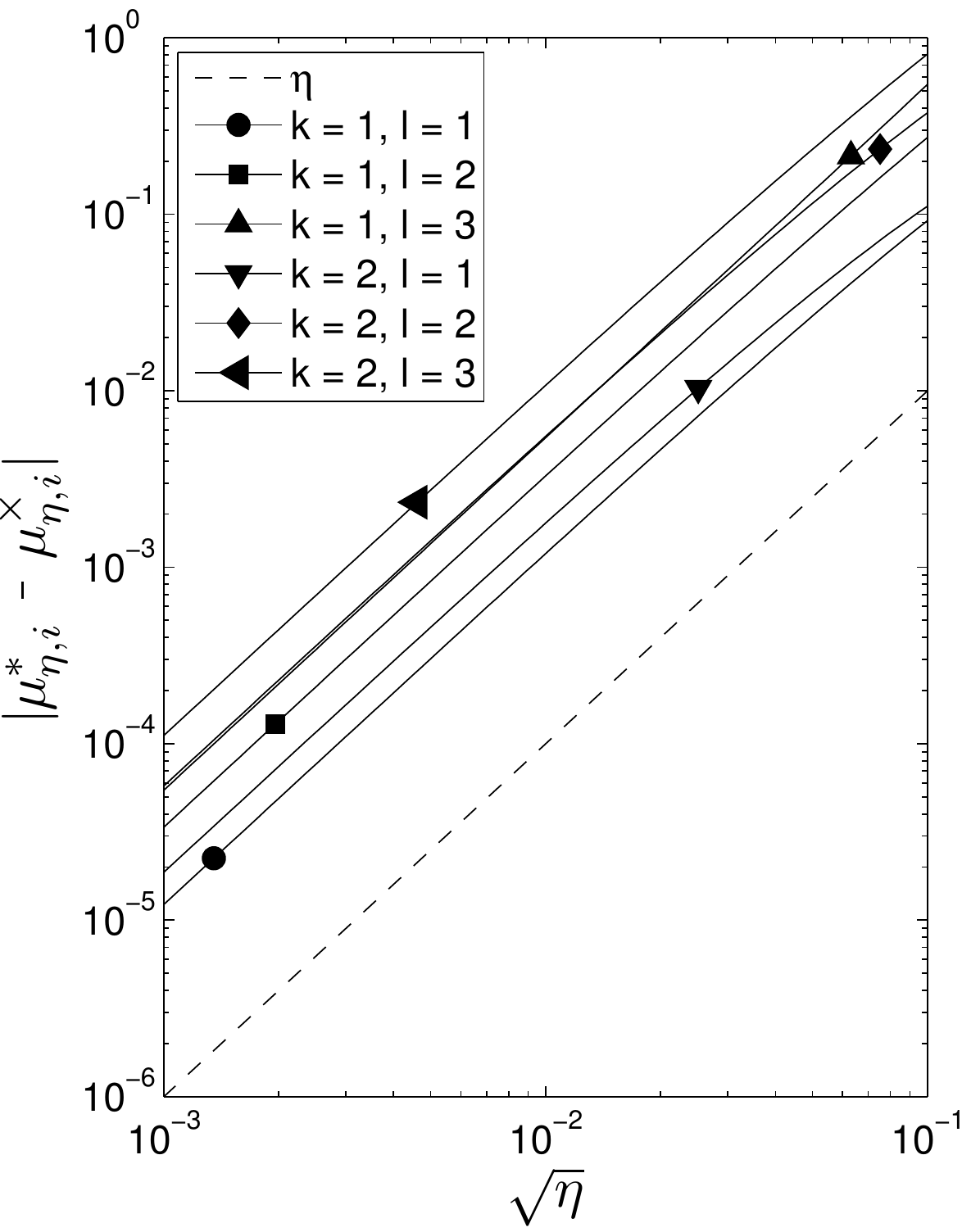}
\includegraphics[width=0.32\columnwidth]{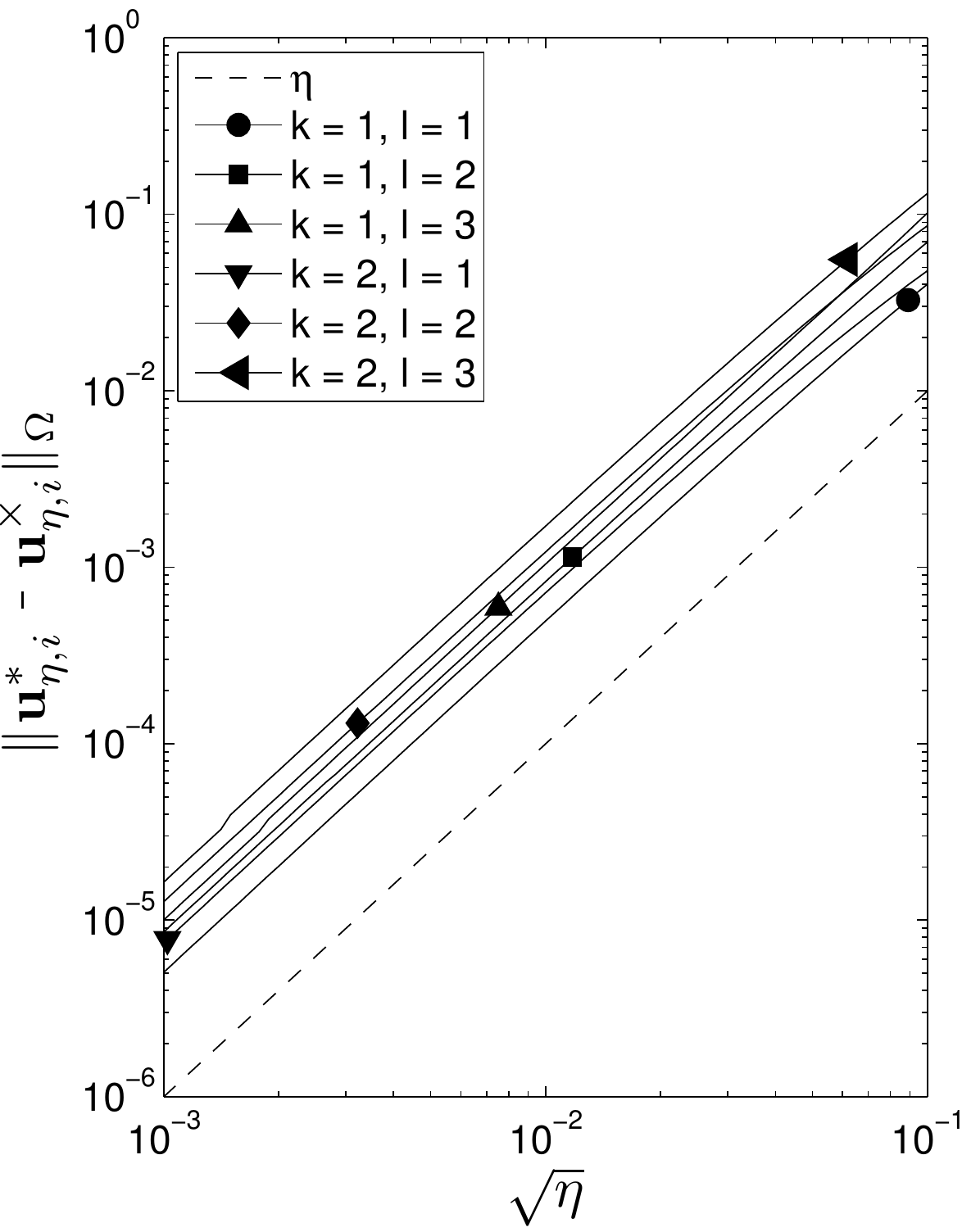}
\end{center}
\caption{
\label{fig:convergence_stokes_navier}
Left: $\left\vert\alpha^*_{\eta,i}/\sqrt{\eta}-1\right\vert$, where $\alpha^*_{\eta,i}$ is the effective slip-length as a function of $\sqrt{\eta}$ for various eigenfunctions of the penalized Stokes operator in a channel.
Middle: difference between eigenvalues of the penalized  Stokes operator 
and corresponding eigenvalues of the Stokes operator in a channel with Navier boundary conditions and a slip length $\alpha = \sqrt{\eta}$, as functions of $\eta$,
for $k=1,2$ and $l=1,2,3$.
Right: same for eigenvalues in $L^2(\Omega)$.
}
\end{figure}

The so-called Navier boundary conditions, already introduced by Navier in the historical paper 
where the equations later known as the Navier-Stokes equations were proposed \cite{Navier1823}, can be written
\begin{subequations}
\label{eq:navier_bc}
\begin{align}
\label{eq:navier_bc_tangential}
& \uu_{\partial \Omega}\cdot \ttau  + \alpha^\dagger \frac{\partial \uu}{\partial \nn} \cdot \ttau = 0 \\
\label{eq:navier_bc_normal}
& \uu_{\partial \Omega} \cdot \nn = 0,
\end{align}
\end{subequations}
where $\nn$ is the outward normal unit vector, $\ttau$ is any tangent unit vector, both along $\partial\Omega$,
and $\alpha^\dagger$ is a positive real number known as the slip length.
Dirichlet boundary conditions, which we have been considering up to now, correspond to the special case $\alpha^\dagger = 0$ (no slip).
Coming back to the penalized problem, we can thus define, by analogy with (\ref{eq:navier_bc_tangential}), an effective slip length $\alpha^*_{i,\eta}$ for its eigenfunctions:
\begin{equation}\label{alpha_def}
\alpha^*_{i,\eta} = - \frac{u^*_{y,i,\eta}(\pi)}{\partial_x u^*_{y,i,\eta}(\pi)},
\end{equation}
assuming that the denominator does not vanish.
The surprising fact that we have found is that $\alpha^*_{i,\eta}$ is close to $\sqrt{\eta}$ independently of $i$.
This can be seen for example by monitoring the difference $(\alpha^*_{i,\eta}-\sqrt{\eta})/\sqrt{\eta}$, which is found to scale like $O(\sqrt{\eta})$ (Fig. \ref{fig:convergence_stokes_navier}, left).

This finding suggests a direct comparison between the eigenfunctions of the penalized Stokes operator
on the one hand, and those of the Stokes operator inside $\Omega$ supplemented with Navier boundary conditions on the other hand.
For this, define $\VV^\dagger_\eta$, the closure in $H^1(\Omega)^d$ of 
$$
\left\{ \uu \in \mathcal{C}^\infty(\Omega) \mid \nab \cdot \uu = 0 ,  \uu_{\partial \Omega}\cdot \nn = 0,  \uu_{\partial \Omega}\cdot \ttau  + \sqrt{\eta} \frac{\partial \uu}{\partial \nn} \cdot \ttau = 0  \right \},
$$
and consider the operator ${A}^\dagger_\eta : (H^2(\Omega))^d \cap \VV_\eta \to \HH $,
which is defined exactly as ${A}^o$ but by replacing $\VV_0$ by $\VV^\dagger_\eta$ in the definition of the starting space.
The eigenvalue problem for ${A}^\dagger_\eta $ is then well defined, and 
the corresponding eigenvalues and eigenfunctions will be denoted respectively by $\mu^\dagger_{i,\eta}$ and $\uu^\dagger_{i,\eta}$.
They can be computed semi-analytically in the same manner as was done for $A^o$ above,
simply by enforcing the Navier conditions (\ref{eq:navier_bc}) at the boundary instead of Dirichlet conditions,
thus yielding both different eigenvalues and different eigenfunctions.
The corresponding eigenvalue equations are:
\begin{subequations}
\label{eq:navier_dispersion_relation}
\begin{align}
G^\dagger_{-}(k,\mu,\eta)  & {} = {k} \,\mathrm{cotanh}\!\left(\frac{\pi {k}}{2}\right)  - q_0 \,\mathrm{cotan}\!\left(\frac{\pi {q_0}}{2}\right) - \sqrt{\eta}\mu  = 0 ,\\
\label{eq:navier_dispersion_relation_b}
G^\dagger_{+}(k,\mu,\eta) & {} = {k}\, \mathrm{tanh}\!\left(\frac{\pi {k}}{2}\right) + q_0 \tan\!\left(\frac{\pi {q_0}}{2}\right) + \sqrt{\eta}\mu = 0,
\end{align}
\end{subequations}
to be compared with (\ref{eq:stokes_exact_determinant}) and (\ref{eq:stokes_penalized_determinant}).

As we conjectured, the numerical evaluation of $\vert \mu^*_{i,\eta} - \mu^\dagger_{i,\eta} \vert$ as a function of ${\eta}$ suggests that it is proportionnal to $\eta$
(Fig.~\ref{fig:convergence_stokes_navier}, middle), and the distance between the eigenfunctions in $L^2(\Omega)$ is of the same order (Fig.~\ref{fig:convergence_stokes_navier}, right).
This supports a leading order description of the error associated to the penalized eigenfunctions by the error coming from the residual slip.
This description has important consequences for applications of the penalization method in computational fluid dynamics,
and it would be of interest to derive a proof for general geometries. 
In the present case, the leading order estimate obtained by the implicit function theorem is the
same as for the penalized eigenvalues (Eq.~\ref{eq:stokes_eigenvalue_convergence}), 
which proves that
\begin{equation}
\label{eq:navier_eigenvalue_convergence}
\mu^\dagger_{i,\eta} =  \mu^*_{i,\eta} + O(\eta).
\end{equation}

	\section{Discretization}

In this section we compare several numerical methods
that can be used to discretize the eigenvalue problems for the penalized Laplace and Stokes operators.

	\subsection{Laplace operator}

	\subsubsection{Numerical methods}

The discretization of penalized operators is not an easy problem because they involve discontinuous mask functions.
In particular, when considering the eigenvalue problem in Fourier space,
a way must be found to truncate the Fourier series of the mask function, or regularize it somehow.
Here, we compare the following approaches:
\begin{enumerate}[label=(\roman{*}), ref=(\roman{*})]
 \item a Fourier collocation method, where the product between the (discontinuous) mask
and the function is evaluated over a regular grid in physical space,
 \item a Fourier-Galerkin method, with a sharp truncation of the discrete Fourier series of the mask function similar to \cite{Schneider2005},
 \item a Fourier-Galerkin method, with a smooth, positivity-preserving truncation of the discrete Fourier series of the mask function (as detailed in Appendix~\ref{appendix_mollification}),
 \item second or fourth order centered finite differences.
\end{enumerate}

Note that the convergence of the collocation scheme (i) for an equivalent eigenvalue problem (but only for fixed $\eta$) was already studied in \cite{Min2003}. 
In order to understand the convergence rate of the collocation scheme, the authors proved estimates
concerning an ``ideal'' Fourier-Galerkin scheme, using the exact Fourier coefficients of the mask function,
and obtained a convergence rate of $-2.5$.
This should not be confused with the problems (ii) and (iii) that are under consideration here,
which rely only on the knowledge of the discrete Fourier coefficients of the mask function.

The Fourier-Galerkin methods are implemented by the pseudo-spectral approach,
which consists in evaluating the product $\chi^\# \psi $ in physical space.
To allow a computation of the Fourier coefficients of $\chi^\# \psi$ without any aliasing error, we take a grid having $N=4K$ points in each direction,
where $K$ is the cut-off wavenumber.
The FFTs are performed using the FFTW library \cite{Frigo2005}.
Whatever the discretization method, we obtain a finite dimensional eigenvalue problem,
which we then solve numerically by a Krylov iteration method,
as efficiently implemented in the Petsc and Slepc libraries \cite{Hernandez2005,Balay2009}.

In the following, we present the results of a parametric study 
where $81$ values of $\eta$, logarithmically equidistant between $10^{-2}$ and $10^{-6}$,
and $16$ values of $N$, of the form $N = 2^i$, $N = 3 \times 2^{i-1}$ where $6 \leq i \leq 13$,
have been screened.

	\subsubsection{Results}

\begin{figure}
\begin{center}
\includegraphics[width=0.32\columnwidth]{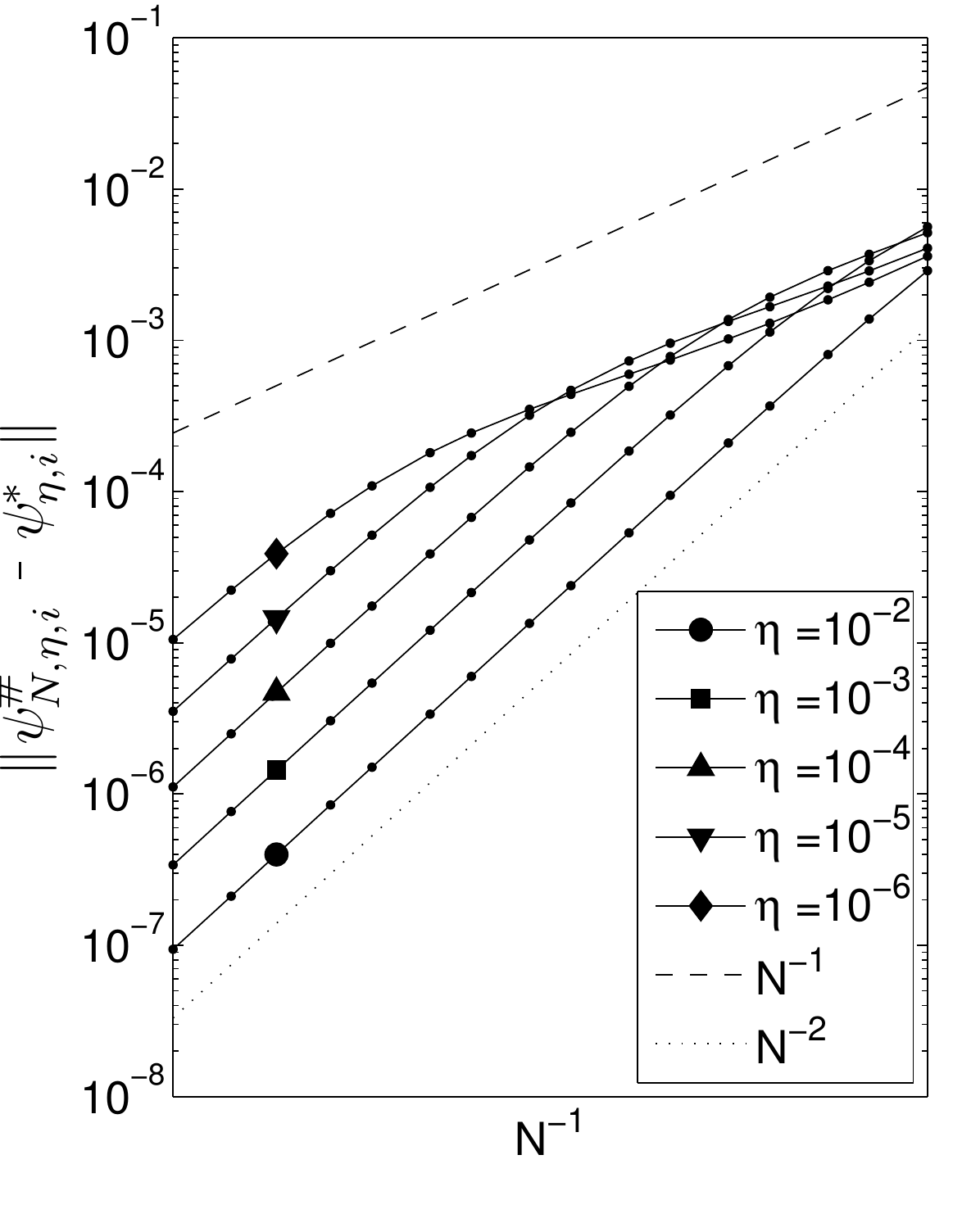}
\includegraphics[width=0.32\columnwidth]{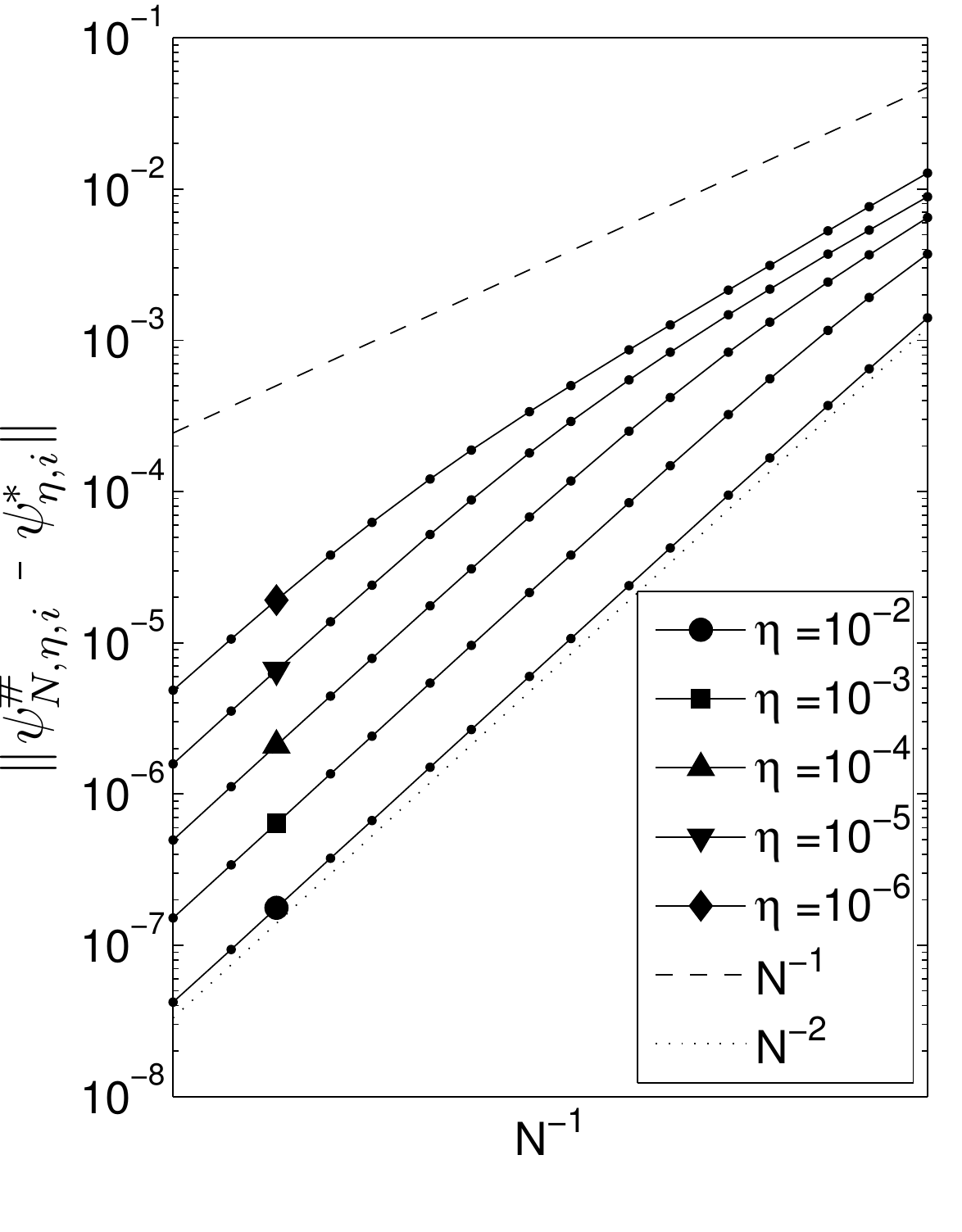}
\includegraphics[width=0.32\columnwidth]{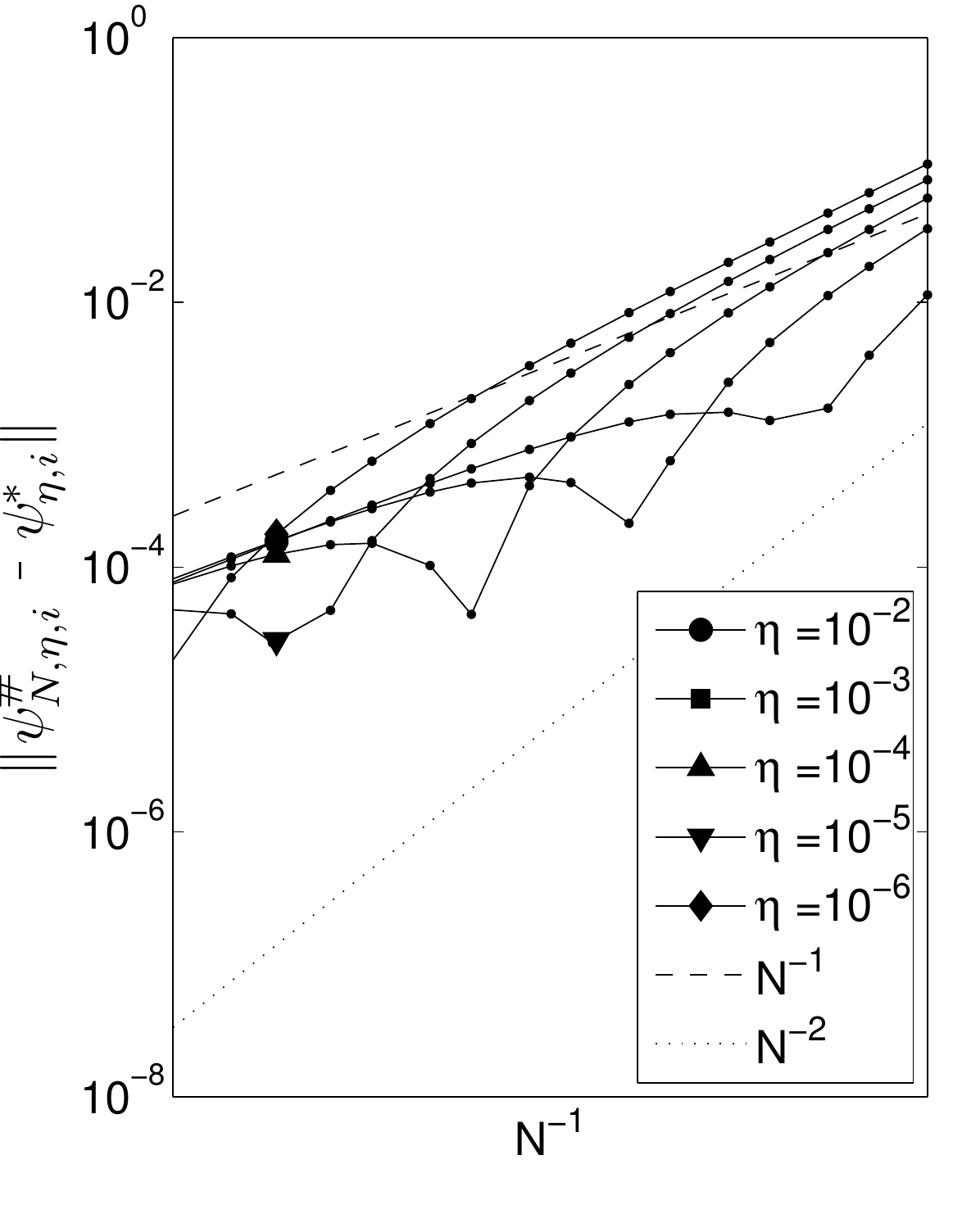}
\end{center}
\caption{
\label{fig:laplace_discrete_error_sol_pen}
$L^2$ distance between the first eigenfunctions of the discrete-penalized Laplace operator
and of the penalized Laplace operator, as a function of $N$ and for varying $\eta$.
Top: Fourier collocation.
Middle: Fourier-Galerkin with sharp truncation.
Right: Fourier-Galerkin with positive mollified mask.                                                                                                  
}
\end{figure}

First, we examine the relationship between the discretized-penalized Laplace operator on the one hand,
and the continuous penalized Laplace operator on the other hand.
In the collocation case, the first eigenfunction converges at a rate $N^{-2}$ (Fig.~\ref{fig:laplace_discrete_error_sol_pen}, left), consistent with the results proved in \cite{Min2003}.
This convergence rate is connected to the regularity of the limiting eigenfunction (see also Appendix~\ref{appendix_poisson}).
The same convergence rate is observed in the sharp Galerkin truncation case (Fig.~\ref{fig:laplace_discrete_error_sol_pen}, left).
Hence the optimal $N^{-2.5}$ convergence rate which would correspond to the ideal Fourier-Galerkin scheme without
truncation of the mask function, and using its exact Fourier coefficients (see \cite{Min2003}), is not attained here.
We have checked that the same behavior is observed for higher order eigenfunctions (up to $i=5$).
In the smooth Galerkin-truncation case, the mollification of the mask introduces perturbations of order $N^{-1}$ in the equation,
which results in perturbations of the same order on the eigenfunctions (Fig.~\ref{fig:laplace_discrete_error_sol_pen}, right).
The error displays a faster decay for sufficiently small values of $N$.

\begin{figure}
\begin{center}
\includegraphics[width=0.32\columnwidth]{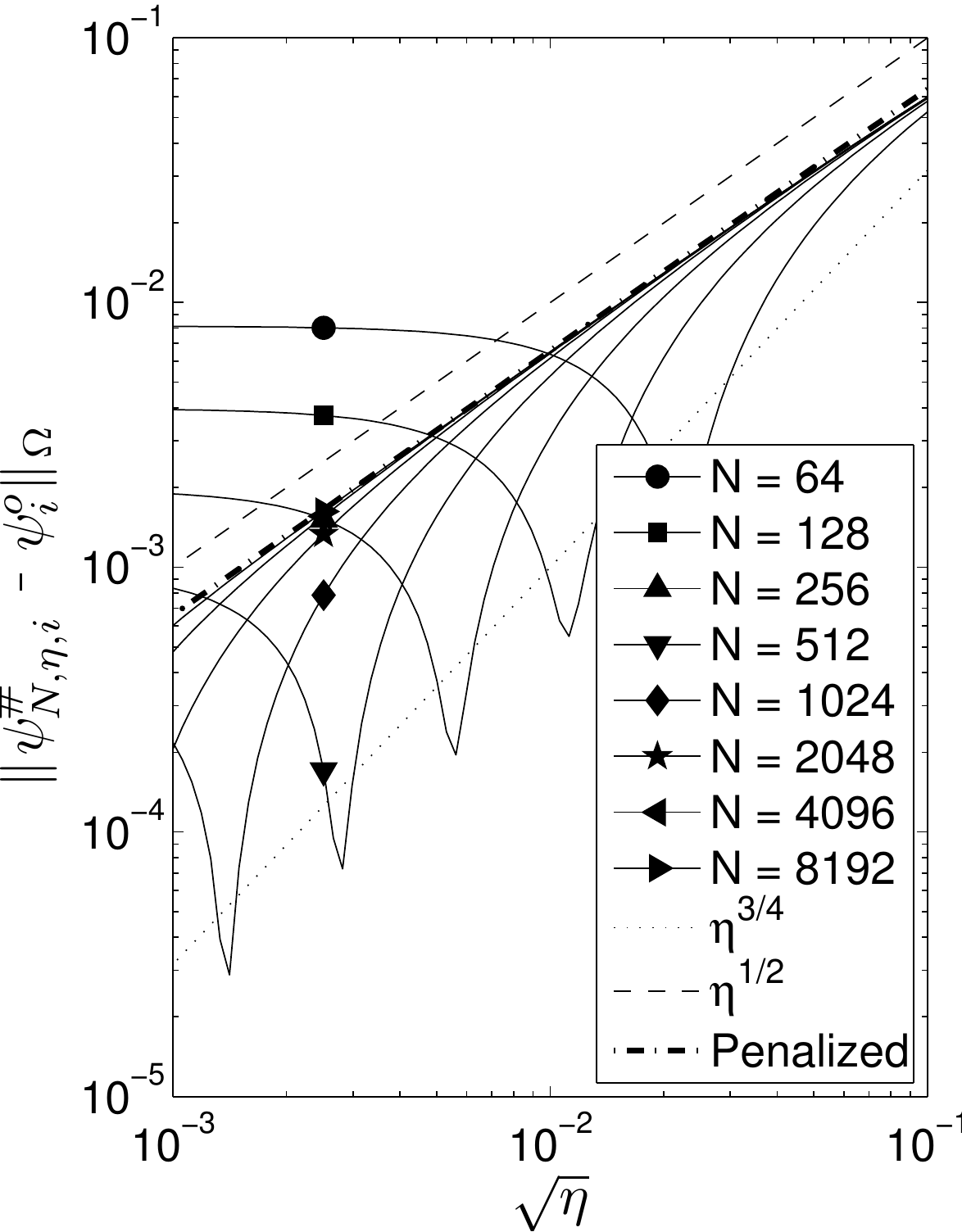}
\includegraphics[width=0.32\columnwidth]{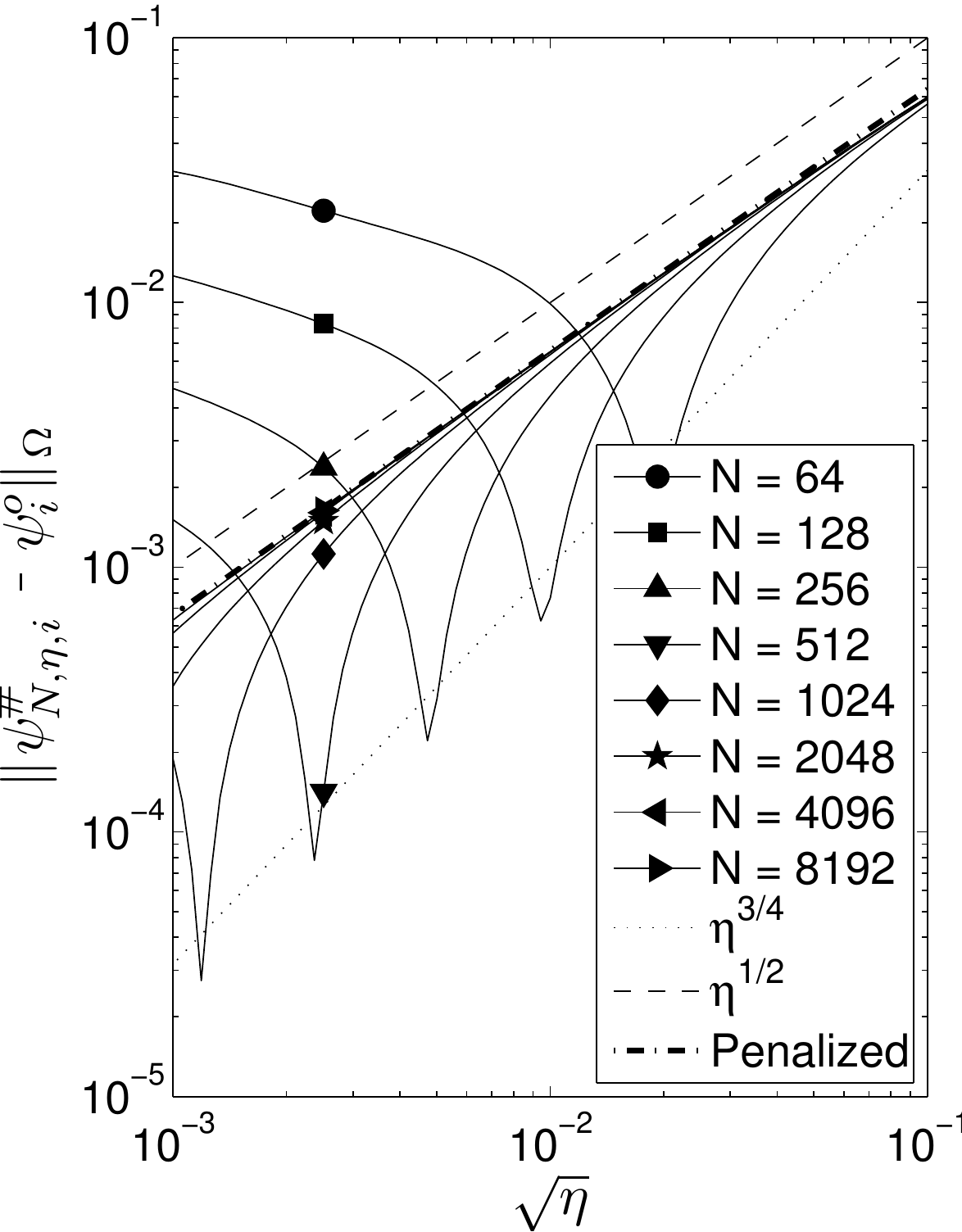}
\includegraphics[width=0.32\columnwidth]{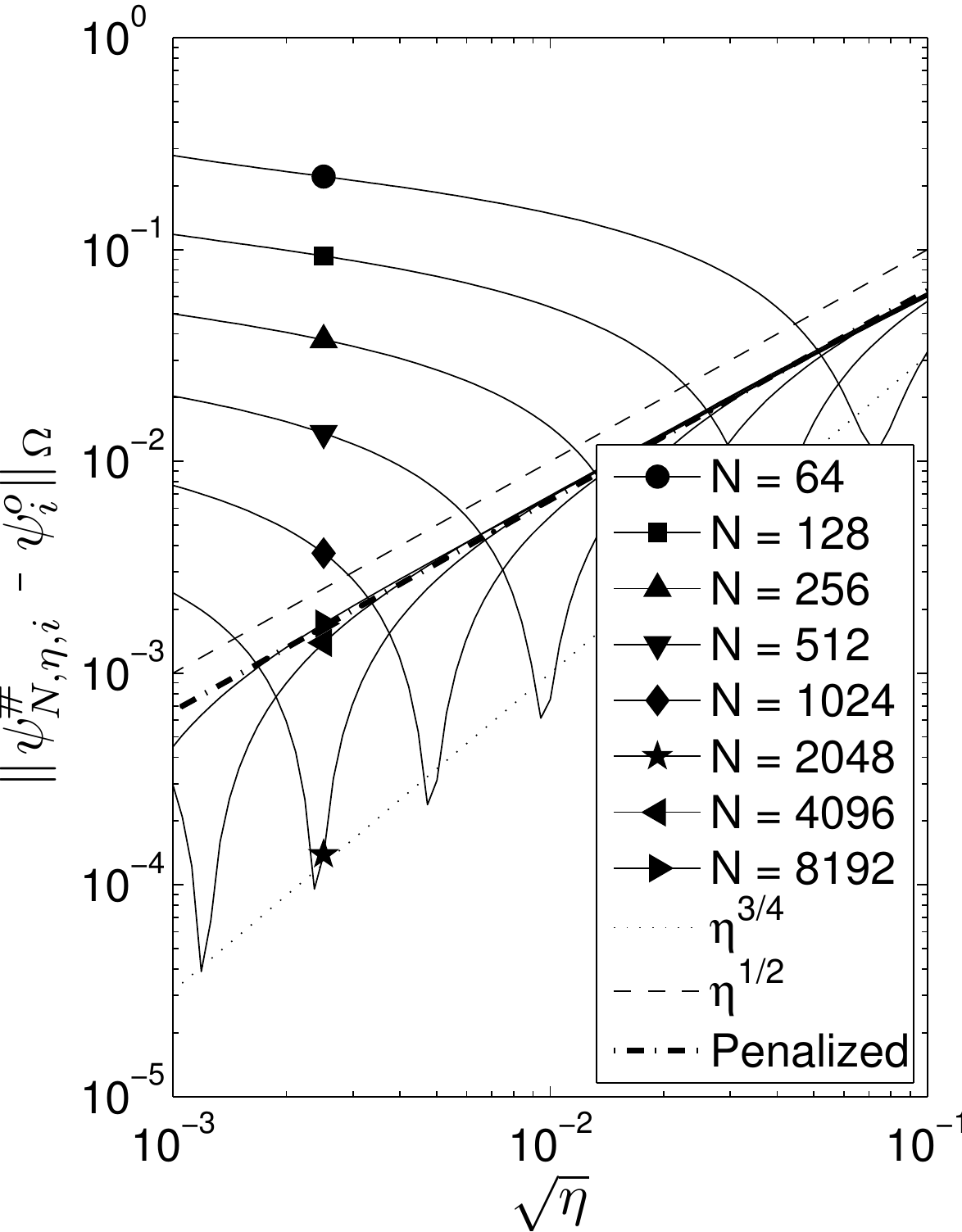}
\end{center}
\caption{
\label{fig:laplace_discrete_error_sol_exa}
$L^2$ distance between the first eigenfunctions of the discrete-penalized Laplace operator
and of the exact Laplace operator with Dirichlet b.c., as a function of $\sqrt{\eta}$ and for varying $N$.
For comparison, the dash-dotted line recalls the error for the exact penalized eigenfunction.
Left: Fourier collocation.
Middle: Fourier-Galerkin with sharp truncation.
Right: Fourier-Galerkin with positive mollified mask. 
}
\end{figure}

We now turn to the distances between the first eigenfunction of the discretized-penalized Laplace operator
and of the exact Laplace operator with Dirichlet b.c., as depicted in Fig.~\ref{fig:laplace_discrete_error_sol_exa}
as a function of $\sqrt{\eta}$.
As expected, for small $\eta$ the $O(\sqrt{\eta})$ behavior of the penalization error dominates.
Interestingly, for all three schemes the error locally drops down to a minimum which is of order $\eta^{3/4}$.
Then, for the collocation scheme a plateau is observed, where the error saturates,
whereas for the two other methods, the error seems to increase without bound for smaller and smaller $\eta$.
The presence of the $\eta^{3/4}$ minimum came as a surprise to us, since it cannot be explained by naive bounds such as (\ref{eq:total_norm_error_asy0}) given by the triangle inequality, and must be due to unexpected cancellations between the penalization and discretization errors.
We comment on its significance further down.

\begin{figure}
\begin{center}
\includegraphics[width=0.32\columnwidth]{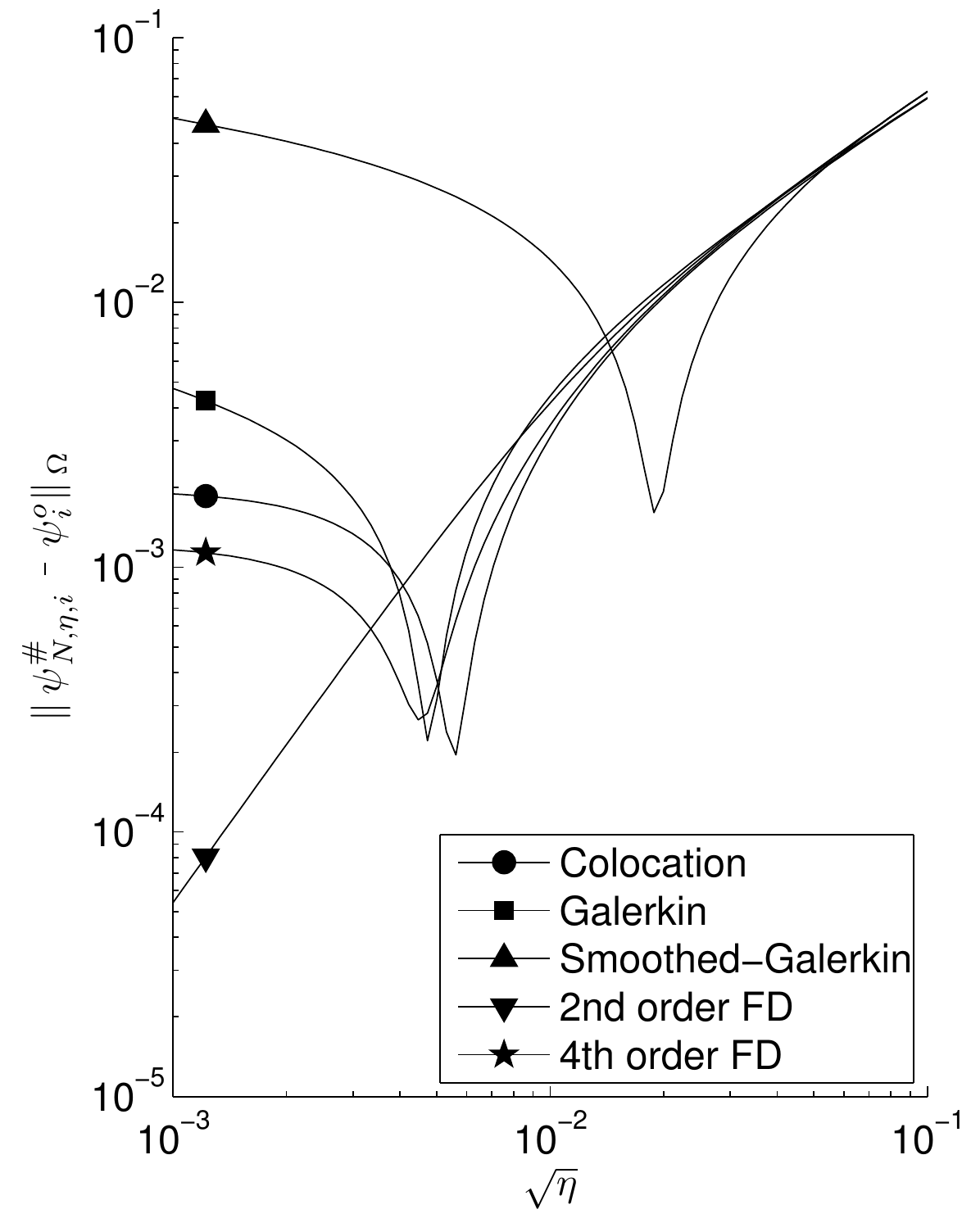}
\includegraphics[width=0.32\columnwidth]{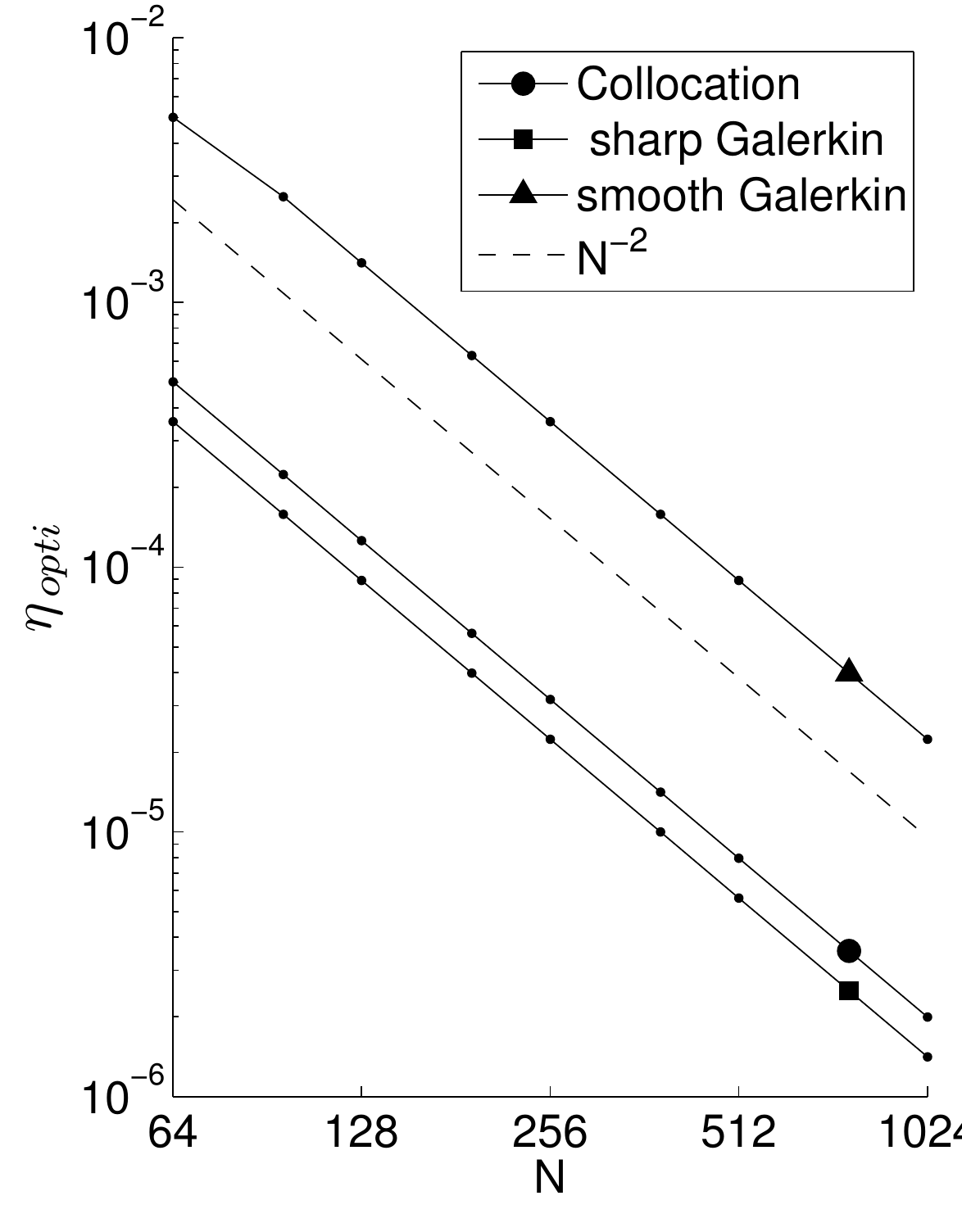}
\includegraphics[width=0.32\columnwidth]{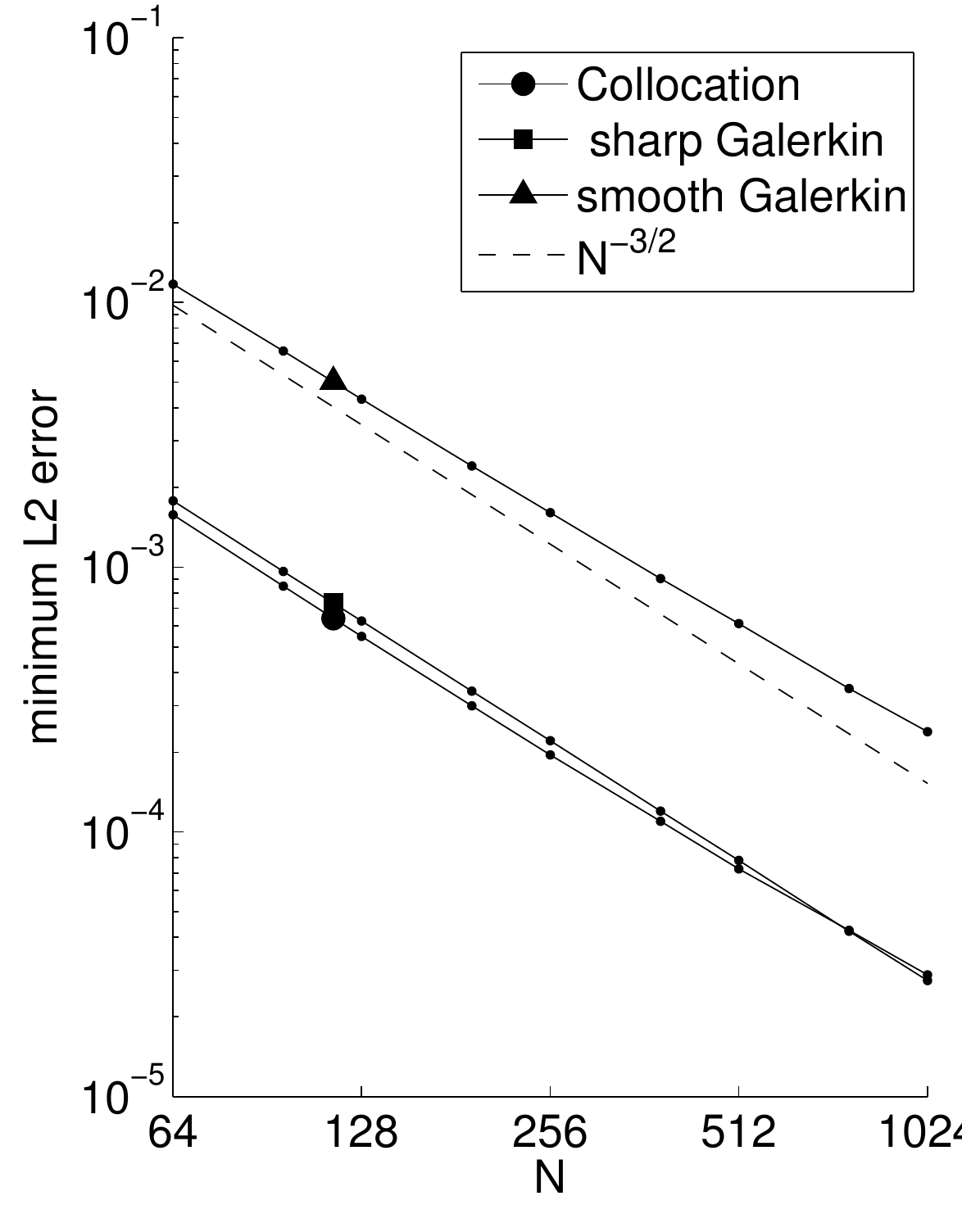}
\end{center}
\caption{
\label{fig:laplace_discrete_error_sol_comparison}
Comparison of various discretization schemes.
Left: $L^2$ distance between the first eigenfunctions of the discrete-penalized Laplace operator and of the exact Laplace operator with Dirichlet b.c., as a function of $\sqrt{\eta}$, for $N = 256$.
Middle: Value $\eta_{opti}$ of $\eta$ at which the minimum error between the first eigenfunctions of the discrete-penalized and exact Laplace operators is attained,
as a function of $N$.
Right: corresponding minimal $L^2$ error.
}
\end{figure}

In Fig.~\ref{fig:laplace_discrete_error_sol_comparison} (left) the error associated to all considered schemes is plotted as a function of $\sqrt{\eta}$ for a fixed value of $N$,
which allows for a direct and quantitative comparison. 
In addition, the errors obtained when using 2nd and 4th order finite differences schemes are shown.
The collocation, truncated-Galerkin, and 4th order finite differences schemes have similar properties,
while the smoothed-Galerkin scheme saturates at a value of the error which is roughly $10$ times larger.
The behavior of the 2nd order finite difference scheme stands out, since the range of $\eta$ where
the error saturates is not even reached in our study.

In practice, $\eta$ is an adjustable parameter which should be chosen in order to minimize the error with the eigenfunction
one wishes to approximate.
We denote by $\eta_{opti}$ such a well-chosen value of $\eta$.
Its behavior as a function of $N$, and the actual value of the error attained for $\eta = \eta_{opti}$,
are shown in Fig.~\ref{fig:laplace_discrete_error_sol_comparison} (middle) and Fig.~\ref{fig:laplace_discrete_error_sol_comparison} (right), respectively.

It appears that $\eta_{opti}$ scales like $N^{-2}$, similarly to what we found for the Poisson problem
presented in the introduction.
However, the resulting error when choosing such an optimal $\eta$ scales like $N^{-3/2}$
and not $N^{-1}$ as we had expected from the Poisson example.
Note that the range of $\eta$ for which these cancellations occurs
is relatively narrow and may be difficult to estimate in many cases of practical interest.

	\subsection{Stokes operator}

	\subsubsection{Numerical method}

In the case of the Stokes operator, we leave aside finite differences due to technical constraints,
and we concentrate on the collocation, truncated-Galerkin and smoothed-Galerkin approaches.
The problem is first reduced to a scalar problem by taking the curl of the eigenvalue equation,
which yields an equation for the vorticity $\omega$ of the eigenfunction:
$$
-\Delta \omega + \frac{1}{\eta} \left(\partial_x (\chi^\# u_y) - \partial_y (\chi^\# u_x)\right) = \mu \omega
$$
The dependency on $y$ is then eliminated by taking the Fourier transform in the $y$ direction,
after which all modes with different wavenumbers become decoupled, and $k$ becomes simply a parameter.
We thus obtain a set of uncoupled one-dimensional problems which, although not Hermitian symmetric, 
can be solved for up to $N=2^{12}$ on a standard personnal computer within a few hours
using, as above, the Petsc/Slepc libraries.
Once this is done, the velocity field is reconstructed from $\omega$ by applying the Biot-Savart operator in Fourier space.
Otherwise, the procedure and the values of $N$ and $\eta$ are the same as in the Laplace case as reported above.

	\subsubsection{Results}

\begin{figure}
\begin{center}
\includegraphics[width=0.32\columnwidth]{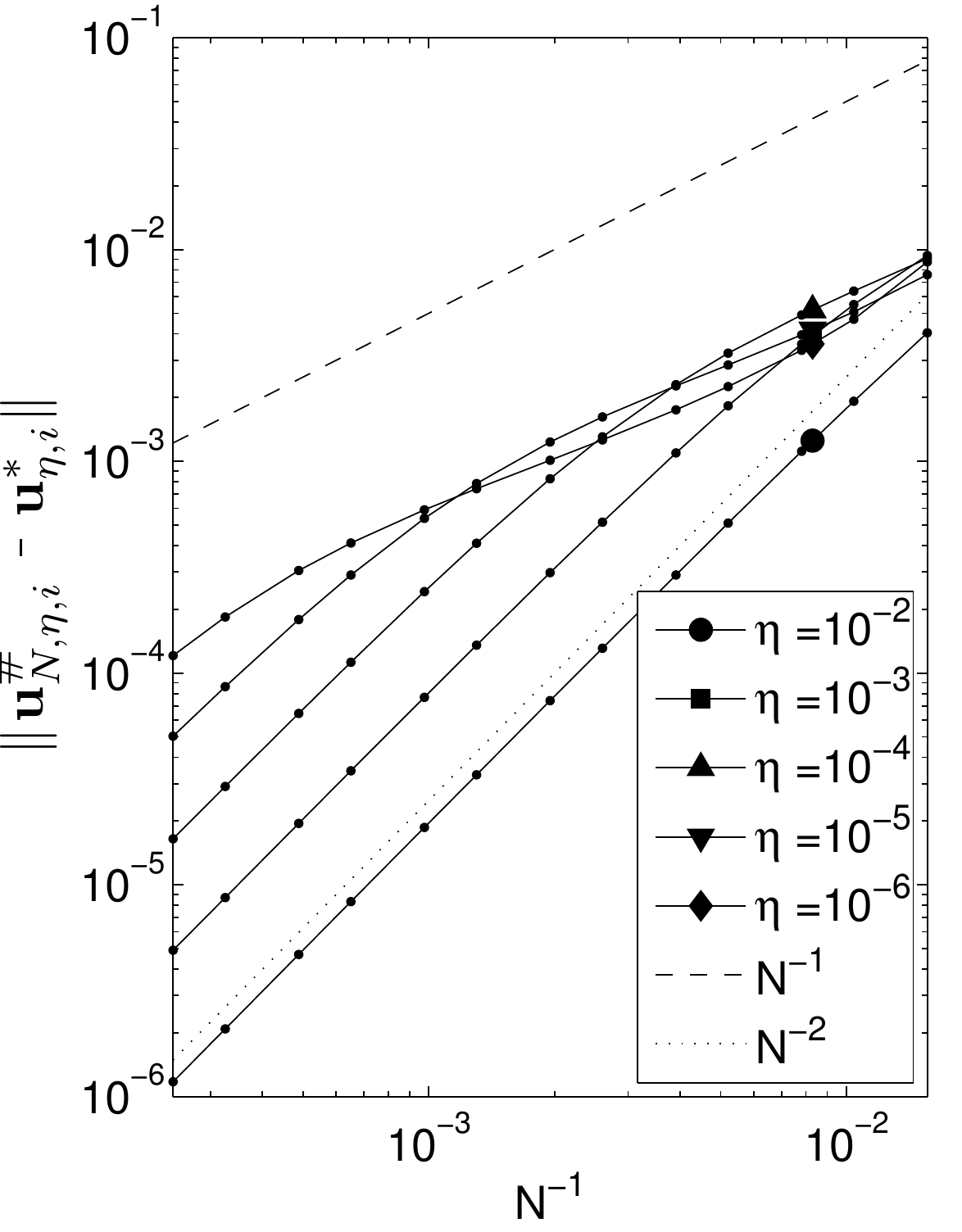}
\includegraphics[width=0.32\columnwidth]{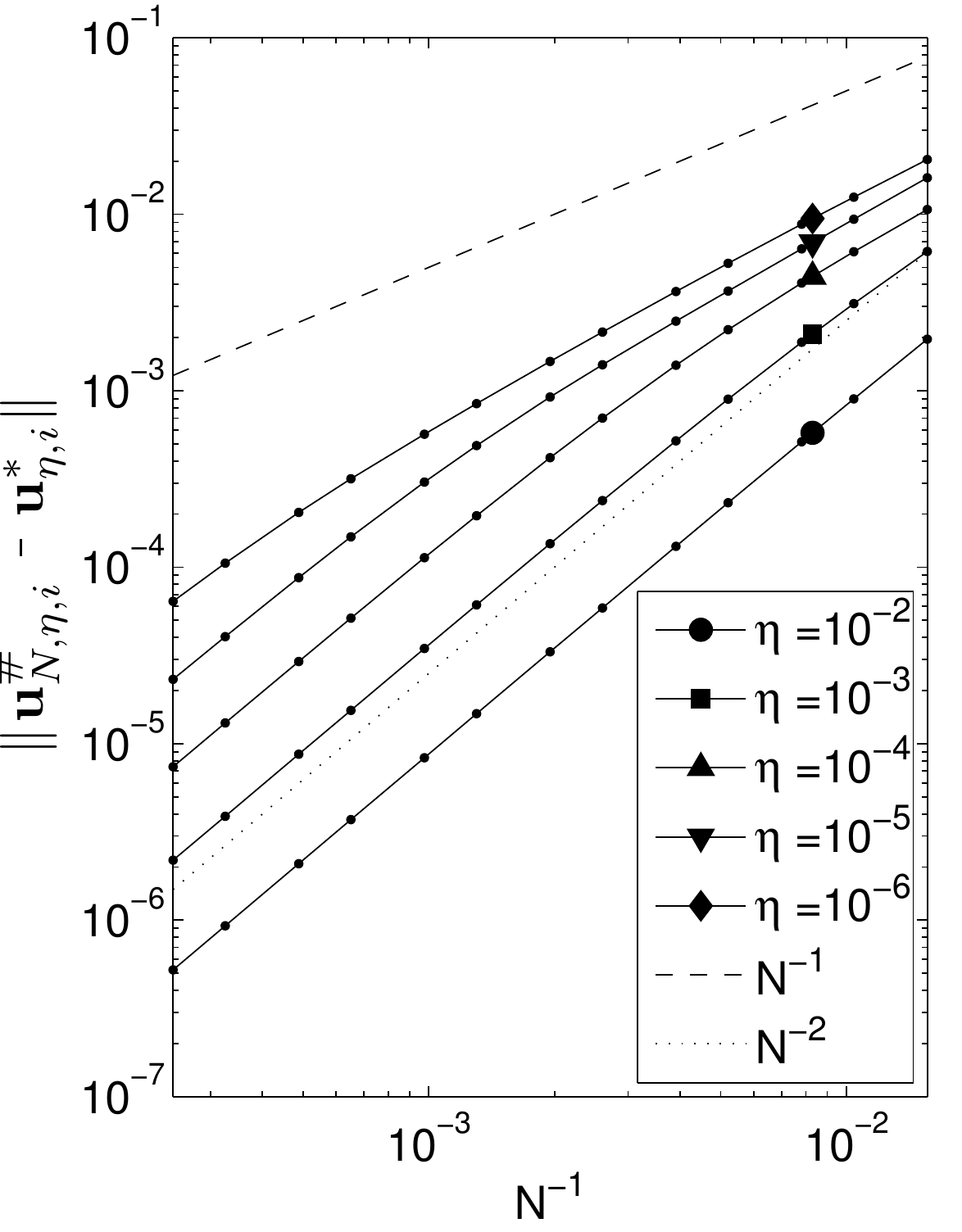}
\includegraphics[width=0.32\columnwidth]{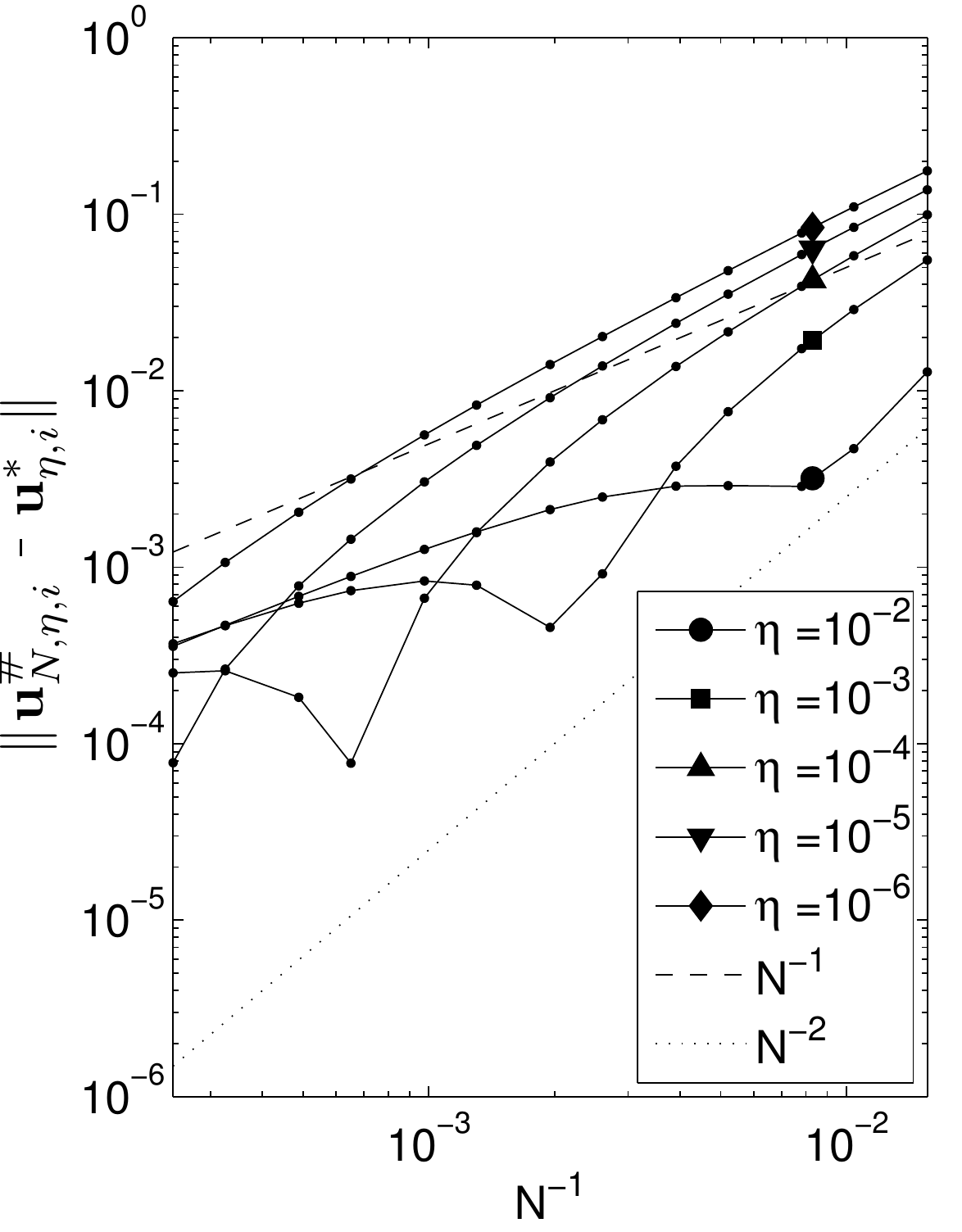}
\end{center}
\caption{
\label{fig:stokes_discrete_error_sol_pen}
$L^2$ distance between the first eigenfunctions of the discrete-penalized Stokes operator
and of the penalized Stokes operator, as a function of $N$ and for varying $\eta$.
Left: Fourier collocation.
Middle: Fourier-Galerkin with sharp truncation.
Right: Fourier-Galerkin with positive mollified mask.                                                                                                  
}
\end{figure}

\begin{figure}
\begin{center}
\includegraphics[width=0.32\columnwidth]{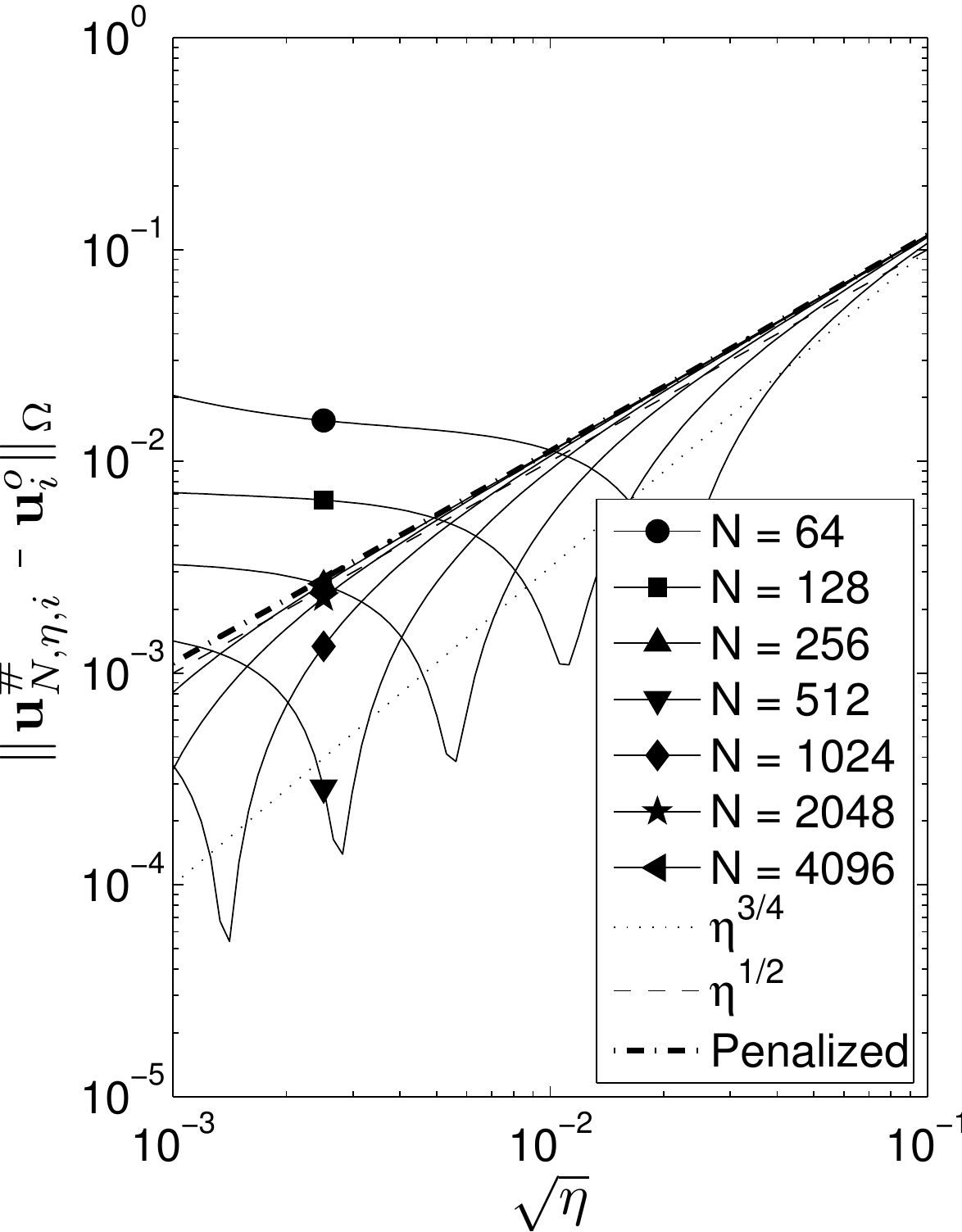}
\includegraphics[width=0.32\columnwidth]{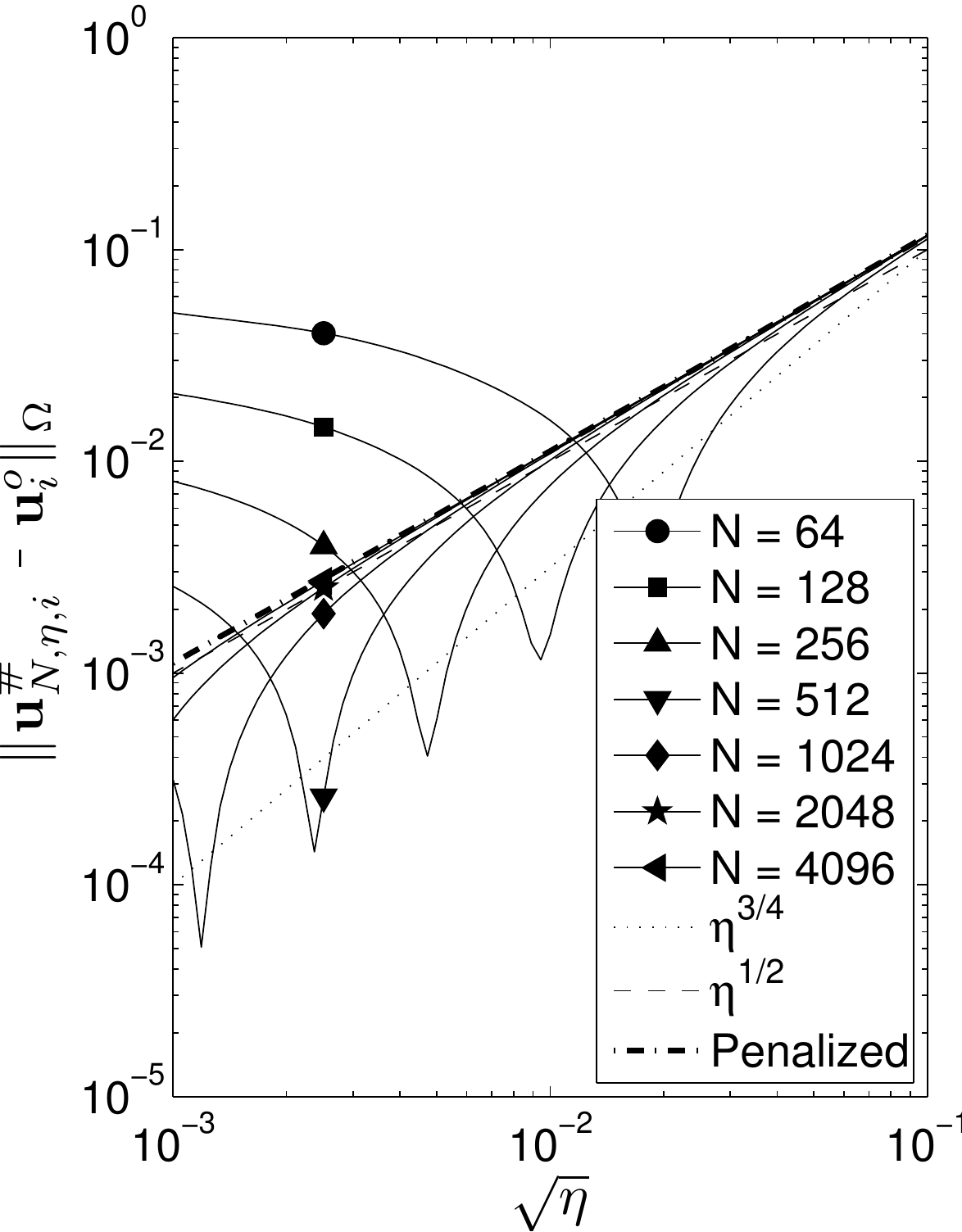}
\includegraphics[width=0.32\columnwidth]{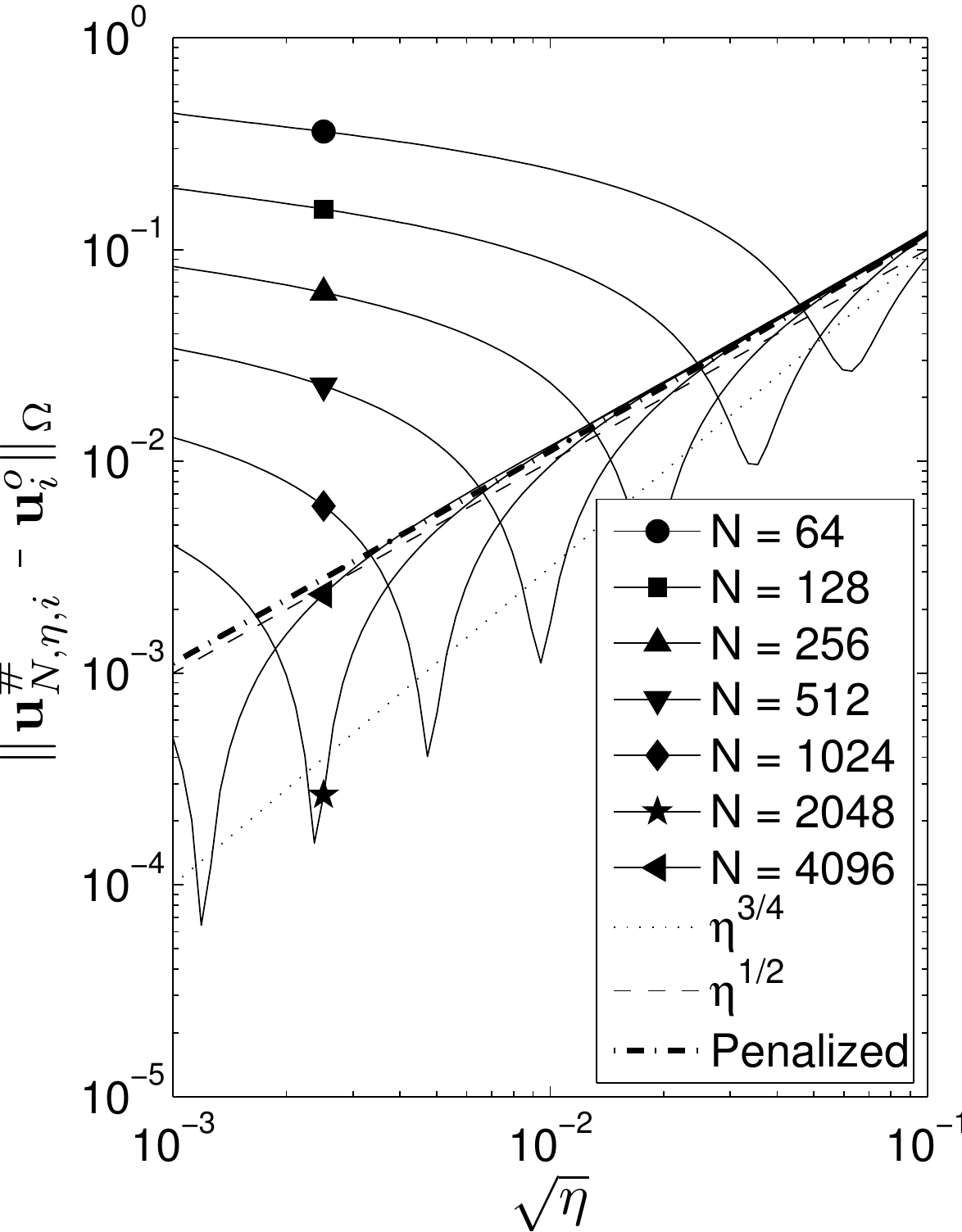}
\end{center}
\caption{
\label{fig:stokes_discrete_error_sol_exa}
$L^2$ distance between the first eigenfunctions of the discrete-penalized Stokes operator
and of the exact Stokes operator with Dirichlet b.c., as a function of $\sqrt{\eta}$ and for varying $N$.
For comparison, the dash-dotted line recalls the error for the exact penalized eigenfunction.
Left: Fourier collocation.
Middle: Fourier-Galerkin with sharp truncation.
Right: Fourier-Galerkin with positive mollified mask. 
}
\end{figure}

Concerning the discretization error (Fig.~\ref{fig:stokes_discrete_error_sol_pen}),
as well as the total error (Fig.~\ref{fig:stokes_discrete_error_sol_exa}),
we do not observe any notable difference with the penalized Laplace operator.
We have checked that this also holds for higher order eigenfunctions (up to $k=3$ and $l=5$).

\begin{figure}
\begin{center}
\includegraphics[width=0.32\columnwidth]{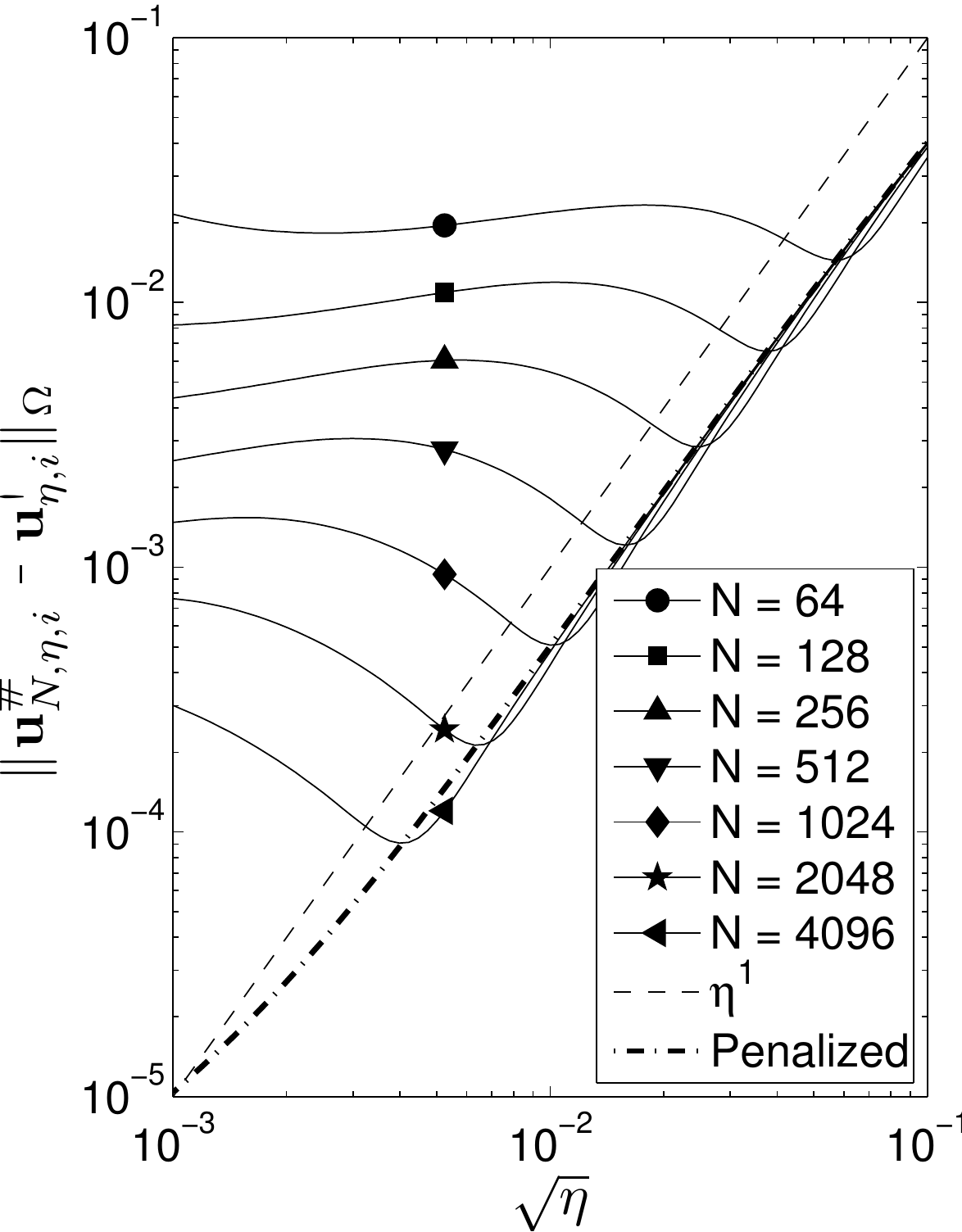}
\includegraphics[width=0.32\columnwidth]{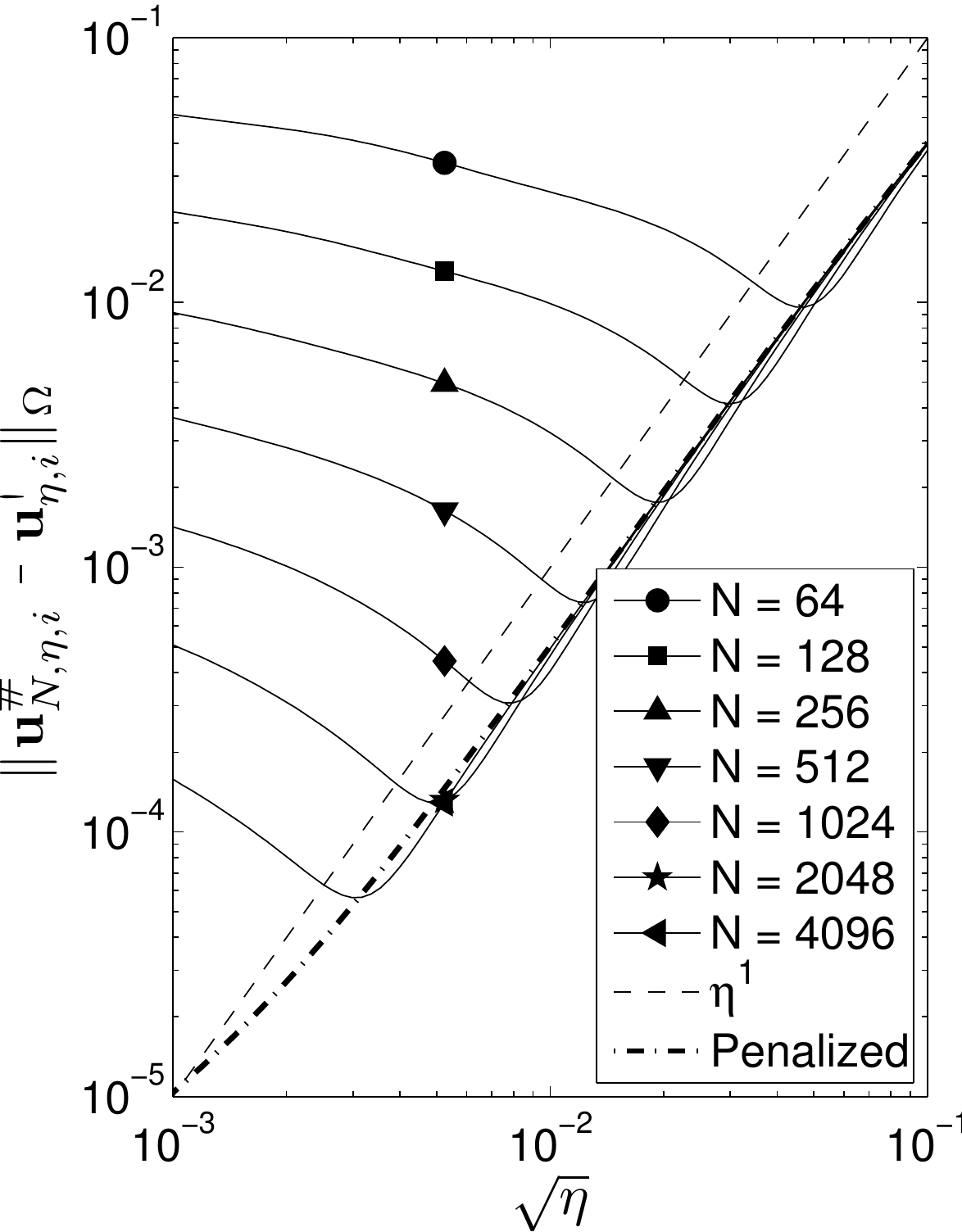}
\includegraphics[width=0.32\columnwidth]{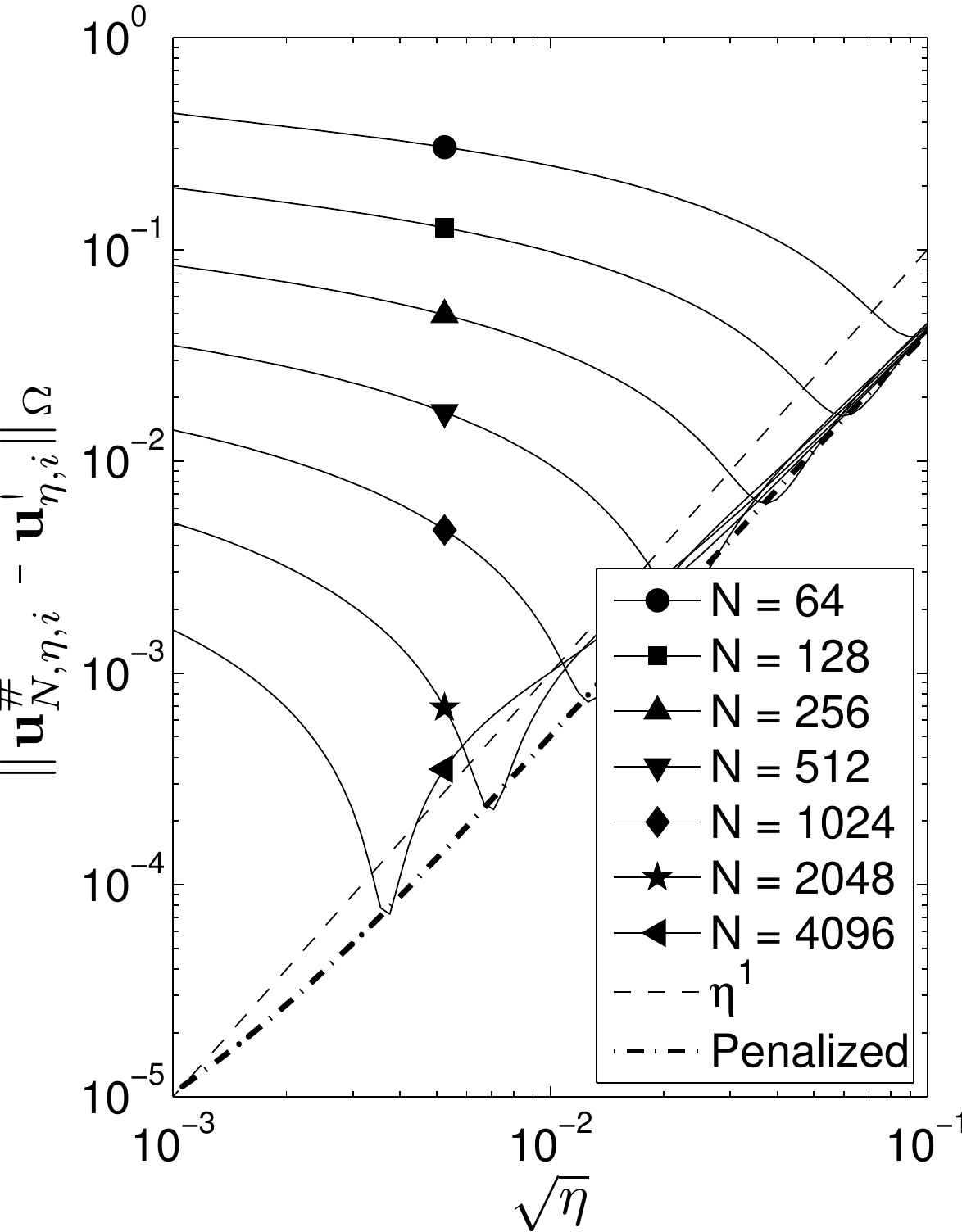}
\end{center}
\caption{
\label{fig:stokes_discrete_error_sol_nav}
$L^2$ distance between the first eigenfunctions of the discrete-penalized Stokes operator
and of the exact Stokes operator with Navier b.c., as a function of $\sqrt{\eta}$ and for varying $N$.
Left: Fourier collocation.
Middle: Fourier-Galerkin with sharp truncation.
Right: Fourier-Galerkin with positive mollified mask. 
}
\end{figure}

The comparison with the first eigenfunction of the exact Stokes operator with Navier b.c. with slip length $\sqrt{\eta}$ is depicted in Fig.~\ref{fig:stokes_discrete_error_sol_nav}.
The error scales like $O(\eta)$ for large $\eta$, as expected from the analysis of the penalized eigenfunctions in the previous section,
and does not have such a sharp minimum as in the Dirichlet case.

\begin{figure}
\begin{center}
\includegraphics[width=0.32\columnwidth]{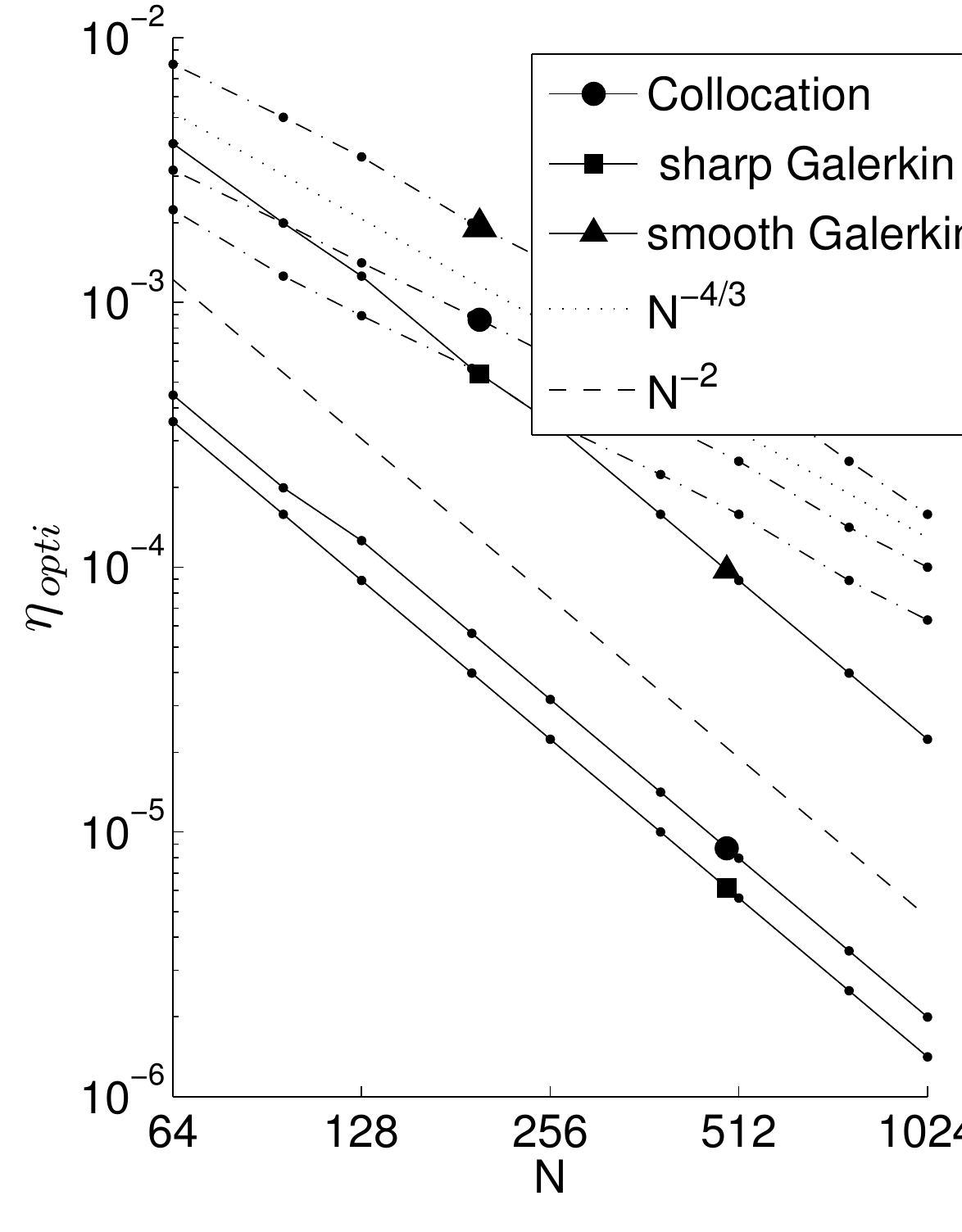}
\includegraphics[width=0.32\columnwidth]{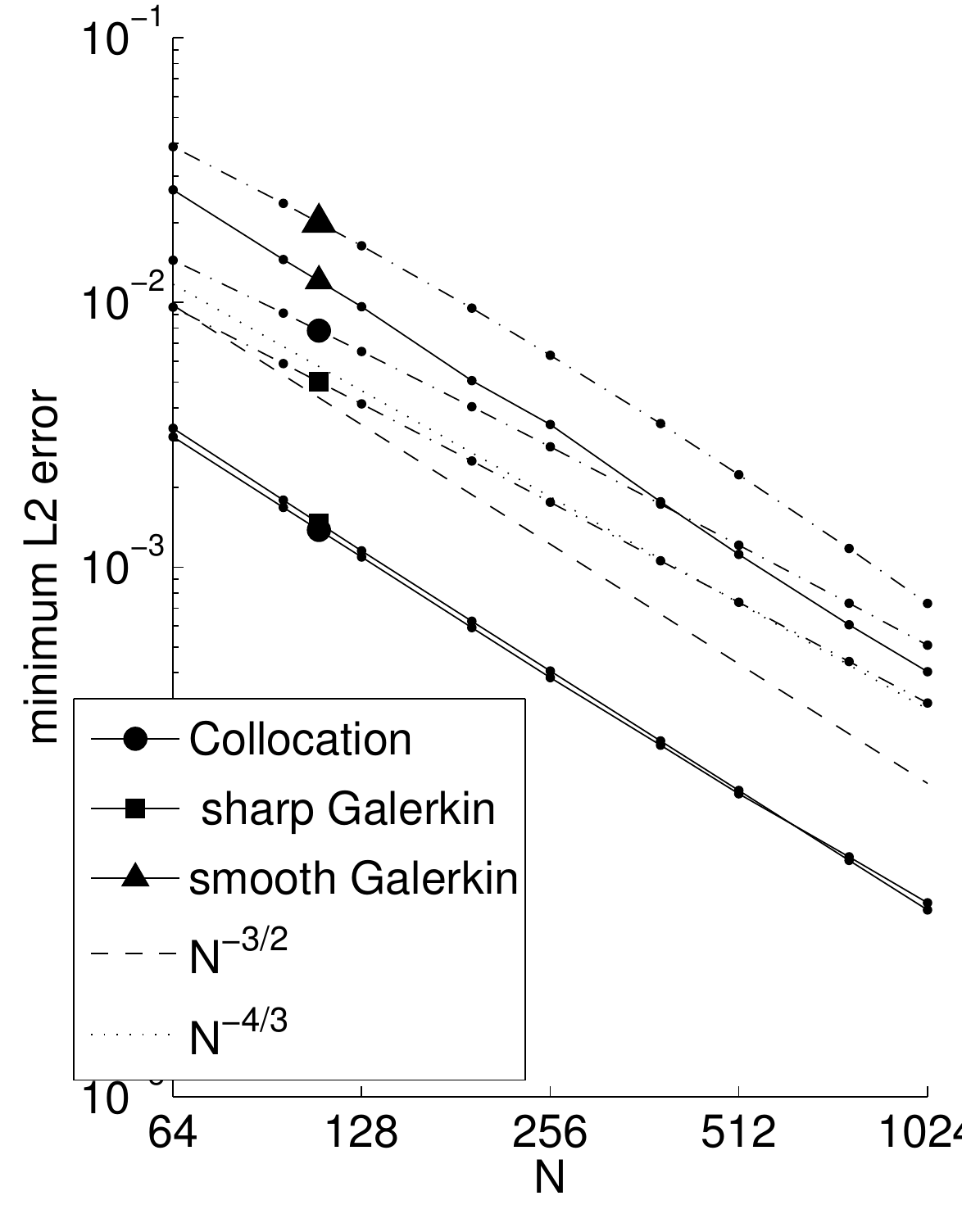}
\end{center}
\caption{
\label{fig:stokes_discrete_error_sol_comparison}
Comparison of various discretization schemes.
Left: 
value $\eta_{opti}$ of $\eta$ at which the minimum error between the first eigenfunctions of the discrete-penalized and exact Stokes operators is attained,
as a function of $N$.
Right: corresponding minimal $L^2$ error.
The solid lines corresponds to Dirichlet boundary conditions,
and the dash-dotted lines correspond to Navier boundary conditions with a slip-length $\sqrt{\eta}$.
}
\end{figure}

In the case of Dirichlet b.c., the behaviors of $\eta_{opti}$ and of the minimal error (Fig.~\ref{fig:stokes_discrete_error_sol_comparison})
are analogous to what we have seen above in the case of the Laplace operator,
namely $\eta_{opti}$ scaling like $N^{-2}$ and the optimal error like $N^{-3/2}$.
For Navier boundary conditions, we find the much weaker scaling $N^{-4/3}$ for $\eta_{opti}$,
which means that the optimal error also scales like $N^{-4/3}$.
Although penalized eigenfunctions for a given $\eta$ are much closer to Navier eigenfunctions with slip length $\sqrt{\eta}$ than to Dirichlet eigenfunctions,
this advantage seems to be destroyed by the discretization when $\eta$ gets smaller than $N^{-4/3}$.

The fact that the error observed with respect to Navier b.c. is higher than with respect to Dirichlet b.c. is 
somewhat disappointing.
It is related to the occurence of unexpected cancellations in the Dirichlet case, which do not occur in the Navier case.
A possibility which remains, to be investigated in the general case, would be that the discrete eigenfunctions are better described
by Navier boundary conditions with a slip length different from $\sqrt{\eta}$.


	\section{Application: dipole-wall collision}

In this section we briefly illustrate the connection between the behavior described above for the linear Stokes eigenvalue problem,
and the initial-boundary value problem for the incompressible Navier-Stokes equations:
\begin{equation}
\label{eq:navier_stokes}
\mathrm{(NSE)}:
\begin{cases}
\partial_t \uu + (\uu \cdot \nab) \uu = -\nab p + \nu \Delta \uu \\
\nab \cdot \uu = 0 \\
\uu_{\mid \partial \Omega} = \mathbf{0}, \quad \uu(0, \xx) = \uu_0, 
\end{cases}
\end{equation}
where $\nu$ is the kinematic viscosity, and the spatial domain $\Omega = ]0,\pi[ \times \TT$ is as before.
The penalized version of (\ref{eq:navier_stokes}) is commonly written as:
\begin{equation}
\label{eq:penalized_navier_stokes}
\mathrm{(PNSE)}:
\begin{cases}
\partial_t \uu + (\uu \cdot \nab) \uu = -\nab p + \nu \Delta \uu - \frac{1}{\eta'} \chi \uu \\
\nab \cdot \uu = 0 \\
\uu(0, \xx) = \uu_0, 
\end{cases}
\end{equation}
where $\xx$ now varies in $\TT^2$.
Note that $\eta'$ in this equation has the dimension of an inverse time, whereas
$\eta$, the penalization parameter as we have considered it up to now, is a squared length. 
In order to connect the two quantities, we have to rewrite (\ref{eq:penalized_navier_stokes}) as follows:
\begin{equation}
\partial_t \uu + \mathbb{P}[(\uu \cdot \nab) \uu] = \nu A^*_{\eta} \uu ,
\end{equation}
where $\eta := \eta'\nu$ is now the appropriately scaled parameter for the penalized Stokes operator as defined by (\ref{eq:penalized_stokes_operator}),
and $\mathbb{P}$ is the Leray projector on $\TT^2$.

\begin{table}
\begin{center}
 \begin{tabular}{c|cccc}
& $N$ & $\eta'$ & $\eta := \nu\eta'$ & $\sqrt{\eta}$\\
\hline
I & $1024$ & $0.63$ & $9.87 \cdot 10^{-5}$ & $9.93 \cdot 10^{-3}$  \\      
II & $2048$ & $0.25$ & $3.95 \cdot 10^{-5}$ &$6.28 \cdot 10^{-3}$ \\
III & $4096$ & $0.10$ & $1.58 \cdot 10^{-5}$ &$3.97 \cdot 10^{-3}$ \\
IV & $8192$ & $0.040$ & $6.32 \cdot 10^{-6}$ & $2.51 \cdot 10^{-3}$
 \end{tabular}
\end{center}
\caption{
\label{tab:parameters}
Parameters used for the four considered Navier-Stokes test cases.
}
\end{table}

For spatial discretization, we use the Fourier-Galerkin scheme with positively smoothed mask function, with a pseudo-spectral implementation on an $N \times N$ grid for efficiency.
Time discretization is achieved with a third order low storage Runge-Kutta method \cite[page 20]{Orlandi2000}, and the timestep is adjusted dynamically to satisfy the CFL condition.
To analyze the influence of the penalization and discretization parameters on the solution of (PNSE),
we shall vary $\eta$ and $N$ only and keep all the other parameters fixed.
$N$ is varied by powers of $2$ from $1024$ to $8192$ (see Table~\ref{tab:parameters}),
and the values of $\eta$ are chosen according to the expected behavior of the optimal $\eta$ according to Fig.~\ref{fig:stokes_discrete_error_sol_comparison},
\textit{i.e.} with the scaling $\eta \propto N^{-4/3}$.

\begin{figure}
\begin{center}
\includegraphics[width=0.49\columnwidth]{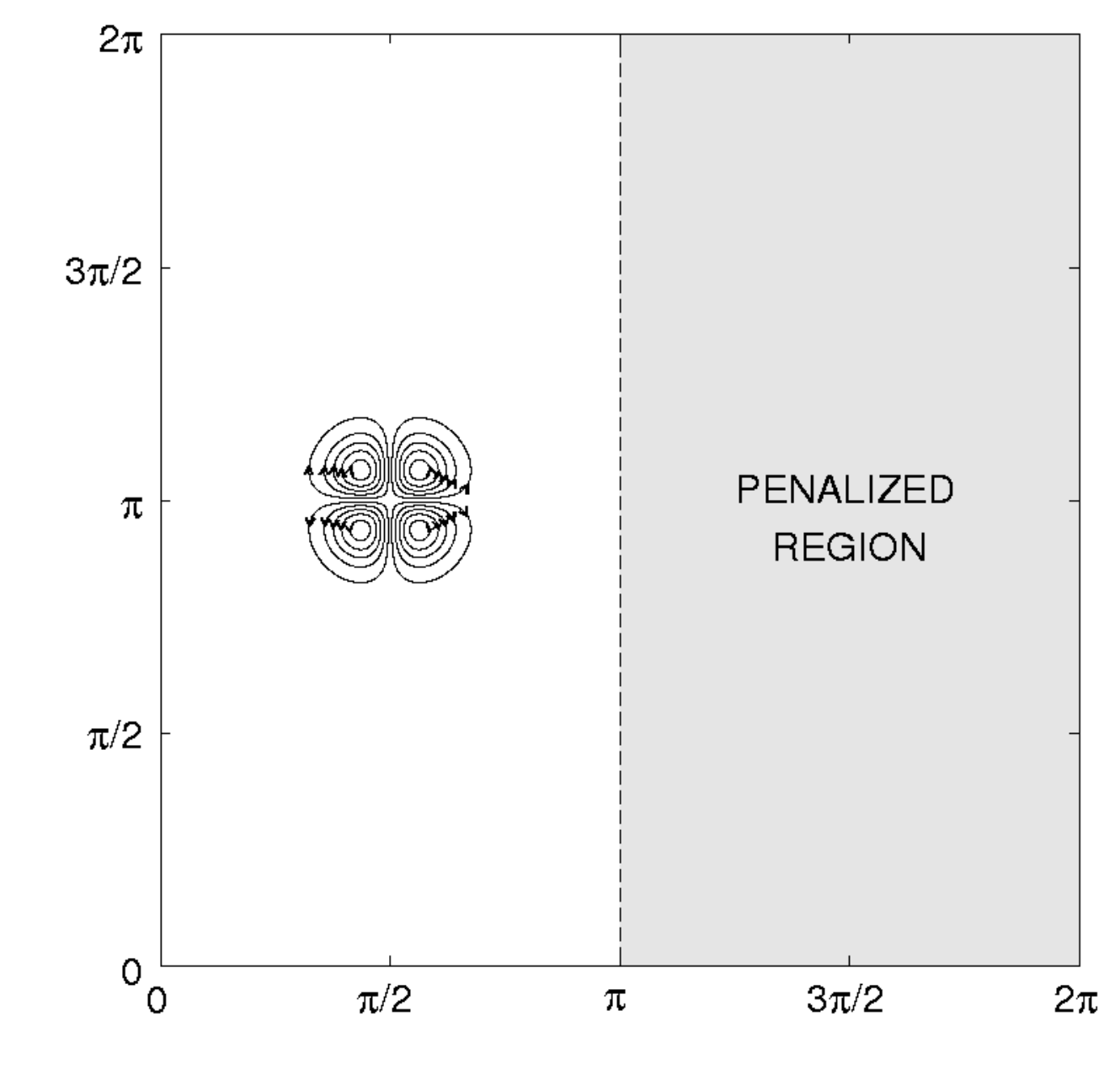}
\includegraphics[width=0.49\columnwidth]{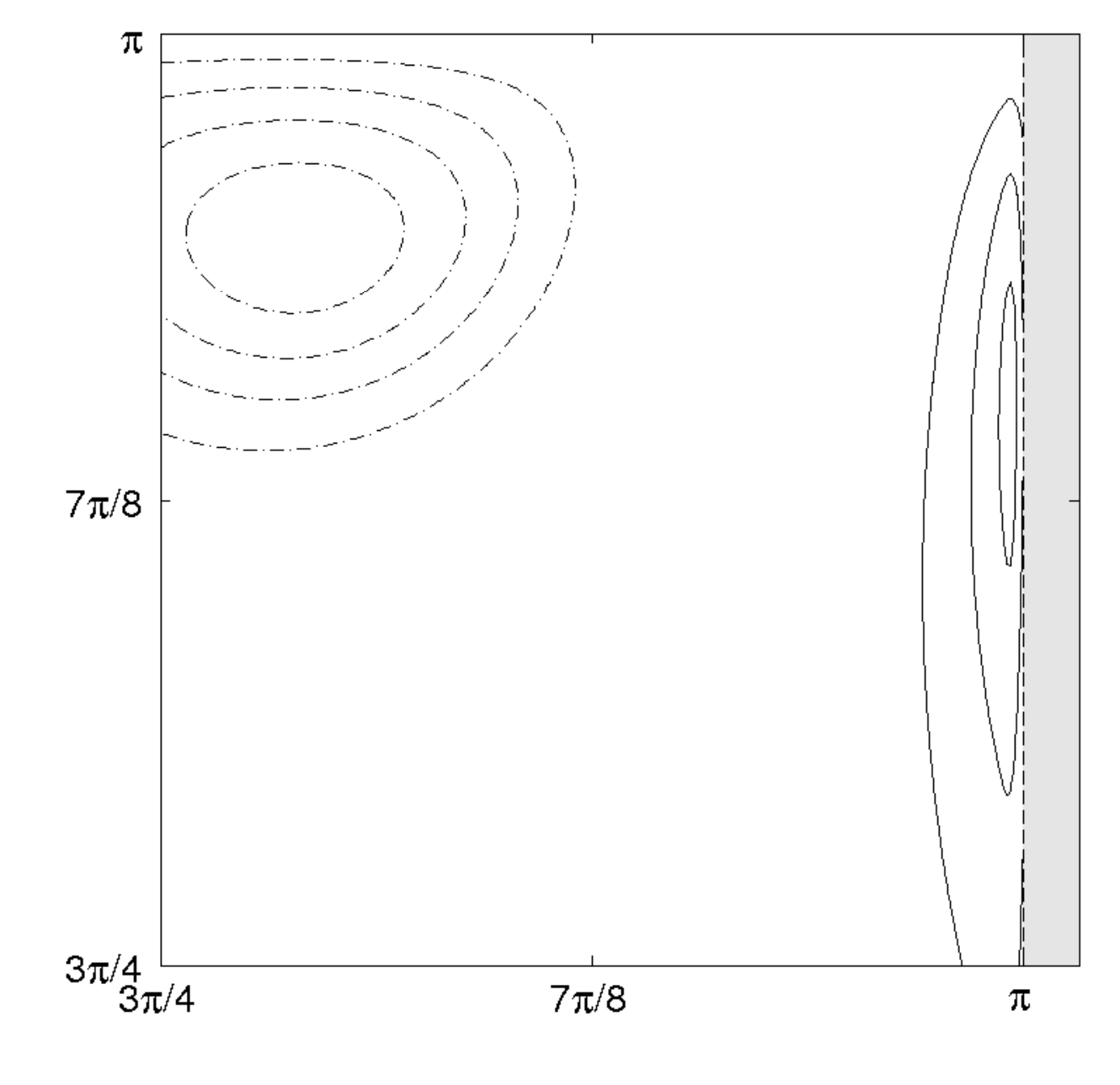} \\
\includegraphics[width=0.49\columnwidth]{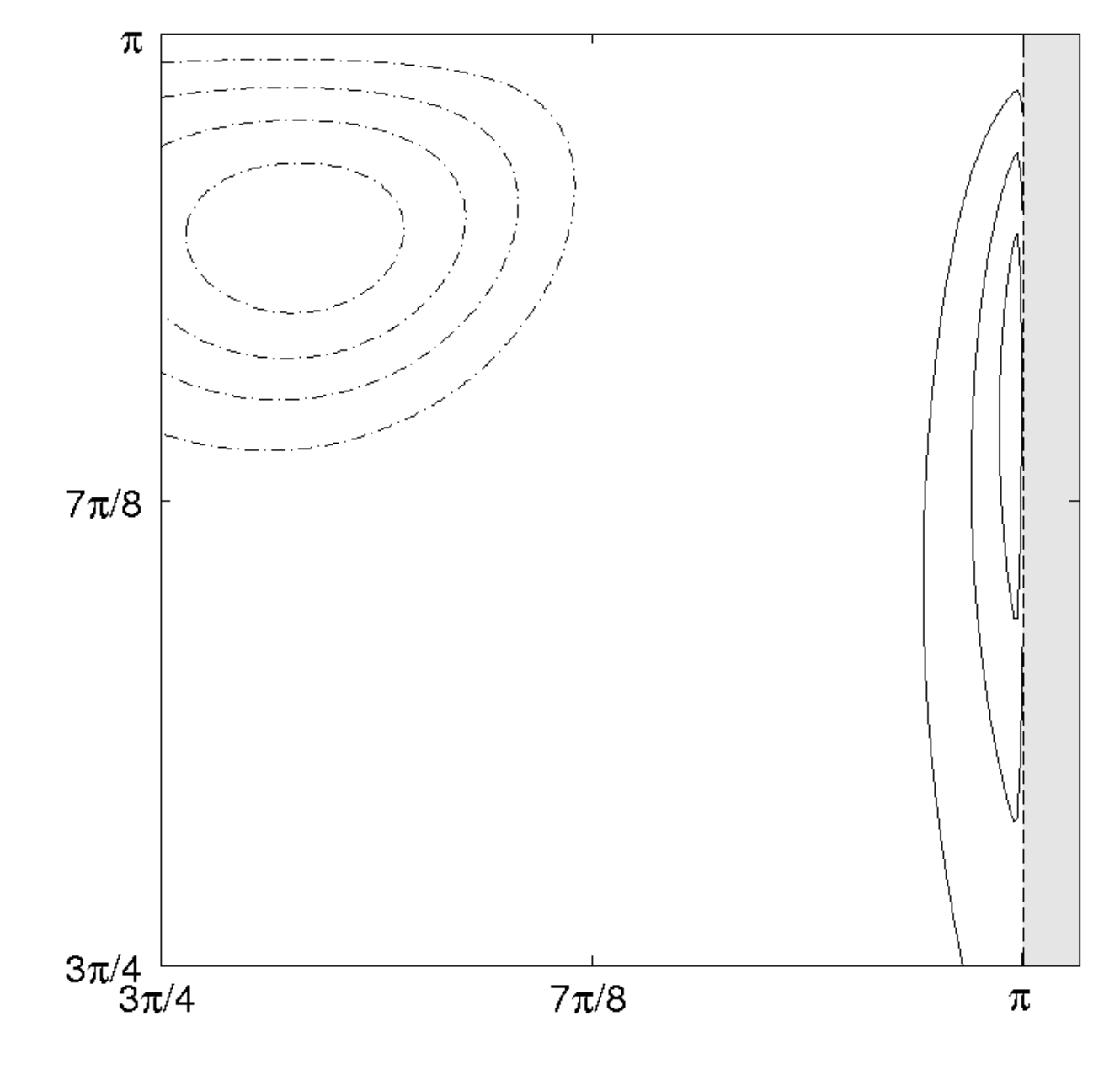}
\includegraphics[width=0.49\columnwidth]{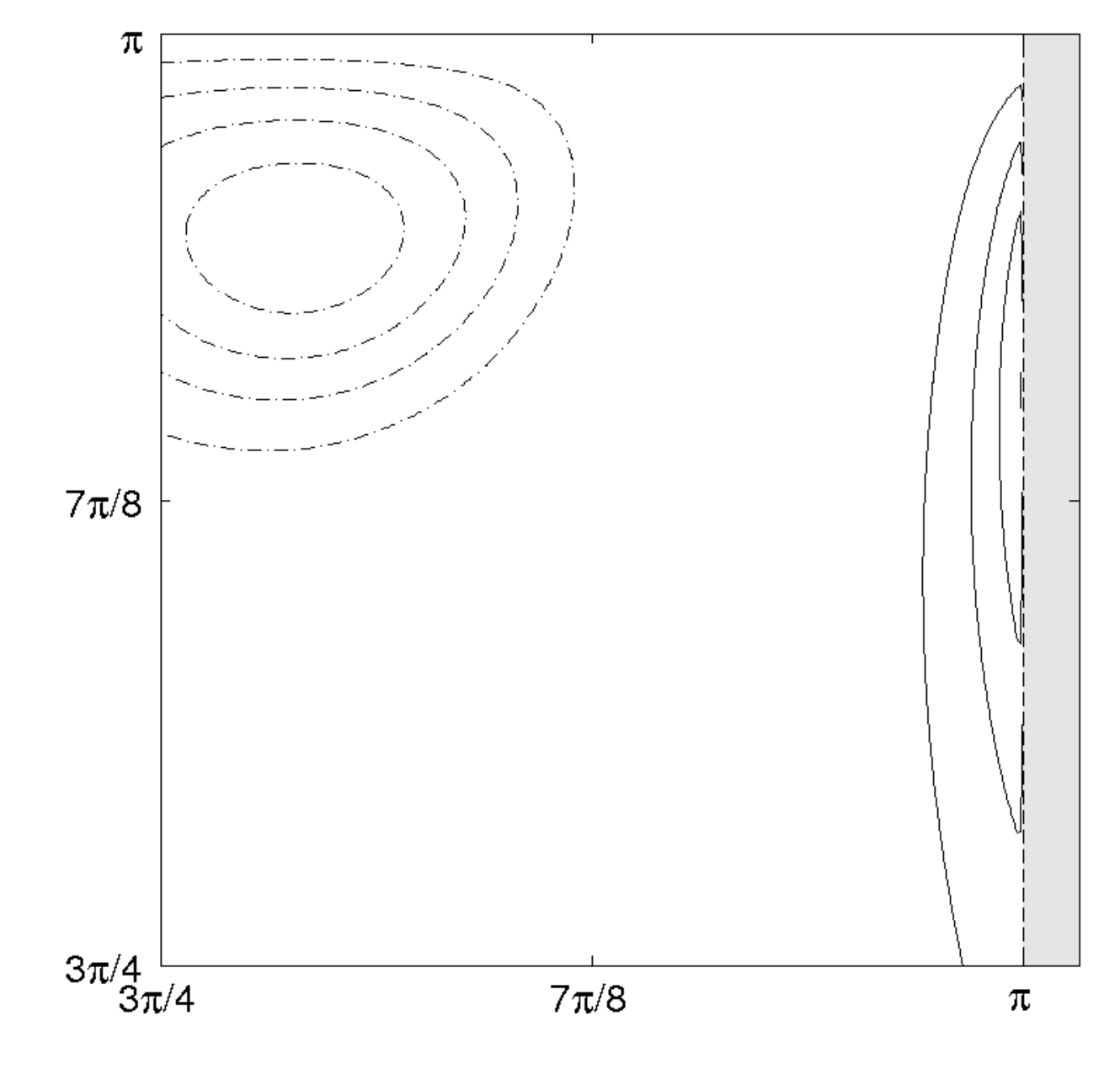}
\end{center}
\caption{
\label{fig:dipole_wall_collision}
Dipole-wall collision at $Re = 122$.
Top left: contour plot of the initial stream function (which is identical in cases I-IV), with $12$ contours equally spaced between $-0.01$ and $0.01$.
Top right: contour plot of vorticity at $t=50$ for $\eta'=0.25$ and $N=2048$ (case II), restricted to the subdomain $[3\pi/4,\pi]\times[3\pi/4,\pi]$, with the following contours lines: $-0.28$, $-0.21$, $-0.14$, $-0.07$ (dash-dotted lines),
and $0.02$, $0.04$, $0.06$ (full lines).
Bottom left: same but for $\eta'=0.1$ and $N=4096$ (case III).
Bottom right: same but for $\eta'=0.04$ and $N=8192$ (case IV).
}
\end{figure}

As initial condition, we choose a quadrupole centered around $x=\pi/2$, $y=\pi$,
for which the vorticity field is given by:
\begin{equation}
\label{eq:initial_condition}
\omega_0(x,y) := \partial_x u_{0,y} - \partial_y u_{0,x} := \frac{A}{s^4} x y (x^2 + y^2 - 6s^2) e^{-\frac{x^2 + y^2}{2s^2}},
\end{equation}
where $A \simeq 0.6258473$ and $s = 0.2$, as shown in Fig.~\ref{fig:dipole_wall_collision} (top, left).
Its initial enstrophy is $Z_0 = \frac{1}{2}\int_\Omega \omega_0^2 \simeq 7.38309 \cdot 10^{-2}$,
and its initial energy is  $E_0 =  \frac{1}{2}\int_\Omega \uu_0^2 \simeq 1.87016 \cdot 10^{-5}$.
Oriented contours of the initial stream function $\psi_0$, define by $\nabla^2 \psi_0 = \omega_0$, are shown in Fig.~\ref{fig:dipole_wall_collision} (top, left).

As viscosity we take $\nu \simeq 1.57914 \cdot 10^{-4}$,
which corresponds to an initial Reynolds number based on root mean square velocity and channel width of $\Rey = \sqrt{2 E_0} \pi / \nu \simeq 122$.
Note that this Reynolds number is taken sufficiently low so that the discretization parameter can be varied
over a sufficiently wide range while still properly resolving the solution.

Vorticity contour plots are shown in Fig.~\ref{fig:dipole_wall_collision}.
As expected, the boundary layer becomes sharper as $N$ increases and $\eta$ diminishes.
Note that, thanks to the smoothing of the mask function, spurious oscillations do not arise.

\begin{figure}
\begin{center}
\includegraphics[width=0.49\columnwidth]{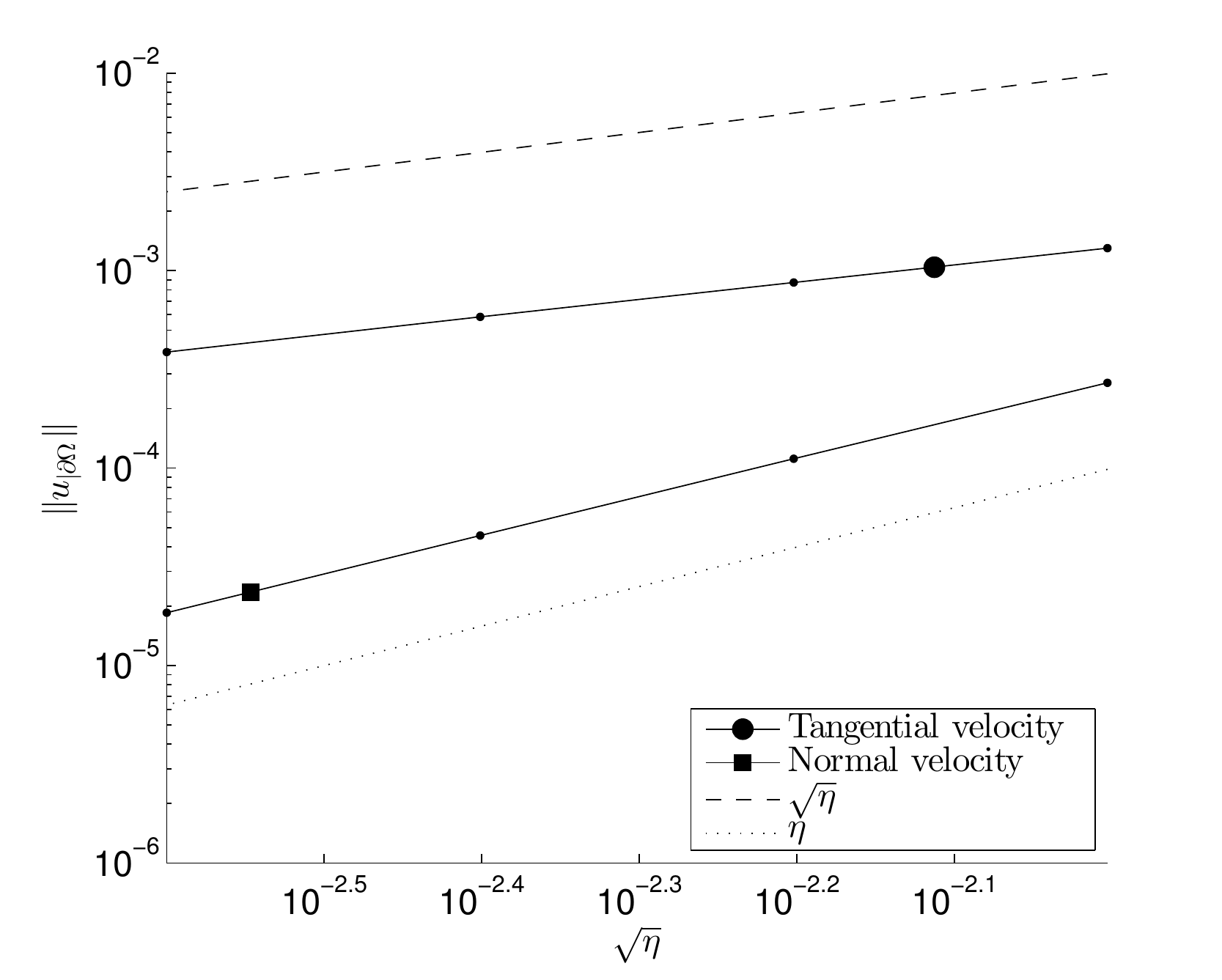}
\includegraphics[width=0.49\columnwidth]{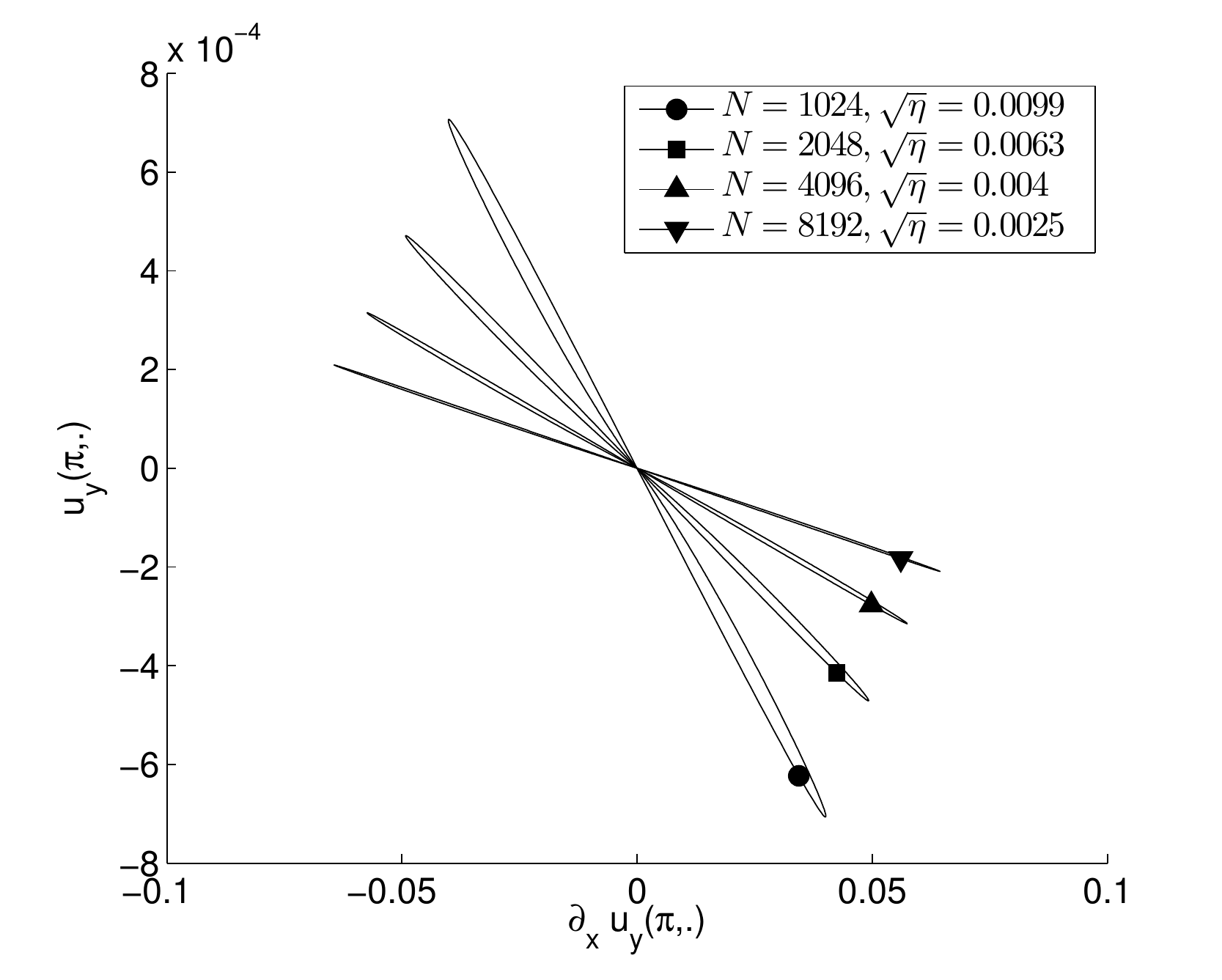} \\
\includegraphics[width=0.49\columnwidth]{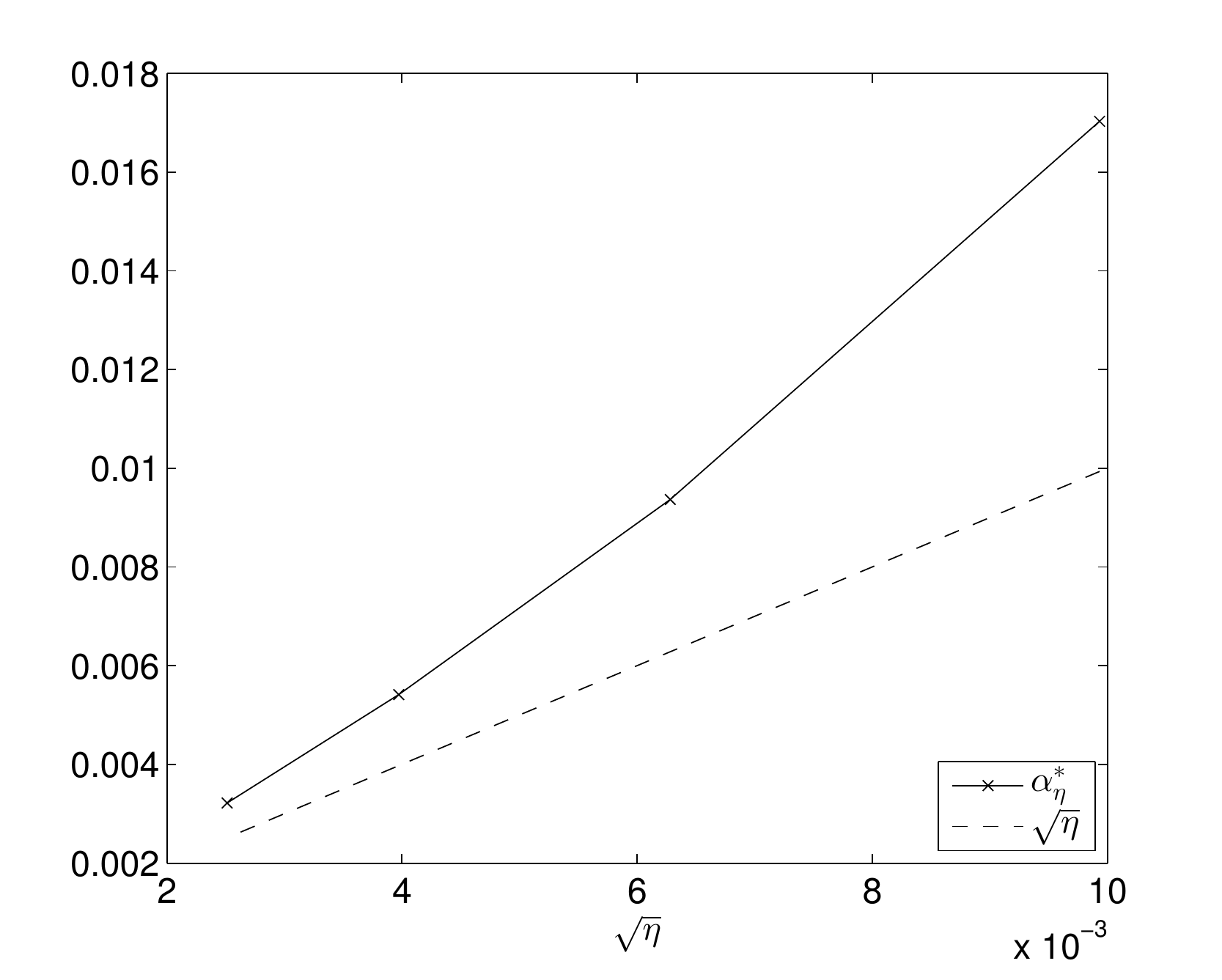}
\includegraphics[width=0.49\columnwidth]{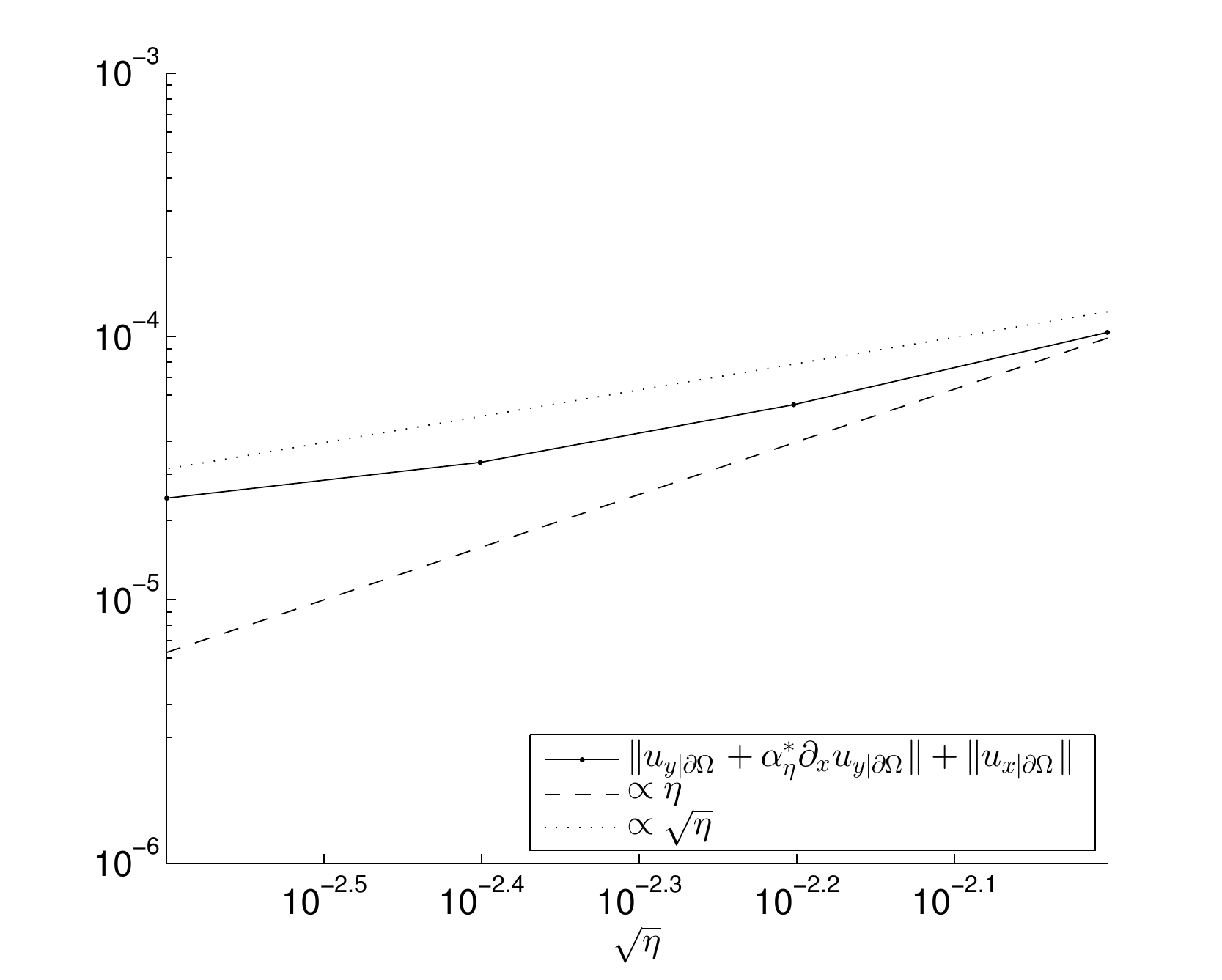}
\end{center}
\caption{
\label{fig:dipole_wall_collision_navier}
A posteriori analysis of boundary conditions for  the penalized Navier-Stokes test cases at $t=50$.
Top left: wall-tangential and wall-normal residual velocities as functions of $\sqrt{\eta}$.
Top right: wall-parallel residual velocity as a function of wall strain rate.
Bottom left: effective residual slip-length $\alpha^*_\eta$ as a function of $\sqrt{\eta}$.
Bottom right: $L^2$ discrepancy along the boundary with respect to Navier boundary conditions with slip-length $\alpha^*_\eta$.
}
\end{figure}

Now turning to more quantitative diagnostics, we compare in Fig.~\ref{fig:dipole_wall_collision_navier} (top left) 
the residual tangential and normal velocities, in the spirit of Fig.~\ref{fig:convergence_u_x_u_y_pi}.
The same scalings are obtained as in the linear Stokes case, namely $\Vert u_{x\mid\partial\Omega} \Vert \propto \eta$ for the normal velocity and $\Vert u_{y\mid\partial\Omega} \Vert \propto \sqrt{\eta}$ 
for the tangential velocity.
By plotting the residual tangential velocity at the wall $u_{y\mid\partial\Omega}$ as a function of the wall strain rate $\partial_x u_{y\mid\partial\Omega}$ (Fig.~\ref{fig:dipole_wall_collision_navier}, top right)
we observe a linear correlation, as expected if Navier boundary conditions are approximately satisfied (see (\ref{eq:navier_bc_tangential})).
The effective slip-length $\alpha^*_\eta$, as obtained from a least-squares fit, is then plotted as a function of $\sqrt{\eta}$ in Fig.~\ref{fig:dipole_wall_collision_navier} (bottom left).
Recall that we have shown in the linear Stokes case that $\alpha^*_\eta \simeq \sqrt{\eta}$.
Here, this relationship seems to be approached in the limit $\eta \to 0$.

All these diagnostics suggest a very good agreement between the picture we have got from the linear case
on the one hand, and the Navier-Stokes test case on the other hand.
The limits of this approach are however seen when measuring the higher order residual $\Vert u_{y\mid \partial \Omega} + \alpha^*_\eta \partial_x u_{y\mid \partial \Omega} \Vert + \Vert u_{x\mid\partial\Omega} \Vert$,
which is an indicator of the precision with which the solution matches Navier boundary conditions with the self-consistently determined slip length.
The curve (Fig.~\ref{fig:dipole_wall_collision_navier}, bottom right) suggests that this residual saturates for small values of the penalization parameter. 
We attribute this saturation to nonlinear effects.
Note however that, despite this saturation, the residual remains about one order of magnitude smaller than the tangential velocity 
(Fig.~\ref{fig:dipole_wall_collision_navier}, top left), suggesting that the penalized solution
indeed approaches the Navier boundary condition faster than the no-slip one.

	\section{Summary and conclusion}

Through a detailed analysis of the spectral decomposition the penalized Laplace and Stokes operators in simplified geometries,
the following properties were observed.

The $O({\eta}^{1/2})$ bound on the error with respect to the Dirichlet boundary conditions,
which was expected from previous studies,
indeed applies to eigenfunctions of sufficiently low order.
However, the convergence rate also depends on the eigenvalue, which entrains a degradation for higher order eigenfunctions,
and even a break down of convergence when the eigenvalue exceeds a certain value ($O(\eta^{-1})$ in the Laplace case and $O(\eta^{-1}-k^2)$ in the Stokes case, respectively).
On dimensional grounds, this means that the penalization does not control the behavior of the solution below a certain scale ($O(\eta^{1/2})$ in the Laplace case and $O((\eta^{-1}-k^{-2})^{-1/2})$ in the Stokes case, respectively).
In the case of the Stokes operator, the error can be described at lowest order by Navier boundary conditions with a residual slip-length
proportionnal to $O({\eta}^{1/2})$, which may have important consequences for the physical interpretation of results
obtained with the volume penalization method.

The discretization of the penalized Laplace and Stokes operators using spectral schemes brings some new important elements into the story.
In both cases, the optimal penalization parameter $\eta_{opti}$ is of order $N^{-2}$,
in agreement with the notion that the finest resolved scale should be of order ${\eta}^{1/2}$.
Thanks to unexpected cancellations between the penalization and discretization errors, the total error with respect to the Dirichlet solution
for $\eta = \eta_{opti}$ is of order $N^{-3/2}$.
These cancellations occur only for a narrow range of values of $\eta$ very close to $\eta_{opti}$,
and we were able to observe them only by systematically scanning for the right value of $\eta$.
This is usually not possible in practical applications of the method, when the exact solution is unknown.
However, we expect the scaling $\eta_{opti} \propto N^{-2}$ to be very general, 
and we propose it as a guideline for future applications of the penalization method with Fourier-type discretizations.

If one wishes to reach a higher precision with a penalization scheme, one possibility is to use a discretization
which behaves better in the presence of discontinuous solutions.
For example, a second order finite volume scheme was considered in \cite{Sarthou2008}, involving a
careful correction in the penalization term.


\bigskip

\appendix

\section{Analytic solution and discretization of the 1D penalized Poisson problem}
\label{appendix_poisson}

In this appendix, we provide a detailed analysis of the introductory example concerning the 1D penalized Poisson equation,
as well as of its discretization.

\subsection{Analytic solution of the penalized Poisson equation}
\begin{figure}
\begin{center}
\centerline{
\includegraphics[width=0.32\columnwidth,clip]{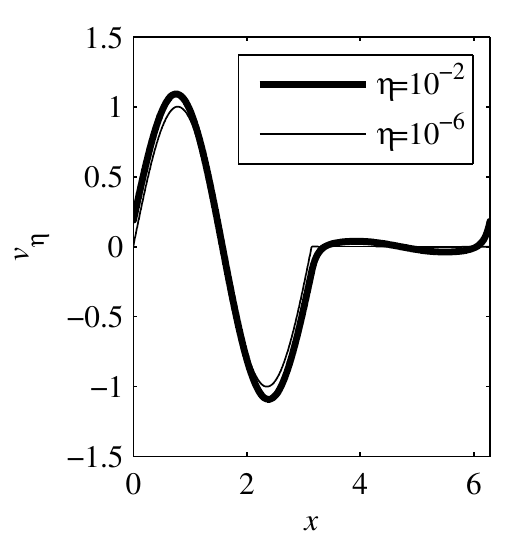}
\includegraphics[width=0.32\columnwidth,clip]{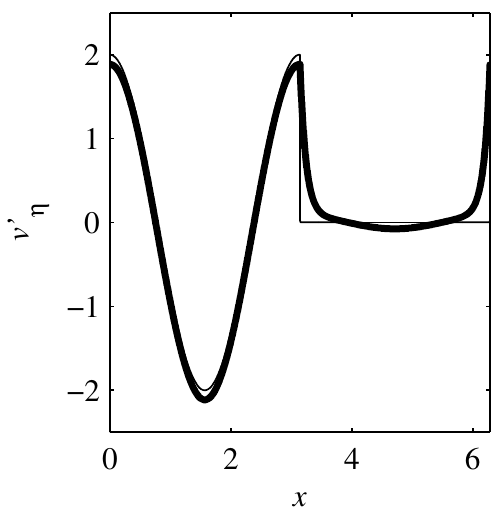}
\includegraphics[width=0.32\columnwidth,clip]{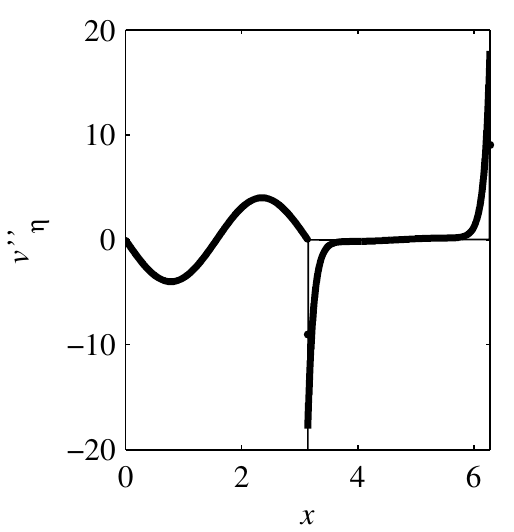}}
\caption{\label{fig:uex} Exact penalized solution (left) for
$m=2$ and its first (middle) and second (right) derivatives.}
\end{center}
\end{figure}

The exact solution to the penalized Poisson problem (\ref{eq:penalized_poisson_eq}) is:
\begin{equation}
  v(x) = \left\{
  \begin{array}{ll}
  \sin mx + A_1 x + A_2, & x \in [0,\pi[ \\
  \frac{m^2\eta}{1+\eta m^2} \sin mx + B_1 e^{-x/\sqrt{\eta}} + B_2 e^{x/\sqrt{\eta}}, & x \in [\pi,2\pi[, \\
  \end{array}
  \right.
\label{eq:penalized_poisson_exact_solution}
\end{equation}
where
\begin{equation}
  \begin{array}{l}
  \displaystyle A_1 = \frac{m}{1+\eta m^2} K, \quad\quad  \displaystyle A_2 = \frac{m \sqrt{\eta}}{1+\eta m^2},
  \frac{(K+1)\cosh{\frac{\pi}{\sqrt{\eta}}}-K-(-1)^m}{\sinh{\frac{\pi}{\sqrt{\eta}}}}, \\
  \displaystyle B_1 = \frac{ e^{2\pi/\sqrt{\eta}} [A_2 (1-e^{-\pi/\sqrt{\eta}})+A_1 \pi] }{ 2 \sinh{\frac{\pi}{\sqrt{\eta}}} },  \quad\quad
  \displaystyle B_2 = - \frac{ e^{-2\pi/\sqrt{\eta}} [A_2(1-e^{\pi/\sqrt{\eta}})+A_1 \pi] }{ 2 \sinh{\frac{\pi}{\sqrt{\eta}}} }, \\
  \displaystyle K = - \frac{ \sqrt{\eta} (1+(-1)^m) }{ \pi \mathrm{cotanh}{\frac{\pi}{2 \sqrt{\eta}}} + 2\sqrt{\eta} } \\
  \end{array}
\label{eq:penalized_poisson_exact_solution_prefactors}
\end{equation}
It is visualized in figure~\ref{fig:uex} for $m=2$. The
second derivative $v''$ is discontinuous at $x=0$ and $x=\pi$. The
penalization boundary layer inside of the solid becomes thinner as
$\eta \to 0$ and, in the limit, the first derivative $v'$ is
discontinuous.

Knowing (\ref{eq:poisson_exact_solution}) and
(\ref{eq:penalized_poisson_exact_solution}), we directly calculate
the penalzation error,
\begin{equation}
  \displaystyle ||\epsilon||_2^2 = \frac{1}{2\pi} \int\limits_0^{2\pi} (v-w^e)^2
  \mathrm{d} x = \frac{||\epsilon||_{L^2(]0,\pi[)}^2 + ||\epsilon||_{L^2(]\pi,2\pi[)}^2}{2},
  \label{eq:penalization_error_norm}
\end{equation}
where $w^e$ is the exact solution
(\ref{eq:poisson_exact_solution}) extended to $\mathbb{T}$,
\begin{equation}
  w^e(x) = \left\{
  \begin{array}{ll}
  \sin mx, & x \in [0,\pi[ \\
  0, & x \in [\pi,2\pi[. \\
  \end{array}
  \right.
\label{eq:poisson_exact_solution_ext}
\end{equation}
The error norm in the fluid domain is
\begin{equation}
  \displaystyle ||\epsilon||_{L^2(]0,\pi[)}^2 = \frac{1}{\pi} \int\limits_0^{\pi} (A_1 x + A_2)^2
  \mathrm{d} x \sim \frac{m^2 \eta}{3}(2-(-1)^m) \quad \mathrm{as}
  \quad \eta \to 0,
\label{eq:penalization_error_norm_fluid}
\end{equation}
and in the solid domain
\begin{equation}
  \begin{array}{l}
  \displaystyle ||\epsilon||_{L^2(]\pi,2\pi[)}^2 = \frac{1}{\pi} \int\limits_{\pi}^{2\pi} \left(\frac{m^2\eta}{1+\eta m^2} \sin mx + B_1 e^{-x/\sqrt{\eta}} + B_2 e^{x/\sqrt{\eta}}\right)^2
  \mathrm{d} x \sim \\
  \displaystyle \quad\quad\quad\quad\quad\quad\quad  \frac{m^2 \eta^{3/2}}{\pi} \quad \mathrm{as} \quad \eta \to
  0.
  \end{array}
\label{eq:penalization_error_norm_solid}
\end{equation}
The penalization error in the fluid decays slower then in the
solid, hence
\begin{equation}
  \displaystyle ||\epsilon||_2 \sim \frac{m \sqrt{2-(-1)^m}}{\sqrt{6}} \sqrt{\eta} \quad \mathrm{as}
  \quad \eta \to 0.
  \label{eq:penalization_error_norm_asy}
\end{equation}

\subsection{Discretization error of the pseudo-spectral solution}

The Fourier coefficients of
(\ref{eq:penalized_poisson_exact_solution}) are given by
\begin{equation}
  \displaystyle \widehat{v}_k = \frac{1}{2\pi} \int\limits_{0}^{2\pi}{v(x) e^{-\iota k x}}\mathrm{d}x,
\label{eq:penalized_poisson_exact_ft}
\end{equation}
where $k \in \mathbb{Z}$ is the wavenumber.
This can be integrated in terms of elementary functions, but herein we only need 
(\ref{eq:penalized_poisson_exact_ft}) expanded in
negative powers of $k$ for large $k$, and then the coefficients of
that series are expanded in $\eta$, giving
\begin{equation}
\begin{array}{l}
  \displaystyle \widehat{v}_k \sim \frac{\iota m}{2\pi} \left[ -(1+(-1)^{m+k})\frac{1}{\sqrt{\eta}} + \frac{(1+(-1)^m) (1+(-1)^k)}{\pi} + O(\sqrt{\eta}) \right]
  \frac{1}{k^3} + \\
  \displaystyle \frac{m}{2\pi} \left[ -\frac{1-(-1)^{m+k}}{\eta} + (1+(-1)^m)(1-(-1)^k)\left(\frac{1}{\pi\sqrt{\eta}}-\frac{1}{\pi^2}\right) +
  O(\sqrt{\eta})\right] \frac{1}{k^4} + \\
  \displaystyle O(\eta^{-3/2})\frac{1}{k^5} + ... +
  O(\eta^{-J/2})\frac{1}{k^{J+2}}, \quad\quad J \in \mathbb{N}.
\end{array}
\label{eq:v_hat_series}
\end{equation}

\begin{figure}
\centerline{
\includegraphics[width=0.49\columnwidth]{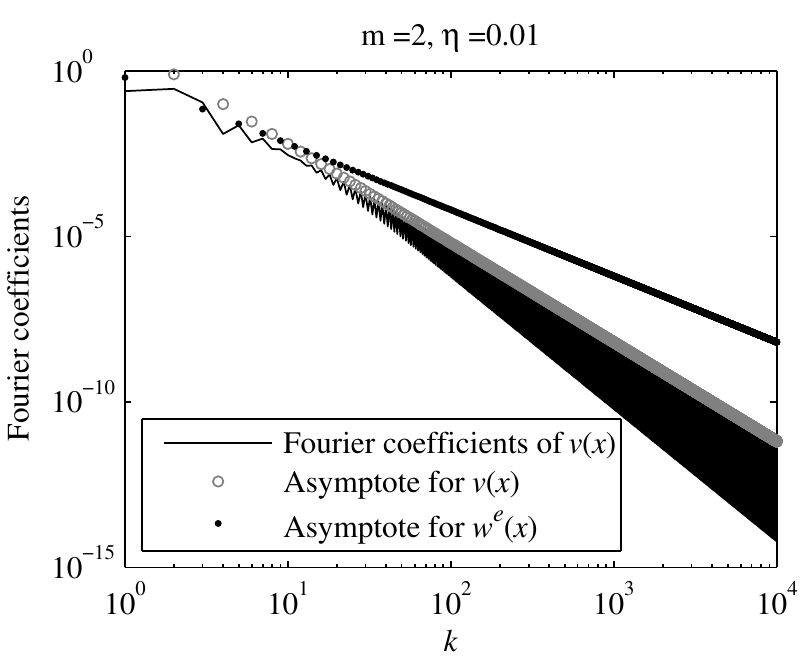}
\includegraphics[width=0.49\columnwidth]{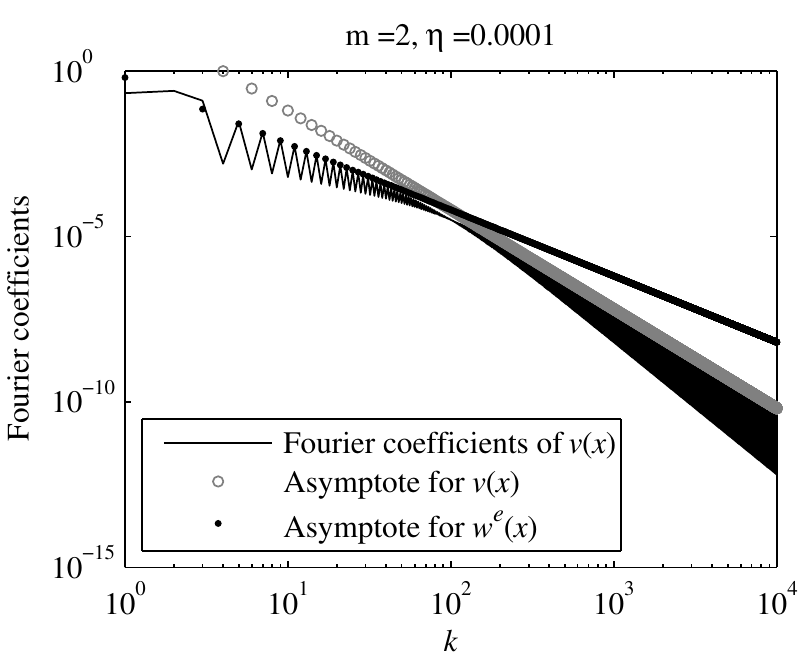}
}
\caption{\label{fig:fourier_coeffs} 
Absolute value of the Fourier coefficients for $m=2$ and two values of the penalization parameter:
$\eta=10^{-2}$ (left) and $\eta=10^{-4}$ (right). 
The solid line depicts the exact values (\ref{eq:penalized_poisson_exact_ft}). 
Gray open circles show the leading-order term in (\ref{eq:v_hat_series}). 
Black dots correspond to (\ref{eq:we_hat_leading}).}
\end{figure}

It is also instructive to consider the Fourier coefficients of
(\ref{eq:poisson_exact_solution_ext}),
\begin{equation}
  \displaystyle \widehat{w}^e_k = \left\{
    \begin{array}{ll}
      \displaystyle \frac{1}{2\pi} \frac{m(1-(-1)^{m+k})}{m^2-k^2}, & k \ne m, k \ne -m \\
      \displaystyle -\iota/4, & k = m \\
      \displaystyle \iota/4, & k = -m. \\
    \end{array}
  \right.
\label{eq:we_hat}
\end{equation}
The leading-order asymptotic approximation to (\ref{eq:we_hat})
for large $k$ is
\begin{equation}
  \displaystyle \widehat{w}^e_k = - \frac{m}{2\pi} \frac{(1-(-1)^{m+k})}{k^2} + o(k^{-2}),
\label{eq:we_hat_leading}
\end{equation}
which shows that $\widehat{w}^e_k$ decays slower than $\widehat{v}_k$, or equivalently that $w^e$ is less regular than $v$. 
Fig.~\ref{fig:fourier_coeffs} shows the Fourier coefficients calculated with two different values of $\eta$. 
The wavenumber corresponding to the slope change is estimated from (\ref{eq:v_hat_series}) and (\ref{eq:we_hat_leading}) as
\begin{equation}
  k_t = 1/\sqrt{\eta}.
\label{eq:k_t}
\end{equation}

The discrete Fourier transform of $v(x)$ is defined as
\begin{equation}
  \begin{array}{l}
  \displaystyle \tilde{v}_k = \frac{1}{N} \sum\limits_{n=0}^{N-1}{v(x_n) e^{-\iota k
  x_n}}, \quad\quad \mathrm{where}~ x_n = 2 \pi n /N, \quad
  k=-N/2...N/2-1
  \end{array}
\label{eq:penalized_poisson_exact_dft}
\end{equation}
$v(x)$ is real-valued, therefore the discrete Fourier transform is
conjugate symmetric, hence $\tilde{v}_{k} = \tilde{v}^*_{-k}$. The inverse of
(\ref{eq:penalized_poisson_exact_dft}) is
\begin{equation}
  \displaystyle v(x_n) = \sum\limits_{k=-N/2}^{N/2-1}{ \tilde{v}_k e^{\iota k
  x_n}}.
\label{eq:penalized_poisson_exact_dft_inv}
\end{equation}
Note that the truncated Fourier series $\sum\limits_{k=-N/2}^{N/2-1}{ \widehat{v}_k e^{\iota k x_n}}$
is not interpolating, but it minimizes the $L^2$ norm of the error
with respect to $v(x)$. Also, if the Fourier coefficients
$\widehat{v}_k$, as defined by (\ref{eq:penalized_poisson_exact_ft}), are known, the discrete Fourier transform can be
obtained by periodization:
\begin{equation}
  \tilde{v}_k =
  \displaystyle\sum\limits_{p=-\infty}^{\infty}{\widehat{v}_{k+Np}}.
\label{eq:v_periodization}
\end{equation}

Let us now solve the penalized Poisson equation
(\ref{eq:penalized_poisson_eq}) using a pseudo-spectral Fourier
method. The spatial domain is discretized with $N$ grid points
$\{x_n = 2 \pi n /N\}$, $n=0...N-1$. The second term on the
left-hand side of (\ref{eq:penalized_poisson_eq}) is evaluated at
these points, then the discrete Fourier transform is performed.
Finally, the pseudo-spectral solution $u$ is obtained from a
linear system of equations for $\tilde{u}_k$
\begin{equation}
  k^2 \tilde{u}_k + \frac{1}{\eta} \widetilde{(\chi u)}_k =
  \tilde{f}_k.
  \label{eq:pseudo-spectral_eqn}
\end{equation}

We are now interested in estimating the error between the
pseudo-spectral solution $u(x)$ and the exact solution $v(x)$ of
the penalized Poisson equation. Let us rewrite
(\ref{eq:pseudo-spectral_eqn}) as
\begin{equation}
  k^2 ( \tilde{u}_k-\tilde{v}_k+\tilde{v}_k ) + \frac{1}{\eta} ( \widetilde{(\chi u)}_k-\widetilde{(\chi v)}_k+\widetilde{(\chi v)}_k ) =
  \tilde{f}_k.
  \label{eq:pseudo-spectral_eqn_rewritten}
\end{equation}
Note that
\begin{equation}
  \tilde{e}_k = \tilde{u}_k - \tilde{v}_k
  \label{eq:error_coef}
\end{equation}
are the coefficients of the error between the pseudo-spectral and
the exact solution, \textit{i.e.} the discretization error. Using
(\ref{eq:penalized_poisson_eq}), the right-hand side can be
expressed in terms of the exact solution $v$,
\begin{equation}
  \tilde{f}_k =  - \widetilde{(v'')}_k + \frac{1}{\eta} \widetilde{(\chi
  v)}_k.
  \label{eq:rhs}
\end{equation}
From (\ref{eq:pseudo-spectral_eqn_rewritten}),
(\ref{eq:error_coef}), (\ref{eq:rhs}) follows the equation for the
discretization error,
\begin{equation}
  k^2 \tilde{e}_k + \frac{1}{\eta} \widetilde{(\chi e)}_k =
  - \widetilde{(v'')}_k - k^2 \tilde{v}_k.
  \label{eq:discret_error_eqn}
\end{equation}
It is useful to rewrite (\ref{eq:discret_error_eqn}) as
\begin{equation}
  \tilde{e}_k - \frac{1}{1+\eta k^2} [ \tilde{e}_k - \widetilde{(\chi e)}_k
  ] = \frac{\eta}{1+\eta k^2} [ - \widetilde{(v'')}_k - k^2 \tilde{v}_k
  ].
  \label{eq:discret_error_eqn_rewritten}
\end{equation}

Our objective is to estimate $||e||_2$ without solving
(\ref{eq:discret_error_eqn_rewritten}). We note that
\begin{equation}
  ||e||_2 = K \eta^{-1/4} ||b||_2
  \label{eq:norm_exact}
\end{equation}
where $b$ is the right hand side of (\ref{eq:discret_error_eqn_rewritten}), and $K=2$ for $m$ even and $K \approx 3.8423$ for $m$ odd. 
This relationship was confirmed numerically with arbitrary values of the
parameters, but it is unclear how to prove it analytically. 

Now let us derive the leading-order asymptotic approximation for
\begin{equation}
  ||b||_2^2 = \displaystyle\sum\limits_{k=-N/2}^{N/2-1}{\left|\frac{\eta}{1+\eta k^2} [ - \widetilde{(v'')}_k - k^2
  \tilde{v}_k]\right|^2}.
  \label{eq:norm_rhs}
\end{equation}
Periodization gives
\begin{equation}
\begin{array}{l}
  - \widetilde{(v'')}_k =
  \displaystyle\sum\limits_{p=-\infty}^{\infty}{(k+Np)^2
  \widehat{v}_{k+Np}}, \\
  - k^2 \tilde{v}_k =
  - k^2 \displaystyle\sum\limits_{p=-\infty}^{\infty}{\widehat{v}_{k+Np}}.
\end{array}
\label{eq:v_periodization_dbleprime}
\end{equation}
We assume that $N$ is even, hence $1 \pm (-1)^{k+Np} = 1 \pm
(-1)^k$. Using (\ref{eq:v_hat_series}) we obtain
\begin{equation}
\begin{array}{l}
  \displaystyle \left| - \widetilde{(v'')}_k - k^2 \tilde{v}_k \right|^2 \sim \frac{m^2 \pi^2}{18} \frac{(1+(-1)^{m+k}) \eta^{-1} k^2 + (1-(-1)^{m+k}) \eta^{-2}}{N^4} \\ 
 \quad \displaystyle + \frac{O(\eta^{-1/2})k^2+O(\eta^{-3/2})}{N^4} \\
  \quad \displaystyle + \sum_{\alpha=2}^{J/2} \frac{ O(\eta^{-(2\alpha-1)}) O(k^2) + O(\eta^{-2\alpha}) O(k^0) }{N^{4\alpha}} \\
  \quad \displaystyle + \sum_{\beta=1}^{J/2} \sum_{\gamma=1}^{J/2} \frac{O(\eta^{-(\beta+\gamma-1/2)}) O(k^1)}{N^{2\beta+2\gamma}} \\
  \quad \displaystyle + \sum_{\beta=1}^{J/2} \sum_{\gamma=1;\gamma\ne \beta}^{J/2} \frac{O(\eta^{-(\beta+\gamma-1)}) O(k^2) + O(\eta^{-(\beta+\gamma)}) O(k^0) }{N^{2\beta+2\gamma}}.
\end{array}
\label{eq:modulus_squared_prelim_O}
\end{equation}
We notice that expansion (\ref{eq:v_hat_series}) is only valid for
large $k$ compared to $\eta^{-1/2}$. However, the truncation and
aliasing errors due to the spectral discretization of any function
are determined by its high-wavenumber residual. The
low-wavenumber modes $\widehat{v}_k$, though entering equation
(\ref{eq:norm_rhs}), must cancel out. Therefore the error
committed in the lower part of the spectrum in
(\ref{eq:v_hat_series}) does not enter the large-$N$ asymptotic
estimates (\ref{eq:norm_rhs_asy})-(\ref{eq:norm_error_asy}) below.

\begin{figure}
\centerline{
\includegraphics[width=0.49\columnwidth]{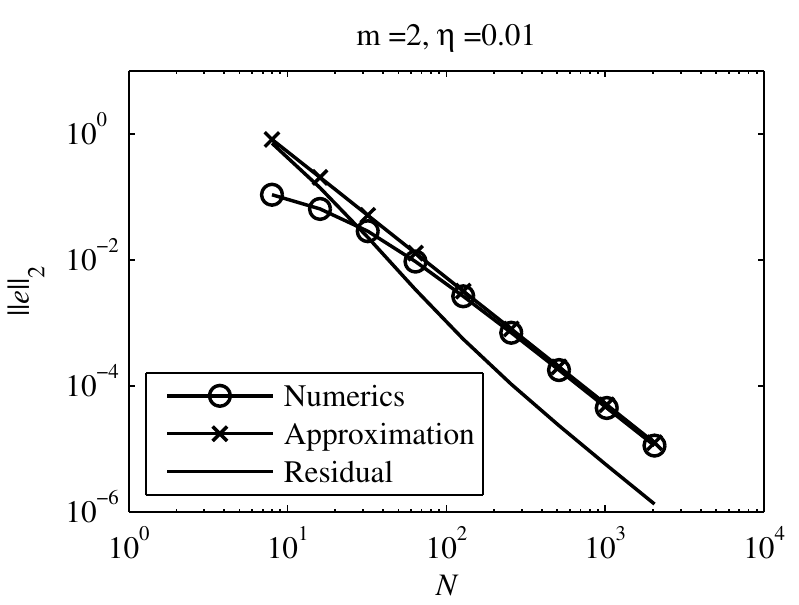}
\includegraphics[width=0.49\columnwidth]{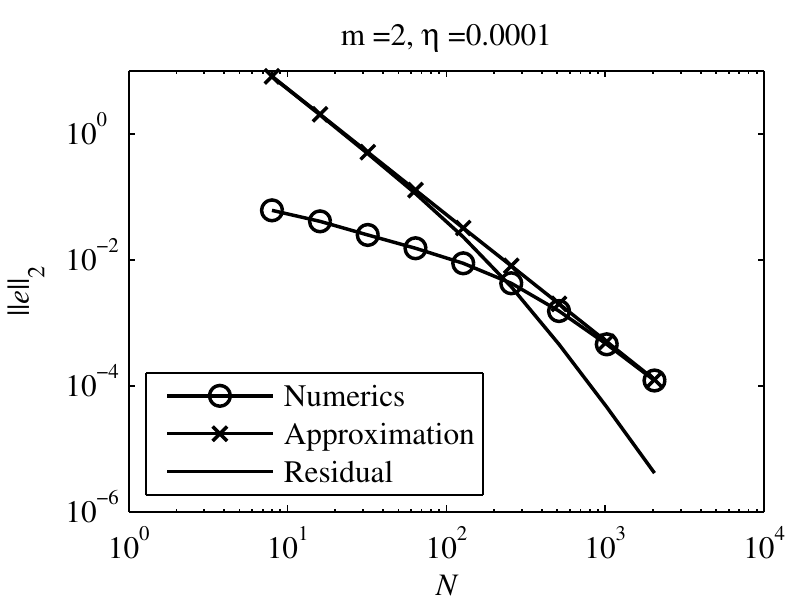}
}
\caption{\label{fig:convergence_error} Absolute value of the
Fourier coefficients for $m=2$ and two values of the penalization
parameter: $\eta=10^{-2}$ (left) and $\eta=10^{-4}$ (right). The line with circular markers is obtained from a numerical solution
discussed in section~\ref{sec:numerical_tests}. Crosses show the first term of the asymptotic
estimate (\ref{eq:norm_error_asy}).
The discrepancy between this estimate and the numerical solution is shown
with a solid line.}
\end{figure}

After multiplying (\ref{eq:modulus_squared_prelim_O}) by
$(k^2+1/\eta)^{-2}$, it is convenient to rewrite its first term as
\begin{equation}
\begin{array}{l}
  \displaystyle\frac{m^2 \pi^2}{18N^4} \frac{(1+(-1)^{m+k}) \eta^{-1} k^2 + (1-(-1)^{m+k})
  \eta^{-2}}{(k^2+1/\eta)^2} = \\
  \quad\quad \displaystyle\frac{m^2 \pi^2}{18 N^4}
  \left[ \frac{1}{\eta} \left( \frac{1}{k^2+1/\eta} + (-1)^m \frac{(-1)^k}{k^2+1/\eta} \right) - \frac{2(-1)^m}{\eta^2} \frac{(-1)^k}{(k^2+1/\eta)^2}
  \right].
\end{array}
\label{eq:modulus_squared_prelim_1_term}
\end{equation}
From (\ref{eq:norm_rhs}), (\ref{eq:modulus_squared_prelim_O}) and
(\ref{eq:modulus_squared_prelim_1_term}), summing up for
$k=-N/2...N/2-1$, we obtain
\begin{equation}
  ||b||_2^2 \sim \left( \frac{m^2 \pi^3}{18 \sqrt{\eta}} + O(\eta^0) \right)
  \frac{1}{N^4} + O(\eta^{-1})\frac{1}{N^5} +
  ... + O(\eta^{-p/2}) \frac{1}{N^{p+3}}, p \in
  \mathbb{N},
\label{eq:norm_rhs_asy}
\end{equation}
and (\ref{eq:norm_exact}) yields the desired estimate:
\begin{equation}
  ||e||_2 \sim \left(K \frac{m \pi^{3/2}}{3\sqrt{2}} \frac{1}{\sqrt{\eta}} + O(\eta^{0}) \right) \frac{1}{N^2} + O(\eta^{-1}) \frac{1}{N^3} +
  ...  + O(\eta^{-q/2}) \frac{1}{N^{q+1}}, q \in \mathbb{N}.
\label{eq:norm_error_asy}
\end{equation}
Figure~\ref{fig:convergence_error} presents a comparison of the
first term in (\ref{eq:norm_error_asy}) with the results of the
numerical pseudo-spectral solution of
(\ref{eq:penalized_poisson_eq}).

\subsection{Numerical tests}\label{sec:numerical_tests}

\begin{figure}
\begin{center}
\includegraphics[width=0.49\columnwidth]{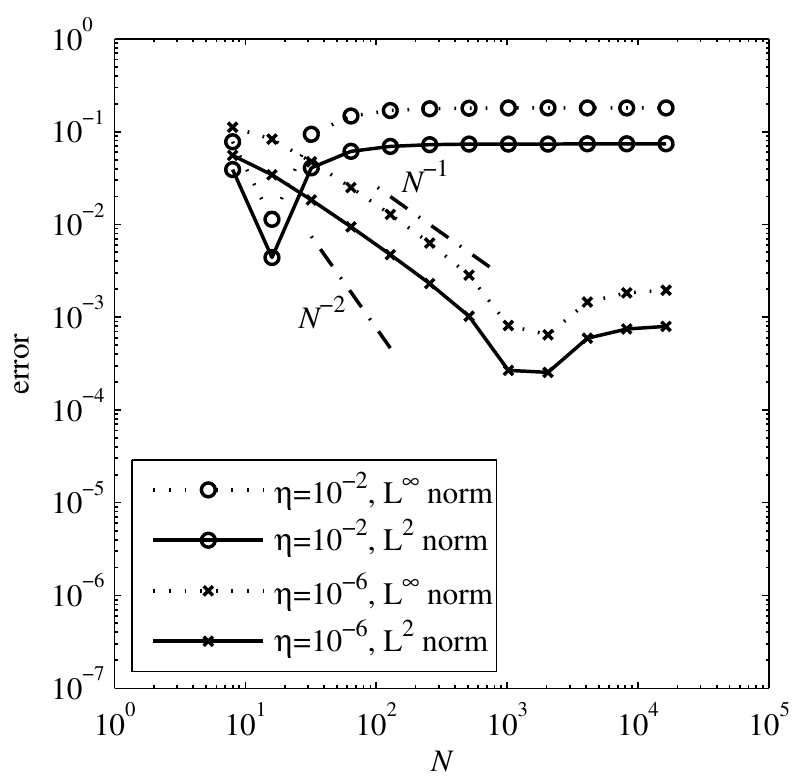}
\includegraphics[width=0.49\columnwidth]{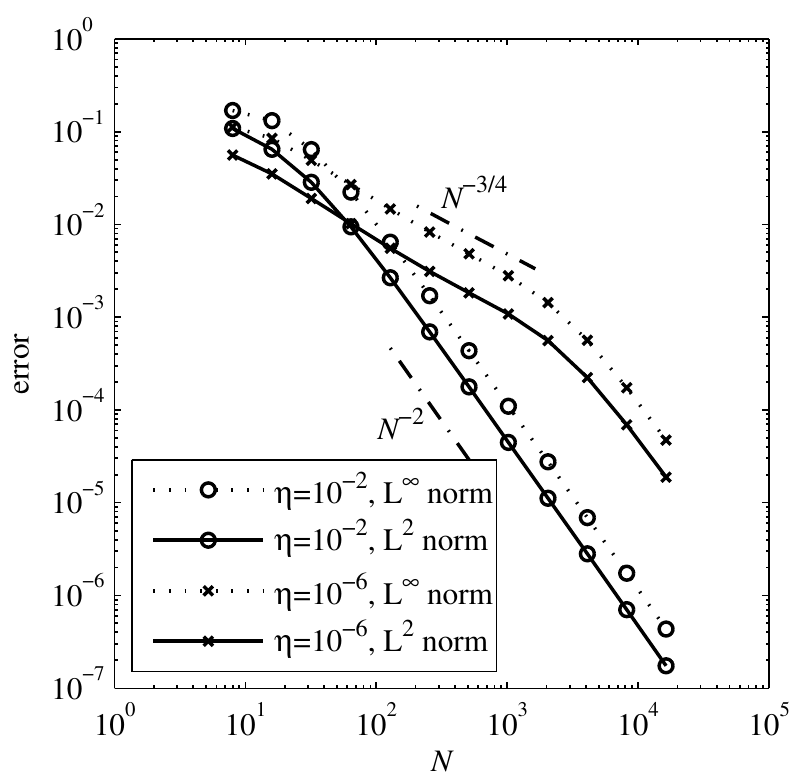}
\caption{\label{fig:fourier_0p5} 
Convergence of the Fourier pseudo-spectral scheme. $m=2$. 
Error with respect to the exact Dirichlet solution in the interior of the fluid domain (left)
and with respect to the the penalized solution in the whole domain (right).
}
\end{center}
\end{figure}

\begin{figure}
\begin{center}
\includegraphics[width=0.49\columnwidth]{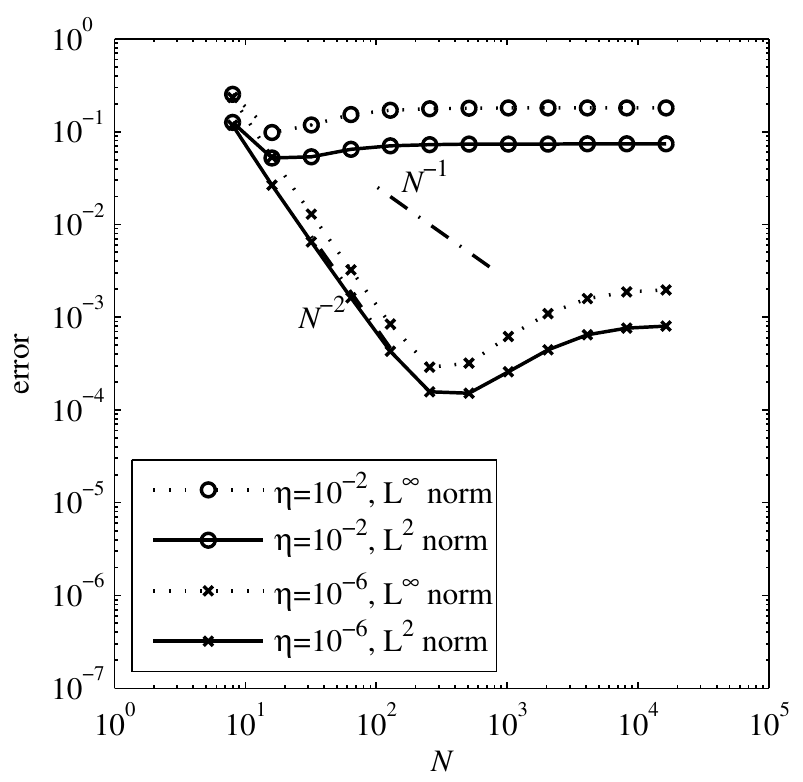}
\includegraphics[width=0.49\columnwidth]{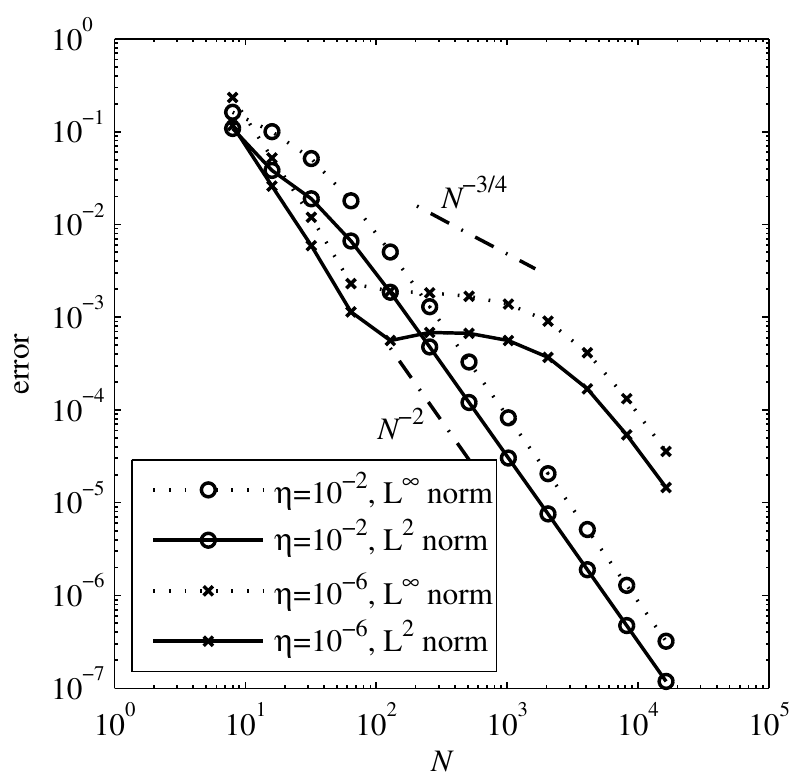}
\caption{\label{fig:fd2_0p5} 
Convergence of the second-order finite difference scheme. $m=2$. 
Error with respect to the exact Dirichlet solution in the interior of the fluid domain (left) 
and with respect to the the penalized solution in the whole domain (right).}
\end{center}
\end{figure}

\begin{figure}
\begin{center}
\includegraphics[width=0.49\columnwidth]{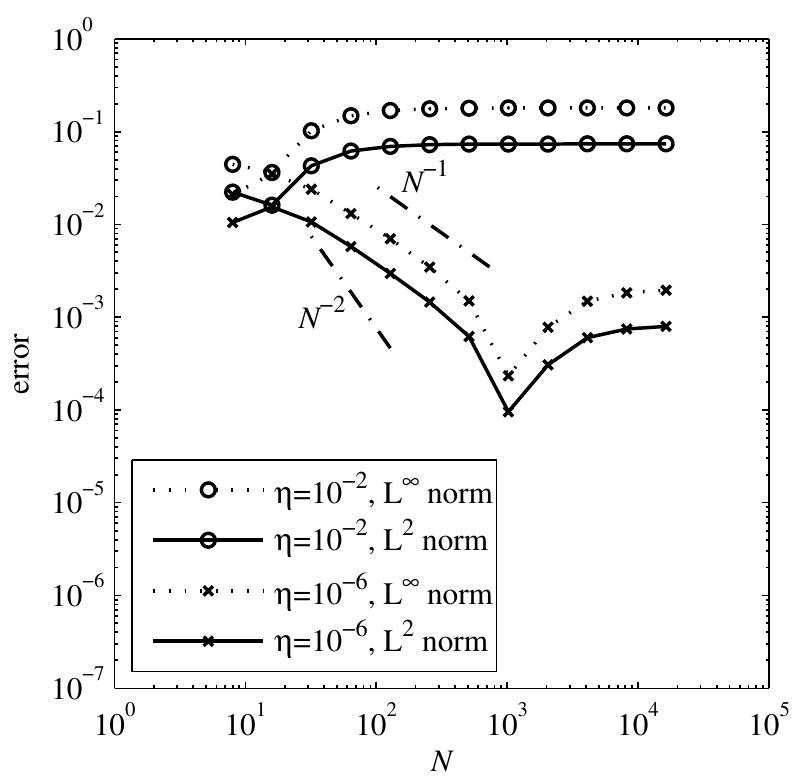}
\includegraphics[width=0.49\columnwidth]{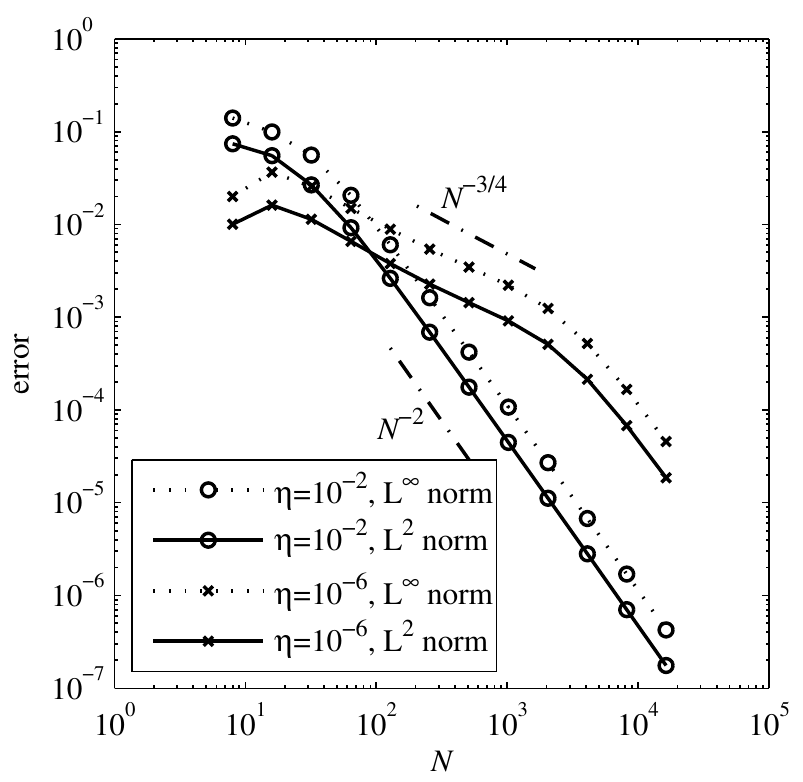}
\caption{\label{fig:fd4_0p5} Convergence of the fourth-order
finite difference scheme. $m=2$. Error with respect to the exact Dirichlet solution in the interior of the fluid domain (left) 
and with respect to the the penalized solution in the whole domain (right).}
\end{center}
\end{figure}

Convergence tests have been performed for three different
discretization schemes employed to solve
(\ref{eq:penalized_poisson_eq}): a pseudo-spectral Fourier,
second- and fourth-order finite difference methods. For the
Fourier spectral method, the second derivative operator is
approximated as
\begin{equation}
  D^2_{F} = \left(\begin{array}{cccccc}
    0 & & & & & -\frac{1}{2}\mathrm{ cotan}{\frac{1h}{2}} \\
    -\frac{1}{2}\mathrm{cotan}{\frac{1h}{2}} & \ddots & & \ddots & & \frac{1}{2}\mathrm{cotan}{\frac{2h}{2}} \\
    \frac{1}{2}\mathrm{cotan}{\frac{2h}{2}} & & \ddots & & & -\frac{1}{2}\mathrm{cotan}{\frac{3h}{2}} \\
    -\frac{1}{2}\mathrm{cotan}{\frac{3h}{2}} & & & \ddots & & \vdots \\
    \vdots & & \ddots & & \ddots & \frac{1}{2}\mathrm{cotan}{\frac{1h}{2}} \\
    \frac{1}{2}\mathrm{cotan}{\frac{1h}{2}} & & & & & 0 \\
  \end{array}\right)^2,
\label{eq:d2_fourier}
\end{equation}
the second-order central finite difference operator is
\begin{equation}
  D^2_{FD2} = \frac{1}{h^2} \left(\begin{array}{ccccc}
    -2 & 1  &   & & 1 \\
    1  & -2 & 1 & & \\
    ~  &    &   & \ddots & \\
    1  &    &   & 1 & -2 \\
  \end{array}\right),
\label{eq:d2_fd2}
\end{equation}
and the fourth-order central finite difference operator is
\renewcommand{\arraystretch}{1.4}
\begin{equation}
  D^2_{FD4} = \frac{1}{h^2} \left(\begin{array}{*{8}{r}}
    -\frac{5}{2}  & \frac{4}{3}  & -\frac{1}{12} &       &       &       & -\frac{1}{12} & \frac{4}{3}   \\
    \frac{4}{3}  & -\frac{5}{2} & \frac{4}{3}   & -\frac{1}{12} &       &       &       & -\frac{1}{12} \\
    -\frac{1}{12} & \frac{4}{3}  & -\frac{5}{2}  & \frac{4}{3}   & -\frac{1}{12} &       &       &       \\
    ~     &      &       &       &       &\ddots &       &       \\
    \frac{4}{3}   & -\frac{1}{12}&       &       &       & -\frac{1}{12} & \frac{4}{3}   & -\frac{5}{2} \\
  \end{array}\right),
\label{eq:d2_fd4}
\end{equation}
where $h=2\pi/N$.
The right-hand side $f(x)$ and the mask function $\chi(x)$ are evaluated at the grid points $x_n = 2 \pi n /N$, $n=0,...,N-1$.
The numerical solution of the penalized equation is then obtained by solving the linear system
\begin{equation}
  ( - D^2 + \frac{1}{\eta} \chi I ) u = f,
\label{eq:num_lin_sys}
\end{equation}
where $I$ is the identity matrix and $D^2$ is the appropriate differentiation matrix. 
Comparison of this numerical solution with the analytical solutions $w$ and $v$ is presented in Figs.~\ref{fig:fourier_0p5},~\ref{fig:fd2_0p5} and~\ref{fig:fd4_0p5}.
The error $e_w$ with respect to the solution of the original boundary value problem $w$ is calculated only in the fluid domain $[0,\pi]$, 
and the error $e$ with respect to the penalized solution $v$ is calculated in the periodic domain $[0,2\pi[$.

In all cases, two regimes are observed, depending on whether the grid is sufficiently dense to approximate the solution in the penalization boundary layer inside the solid. 
For the pseudo-spectral method (figure~\ref{fig:fourier_0p5}), 
the error $e_w$ decays like $N^{-1}$ until $N \approx N_t~=~2/\sqrt{\eta}$, 
when the spectrum is truncated at $k=k_t$ (see (\ref{eq:k_t}) and Fig.~\ref{fig:fourier_coeffs}). 
A faster convergence is observed when the number of grid points is near $N_t$, but with $N$ increasing further on, 
the error starts growing and asymptotically behaves like the penalization error $\epsilon=|v-w^e|$.

The error $e$ also exhibits two distinct regimes, 
but there is no super-convergence in between. 
When $N$ is small, the observations suggest a power law $e \approx \eta^a N^{-3/4}$ with $a \approx 0.071$. 
When $N$ is large enough, $e$ decays like $N^{-2}$ but diverges with $\eta \to 0$ like $1/\sqrt{\eta}$, 
in agreement with (\ref{fig:convergence_error}).

Interestingly, the second-order finite difference enjoys a faster convergence in the intermediate regime (figure~\ref{fig:fd2_0p5}).
Both $e_w$ and $e$ decay like $N^{-2}$ when $N$ is small. 
For large $N$, $e$ decays like $N^{-2}$ and $e_w$ saturates. 

The rapid convergence in this case can be explained as follows.
The second-order finite-difference discretization results
in linear algebraic equations which are exactly the same for the
penalized problem as for the original Dirichlet boundary-value problem
at all points inside the fluid domain, except $x_1$ and $x_{N/2-1}$.
At these two points, an equivalent near-boundary scheme for a
non-homogeneous Dirichlet boundary-value problem can be written as
\begin{equation}
  -\frac{u_2-2u_1+\alpha}{h^2} = f_1 \,, \quad\quad
-\frac{\beta-2u_{N/2-1}+u_{N/2-2}}{h^2} = f_{N/2-1} \,,
\label{eq:fd2_nonhomo_dir}
\end{equation}
where $\alpha=\beta=0$ for the original homogeneous Dirichlet problem and $\max(|\alpha|,|\beta|)<f(h)\sqrt{\eta}$ for the penalized
problem.
Therefore, for a fixed $h$ and sufficiently small $\eta$, the
departure from the homogeneous Dirichlet boundary condition is
negligible
and the rate of convergence of $O(h^2)$ is observed in a wide range of $h$.
The fourth-order finite-difference scheme performs very much like the
spectral method (Fig.~\ref{fig:fd4_0p5}).


\subsection{Concluding remarks}

A Fourier pseudo-spectral discretization of the one-dimensional
penalized Poisson equation has been studied. For a fixed $\eta$,
second-order convergence with $N$ is found with respect to the
exact solution of the penalized problem, as expected from its
regularity. Moreover, the analysis shows that the pre-factor of
this asymptote diverges like $1/\sqrt{\eta}$ as $\eta \to 0$, see
(\ref{eq:norm_error_asy}).

A matter of more practical interest is the convergence of the
numerical solution towards the exact solution of the Dirichlet
boundary value problem. For a fixed $\eta$, the total error (of
penalization and truncation) has a minimum at $N=2/\sqrt{\eta}$.
Varying $\eta$ with $N$ like $\eta=4/N^2$ would allow the error to
decay approximately like $N^{-1}$.


Interestingly, the total error of the second-order
finite-difference discretization decays like $N^{-2}$ in a wide range
of $N$, if there is a grid point coinciding with the boundary and the mask function at
that point is greater than 0.
This happens because the second-order finite-difference discretization
of the penalized problem becomes equivalent to a non-homogeneous
Dirichlet boundary value problem with negligibly small values at the
boundaries.

\section{Positive-definite mollified penalized operators}
\label{appendix_mollification}

Here we explain the details of the smoothing approach which was used in Section~4.
The idea is to approximate $\chi$ by a smooth function prior to discretization
using a Fourier-Galerkin scheme.
We shall denote by $\chi^\#$ smooth approximations of $\chi$. 
We seek one having the following properties:
\begin{enumerate}[label=(\roman{*}), ref=(\roman{*})]
 \item compactly supported Fourier series: $\widehat{\chi^\#}[\kk] = 0$ for $\vert \kk \vert \geq K$,
 \item positivity: $\chi \geq 0$,
 \item consistency: $\chi^\#$ is close to $\chi$ in $L^1(\TT^d)$
 \item fast decay: $\chi^\#$ is of the order of the numerical round-off error inside $\Omega$.
\end{enumerate}
Condition (i) is necessary for a proper Galerkin discretization, as we shall implement further down,
while (ii) preserves the positivity of the mollified operator,
and (iii) keeps the solution close to the one of the original problem.
Condition (iv) avoids oscillations of $\chi^\#$ inside $\Omega$, which could create an unacceptable perturbation of the solution.

To enforce (i), we look for $\chi^\#$ as a convolution 
$\chi^\# = \chi \ast \Phi := \int_{\RR^d} \chi(\yy-\xx) \Phi(\yy) \mathrm{d}\yy$,
where $\Phi$ is in $L^2(\RR^d)$, and the Fourier transform of $\Phi$,
denoted by $\widehat{\Phi}(\kk)$, vanishes for $\vert \kk \vert \geq K$.
Condition (ii) implies that $\Phi$ should be positive, and (iii) means that $\widehat{\Phi}$ should be as close to $1$ as possible.
A first idea would be to take $\widehat{\Phi}(\kk) = 1$ if $\vert \kk \vert \leq K$ and $0$ otherwise,
but that would contradict (ii) and moreover the resulting $\chi^\#$ would have a very bad localization which enters in conflict with condition (iv).
In fact, condition (iv) imposes that $\Phi$ should decay fast at infinity.
In \cite{Ehm2004}, mollifiers having optimal localization (in the sense that their variance 
is minimal in a certain class of functions) were constructed as follows.
Let $J_0$ be the zeroth order Bessel function of the first kind, $j_0$ its smallest positive root, and define
\begin{equation}
\widehat{\phi}(\kk) = 
\begin{cases} 
J_0(j_0 \vert \kk \vert) \rm {\ for\ } \vert \kk \vert \leq 1 \\ 
0 \rm{\ for\ } \vert \kk \vert  > 1.
\end{cases}
\end{equation}
By construction, $\phi$ is well localized in the sense that its variance $\int_{\mathbb{R}^d} \vert \xx \vert^2 \phi(\xx) d\xx$ is as small as possible,
but on the other hand it decays quite slowly at infinity. 
Indeed, $\widehat{\phi}$ is continuous on the circle $\vert \kk \vert = 1$ but not $C^1$,
therefore $\phi(\xx)$ decays at best like $\vert \xx \vert^{-1.5}$.
Moreover, $\phi$ is not positive.
Both issues can be dealt with by taking
\begin{equation}
\Phi(\xx) = C \phi\left(\frac{K\xx}{2b}\right)^{2b}, 
\end{equation}
where $b > 1$ and $C$ is a normalization constant.
Such a $\Phi$ satifies all the conditions that we have imposed, as long as $b$ is chosen properly.
In the following, we call $\Phi$ a Bessel mollifier.

\begin{figure}
\includegraphics[width=0.49\columnwidth,clip=false]{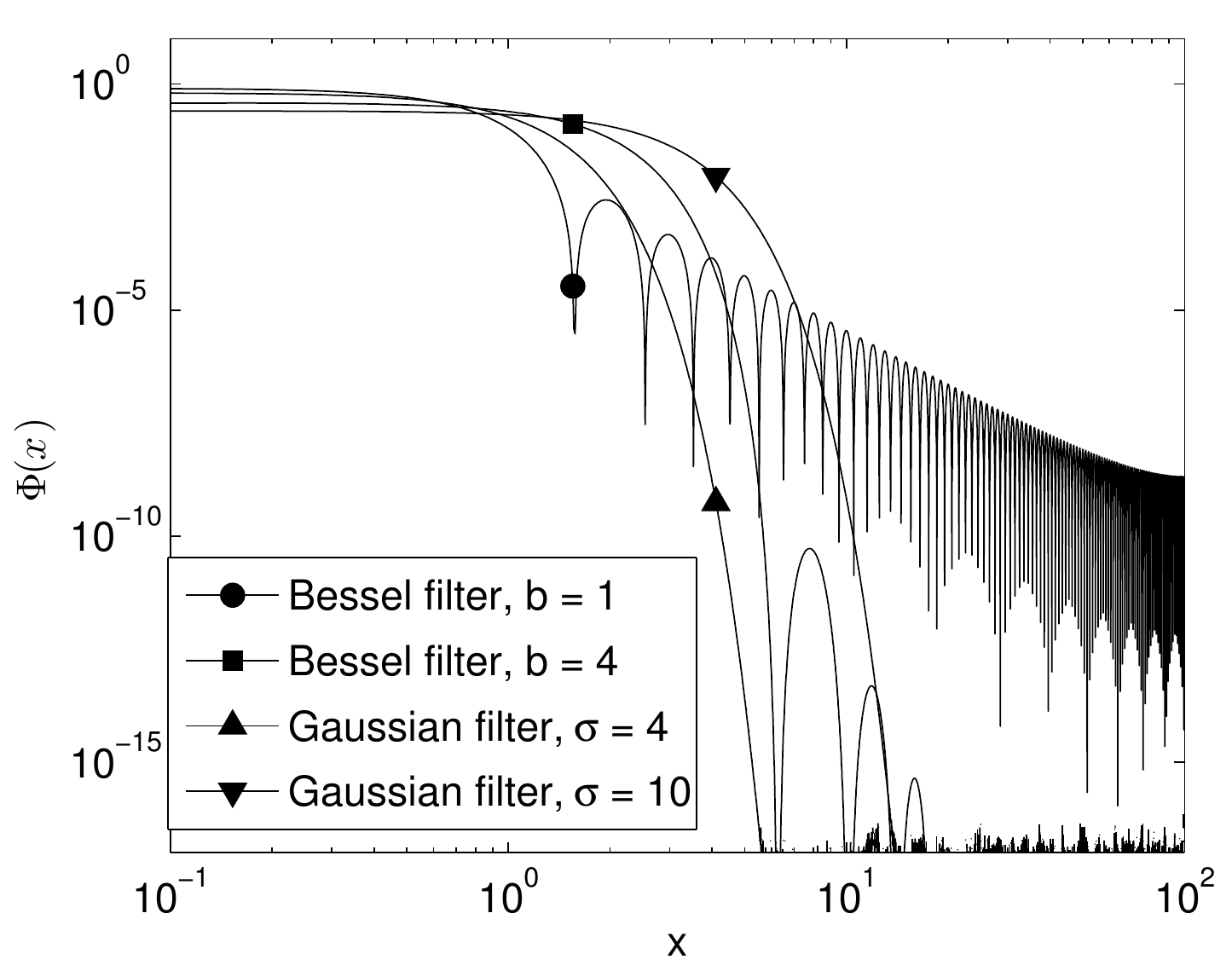}
\includegraphics[width=0.49\columnwidth,clip=false]{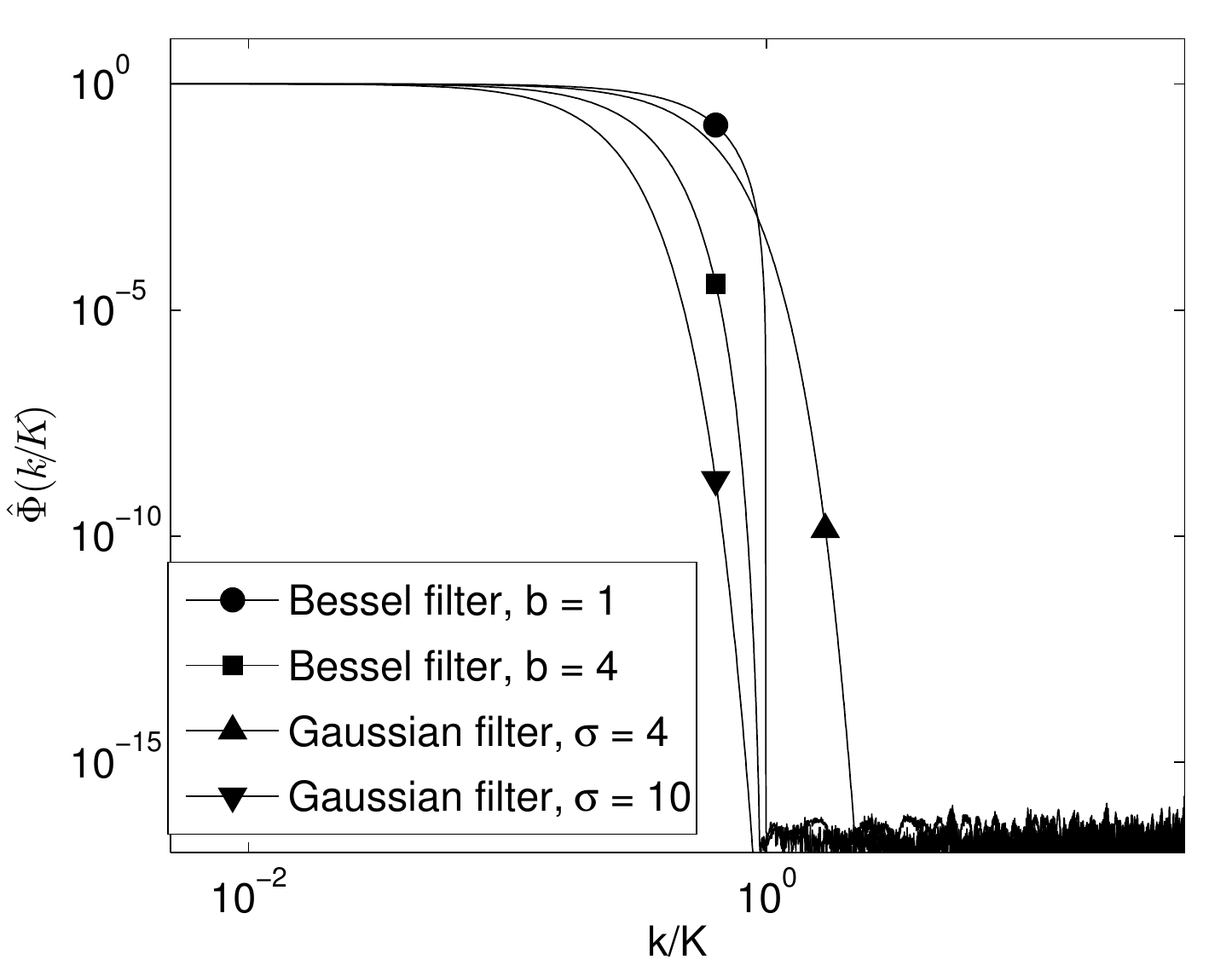}
\caption{
\label{fig:mask_smoothing_filters}
Comparison of Bessel and Gaussian mollifiers, in physical space (left) and Fourier space (right).
In physical space, the horizontal scale is expressed in number of grid points.
}
\end{figure}

To choose $b$ we resort to a qualitative judgement, based on Fig.~\ref{fig:mask_smoothing_filters},
which shows the profiles of $\Phi$ and $\widehat{\Phi}$ for $b=1$ and $b=4$.
For simplicity $K$ has been scaled to $1$.
We see that $b=4$ is a reasonable choice, since $\Phi$ decays quickly below
the round-off error, and the Bessel mollifier with $b=4$ roughly extends over $18$ grid points.
For comparison, Gaussian mollifiers are also shown, for which $\Phi(\kk) = \exp\left(-\frac{1}{2}\sigma^2\kk^2\right)$.
Gaussian mollifiers can be employed, but they are not exactly compactly supported in Fourier space, and moreover their localization in physical space is 
not as good as Bessel mollifiers for a given cut-off wavenumber.

\bibliography{biblio}
\bibliographystyle{siam}

\end{document}